\documentclass[]{IEEEtran}
\IEEEoverridecommandlockouts
\usepackage[space]{cite}
\usepackage{authblk}
\usepackage{amsmath, amsthm, amssymb,amsfonts, bm}
\usepackage{mathtools}
\usepackage{tikz}
\usepackage{caption}
\usepackage{subcaption}
\newtheorem{theorem}{Theorem}
\newtheorem{lemma}[theorem]{Lemma}

\newtheorem{prop}[theorem]{Proposition}
\usepackage{algorithm}
\usepackage{algorithmic}

\usepackage{array}
\usepackage{graphicx}
\usepackage{textcomp}
\usepackage{xcolor}
\usepackage{adjustbox}
\usepackage{fancyhdr}
\definecolor{peachfill}{RGB}{255, 230, 204}
\definecolor{darkorangeborder}{RGB}{215, 155, 0}
\definecolor{pinkfill}{RGB}{248, 206, 204}
\definecolor{darkredborder}{RGB}{184, 84, 80}
\definecolor{bluefill}{RGB}{218, 232, 252}
\definecolor{blueborder}{RGB}{108, 142, 191}
\definecolor{greenfill}{RGB}{213, 232, 212}
\definecolor{greenborder}{RGB}{130, 179, 102}
\definecolor{purplefill}{RGB}{225, 213, 231}
\definecolor{purpleborder}{RGB}{150, 115, 166}
\DeclareRobustCommand{\rchi}{{\mathpalette\irchi\relax}}
\newcommand{\irchi}[2]{\raisebox{\depth}{$#1\chi$}}
\usepackage{nomencl}

\usepackage{hyperref}
\hypersetup{
    colorlinks=true,
    linkcolor=blue,
    filecolor=magenta,      
    urlcolor=cyan,
    pdftitle={Overleaf Example},
    pdfpagemode=FullScreen,
    }

\fancypagestyle{firststyle}
{
   \fancyhf{}
   \fancyfoot[C]{\scriptsize \textcopyright 2023 IEEE. Personal use of this material is permitted.  Permission from IEEE must be obtained for all other uses, in any current or future media, including reprinting/republishing this material for advertising or promotional purposes, creating new collective works, for resale or redistribution to servers or lists, or reuse of any copyrighted component of this work in other works.}
}

\begin{document}
\bstctlcite{IEEEexample:BSTcontrol}
\setlength\tabcolsep{2.4pt}

\title{\textcolor{black}{A Compound Gaussian Least Squares Algorithm and Unrolled Network for Linear Inverse Problems}
}
\author[1, 2]{Carter Lyons\thanks{carter.lyons@colostate.edu}
}
\author[1]{Raghu G. Raj\thanks{raghu.raj@nrl.navy.mil}
}
\author[2]{Margaret Cheney\thanks{margaret.cheney@colostate.edu}\thanks{This work was sponsored by the Office of Naval Research via the NRL base program.}\thanks{This material is based upon research supported in part by, the U. S. Office
of Naval Research under award number  N00014-21-1-2145 and by the Air Force Office of Scientific Research under award number FA9550-21-1-0169.}
}
\affil[1]{U.S. Naval Research Laboratory, Washington, D.C.}
\affil[2]{Colorado State University, Fort Collins, CO}

\maketitle
\thispagestyle{firststyle}

\begin{abstract}
For solving linear inverse problems, particularly of the type that appears in tomographic imaging and compressive sensing, this paper develops two new approaches.   The first approach is an iterative algorithm that minimizes a regularized least squares objective function where the regularization is based on a compound Gaussian prior distribution.   The compound Gaussian prior subsumes many of the commonly used priors in image reconstruction, including those of sparsity-based approaches. The developed iterative algorithm gives rise to the paper’s second new approach, which is a deep neural network that corresponds to an ``unrolling" or ``unfolding" of the iterative algorithm.   Unrolled deep neural networks have interpretable layers and outperform standard deep learning methods. This paper includes a detailed computational theory that provides insight into the construction and performance of both algorithms. The conclusion is that both algorithms outperform other state-of-the-art approaches to tomographic image formation and compressive sensing, especially in the difficult regime of low training.     

\end{abstract}

\begin{IEEEkeywords}
Machine learning, neural networks, inverse problems, nonlinear programming, least squares methods
\end{IEEEkeywords}

\makenomenclature

\nomenclature[01]{$\mathbb{R}$}{Set of real numbers.}
\nomenclature[02]{$\pmb{y}$}{$=[y_i]\in\mathbb{R}^n$. Boldface characters are vectors.}
\nomenclature[03]{$Y$}{$ = [Y_{ij}]\in\mathbb{R}^{n\times m}$. Uppercase characters are matrices.}
\nomenclature[04]{$(\cdot)^T$}{Transpose of vector or matrix $(\cdot)$.}
\nomenclature[05]{$\odot$}{Hadamard product.}
\nomenclature[06]{$D\{\pmb{v}\}$}{Diagonal matrix with $v_1, v_2,\ldots, v_n$ on the diagonal.}
\nomenclature[07]{$A_{\pmb{v}}$}{$=AD\{\pmb{v}\}$ for matrix $A$ of compatible size.}
\nomenclature[08]{$f(\pmb{v})$}{$=[f(v_i)]$ for componentwise function $f:\mathbb{R}\to\mathbb{R}$.}
\nomenclature[09]{$\scalebox{.92}{$\mathcal{P}$}_{a,b}\scalebox{.85}{$(x)$}$}{$=a+\textnormal{ReLU}(x-a)-\textnormal{ReLU}(x-b)$, for $a, b\in\mathbb{R}$, is a modified ReLU (mReLU) activation function.}
\printnomenclature

\section{Introduction}


\IEEEPARstart{M}{otivated} by the success of machine learning (ML) in image recognition and advances in deep learning, ML has been extended to various inverse problems --including the signal reconstruction problems of tomographic imaging and compressive sensing. Improving signal reconstruction, the inverse problem of recovering a signal from undersampled measurements, is an active area of research as it has widespread applicability including radar, sonar, medical, and tomographic imaging. Two prevalent classes of approaches are used in practice for signal reconstruction, namely model-based and data-driven approaches.

First, model-based approaches incorporate certain assumptions, such as the expected prior density or sparsity level of the coefficients, into the signal estimation method. Often, model-based methods apply an iterative algorithm designed to minimize an objective function. Examples include Iterative Shrinkage and Thresholding (ISTA), Bayesian Compressive Sensing (BCS), Basis Pursuit, and Compressive Sampling Matching Pursuit (CoSaMP)~\cite{Compressive_Sensing, fast_ISTA, BCS, CoSamp}. Many previous works model the prior density, of the signal coefficients, as a generalized Gaussian and solve a corresponding least squares problem with $\ell_p$ regularization~\cite{fast_ISTA, ISTA, l1ls, CoSamp, needell2010signal}. Alternatively, some previous works consider a compound Gaussian (CG) prior~\cite{HB-MAP, fast_HB-MAP}, which is a class of densities subsuming generalized Gaussian and other densities as special cases, that better captures statistical properties of the coefficients of images~\cite{Scale_Mixtures, Wavelet_Trees, chance2011information}.

Second, data-driven approaches learn the signal reconstruction mapping directly by training a standard deep neural network (DNN) on pairs of undersampled signal measurements and signal coefficients. Common DNNs in signal reconstruction are structured upon a convolutional neural network (CNN), recurrent neural network, or a generative adversarial network~\cite{reconnet, he2020iRadonMAP, jin2017FBPConvNet, wang2020deep, qin2018convolutional, ganbora, liang2020deep, lucas2018using}.

Algorithm unrolling is a recent approach to signal reconstruction, stemming from the original work of Gregor and LeCun~\cite{learned_ISTA}, which combines the model-based and data-driven methods by structuring the DNN layers based on an iterative algorithm. Unlike a standard DNN, which acts as a black-box process, we have an understanding of the inner workings of an unrolled DNN from understanding the original iterative algorithm. Works utilizing algorithm unrolling have shown excellent performance in signal reconstruction while offering simple interpretability of the network layers~\cite{algorithm_unrolling}.  Examples of iterative imaging algorithms that have been unrolled include: ISTA~\cite{MADUN, zhang2018ista, learned_ISTA, xiang2021FISTANet}, \textcolor{black}{proximal gradient or gradient descent~\cite{learned_proximal_operators, deep_priors, zhang2022learn++}, the inertial proximal algorithm for nonconvex optimization~\cite{su2020iPianoNet}, and the primal-dual algorithm~\cite{adler2018LPD}.} 

\textcolor{black}{While many inverse problems are non-linear, we focus on linear inverse problems as this is frequently the assumed measurement model in many applications -- including compressive sensing, computed tomography (CT), and magnetic resonance imaging (MRI) -- and leave non-linear inverse problem applications to future work.}
Let $\pmb{x}\in\mathbb{R}^n$ be a vectorized signal that has a representation $\pmb{x} = \Phi \pmb{c}$ with respect to (w.r.t) a dictionary, $\Phi\in\mathbb{R}^{n\times n}$, and coefficients, $\pmb{c}\in\mathbb{R}^n.$ An example, $\Phi$ is a wavelet transform and $\pmb{c}$ the wavelet coefficients. The linear measurement model is
\begin{equation}
    \pmb{y} = \Psi\Phi\pmb{c} + \pmb{\nu} \label{eqn:linear_msrmt}
\end{equation}
where $\pmb{y}\in\mathbb{R}^m$ are the measurements produced by observing $\pmb{x}$ through a sensing matrix, $\Psi\in\mathbb{R}^{m\times n}$, with additive white noise, $\pmb{\nu}\in\mathbb{R}^m$. For many applications of interest $m\ll n$. Signal reconstruction, or estimation, \textcolor{black}{aims to recover $\pmb{c}$ given $\pmb{y}$, $\Psi$, and $\Phi$.}

\subsection{Contributions}

\textcolor{black}{In this work, we expand on initial work from~\cite{APSIPAlyonsrajcheney, Asilomarlyonsrajcheney} by:}
\begin{enumerate}
    \item Construct a novel iterative signal estimation algorithm, named compound Gaussian least squares (CG-LS), to solve (\ref{eqn:linear_msrmt}) with general $\Psi$ and $\Phi$ matrices. A regularized least squares (RLS) optimization is the foundation for CG-LS where the regularization enforces a CG prior on the signal coefficients. We provide experimental results illustrating the effectiveness of CG-LS in \textcolor{black}{linear} tomographic imaging.   
    \item \textcolor{black}{Develop a DNN by unrolling CG-LS. The new network, which we name CG-Net, is, to the best of our knowledge, the first DNN for general linear inverse problems to be fundamentally informed by a CG prior. The efficacy of CG-Net to reconstruct images after training is evaluated and shown to outperform other state-of-the-art deep learning methods in the low training scenario.}
    \item Provide fundamental characterization of the existence and location of minimizers of the CG-LS cost function. 
    \item Derive convergence analysis of CG-LS to stationary points of the cost function under a two-block coordinate descent with one block estimated via steepest descent.
\end{enumerate}

\textcolor{black}{Specifically, as compared to~\cite{APSIPAlyonsrajcheney, Asilomarlyonsrajcheney}, our work here generalizes the CG-LS and CG-Net implementations by replacing Newton's descent with a generic steepest descent option providing greater flexibility and learning capacities. Additionally, theoretical analysis and a thorough empirical validation of CG-LS, absent from~\cite{APSIPAlyonsrajcheney, Asilomarlyonsrajcheney}, is provided here. Finally, significant additional validation of CG-Net, including the use of other datasets and evaluating on alternative reconstruction problems, is supplied here with vastly more comparisons to prior art.}

 \subsection{Compound Gaussian Prior}
 
  A fruitful way to formulate inverse problems in imaging is by Bayesian estimation. In particular, consider the RLS estimate of $\pmb{c}$ from (\ref{eqn:linear_msrmt}) $\widehat{\pmb{c}} = \arg\min \,\, ||\pmb{y}-\Psi\Phi\pmb{c}||_2^2 + R(\pmb{c}),$ which can, equivalently, be viewed as a  maximum a posteriori (MAP) estimation when the regularization satisfies $R(\pmb{c})\propto \log(p(\pmb{c}))$. Therefore, the choice of regularization, $R(\pmb{c})$, or prior density, $p(\pmb{c})$, of the coefficients, $\pmb{c}$, is a crucial component to incorporate domain level knowledge into the signal reconstruction problem. Many previous works have employed a generalized Gaussian prior including a Gaussian prior~\cite{bertero1988linear, ying2004tikhonov}, corresponding to a Tikhonov regression, or a Laplacian prior~\cite{Compressive_Sensing, ISTA, fast_ISTA, l1ls, CoSamp,  bioucas2007new}, as is predominant in the compressive sensing (CS) framework. Additional regularizations, not derived from a specific prior density, have been implemented for signal reconstruction including total variation norm~\cite{dahl2010algorithms,  bioucas2007new}, \textcolor{black}{stochastic based regularization~\cite{zhang2022stochastic},} and deep learning-based regularization~\cite{wang2020deep, learned_proximal_operators, deep_priors}.

Through the study of the statistics of image sparsity coefficients, it has been shown that coefficients of natural images exhibit self-similarity, heavy-tailed marginal distributions, and self-reinforcement among local coefficients~\cite{Wavelet_Trees}. Such properties are not encompassed by the generalized Gaussian prior typically assumed for the image coefficients. Instead, a class of densities known as CG densities~\cite{compound_gaussian}, or Gaussian scale mixtures~\cite{Scale_Mixtures, Wavelet_Trees}, better capture statistical properties of natural images and signals from other modalities such as radar~\cite{chance2011information, waveform_opt}. 
A useful formulation of the CG prior lies in modeling the coefficients of signals as
\begin{equation}
    \pmb{c} = \pmb{z}\odot\pmb{u} \label{eqn:CG}
\end{equation}
such that $\pmb{z} = h(\pmb{\rchi})$, where $h:\mathbb{R}\to\mathbb{R}$ is a componentwise, positive, nonlinear function, $\pmb{\rchi}$ follows a multi-scale Gaussian tree process~\cite{Wavelet_Trees,HB-MAP}, $\pmb{u}\sim \mathcal{N}(\pmb{0},\Sigma_u)$, and $\pmb{u}$ and $\pmb{z}$ are independent random variables. In Appendix \ref{apndx:compound gaussian representations}, Proposition \ref{prop:CG subsumes} shows the CG prior subsumes many well-known distributions including the generalized Gaussian. Additionally, Proposition \ref{prop:Laplace as CG} in Appendix \ref{apndx:compound gaussian representations} provides the specific nonlinearity, $h$, producing a Laplace prior allowing an interpretation of the CG prior as a generalization of CS work.

Previously, the CG prior has been used, with $h(x) = \sqrt{\exp(x/\alpha)}$ for $\alpha \in (0,\infty)$, in a hierarchical Bayesian MAP estimate of wavelet and discrete cosine transformation (DCT) coefficients~\cite{HB-MAP, fast_HB-MAP}. This algorithm produces reconstructed images with superior quality, measured by the structural similarity index (SSIM), over other state-of-the-art CS techniques~\cite{HB-MAP}. Furthermore, the CG prior, under a single random scale variable has been successfully used for image denoising~\cite{portilla2003image} and hyperspectral image compressive sensing~\cite{huang2021deep}.

\subsection{Deep Neural Networks and Algorithm Unrolling}

\textcolor{black}{A DNN is a collection of ordered layers, denoted $\bm{L}_0, \bm{L}_1, \ldots, \bm{L}_K$ for $K > 1$, where successive layers feed into one another from the input layer, $\bm{L}_0$, to the output layer, $\bm{L}_K$. Intermediate layers $\bm{L}_1, \ldots, \bm{L}_{K-1}$ are known as hidden layers. Each layer, $\bm{L}_k$, contains $d_k$ hidden units~\cite{goodfellow2016deep}, which are assigned a computed real-value when a signal is transmitted through the DNN.} 

\textcolor{black}{A function $\bm{f}_k:\mathbb{R}^{d_{i_1(k)}}\times \cdots \times \mathbb{R}^{d_{i_j(k)}} \to \mathbb{R}^{d_k}$, that is parameterized by some $\bm{\theta}_k$, defines the computation, i.e. signal transmission, at layer $\bm{L}_k$ where $\mathcal{I}_k\coloneqq\{i_1(k),\ldots, i_j(k)\}\subseteq \{0,1,\ldots, k-1\}$ are the indices of layers that feed into $\bm{L}_k$. That is, given an input signal $\widetilde{\bm{y}}\in\mathbb{R}^{d_0}$ assigned to $\bm{L}_0$, a DNN is a composition of parameterized vector input and vector output functions where
 \[
\bm{L}_k \equiv \bm{f}_k\left(\bm{L}_{i_1(k)}, \ldots, \bm{L}_{i_j(k)}; \bm{\theta}_k\right) \in\mathbb{R}^{d_k}.
\]
}
Fully connected networks, as an example, use $\mathcal{I}_k = \{k-1\}$ and $\pmb{L}_k \equiv \pmb{f}_k(\pmb{L}_{k-1}; \pmb{\theta}_k = [W_k,\pmb{b}_k]) = \sigma(W_k\pmb{L}_{k-1}+\pmb{b}_k)$ where $W_k\in\mathbb{R}^{d_{k-1}\times d_k}$, $\pmb{b}_k\in\mathbb{R}^{d_k}$, and $\sigma$ is a componentwise activation function.

A DNN learns, or trains, its parameters, $\pmb{\Theta} = (\pmb{\theta}_1, \ldots, \pmb{\theta}_K)$, by optimizing a loss function $\mathcal{L}(\pmb{\Theta})$ over a training dataset $\mathcal{D} = \{(\widetilde{\pmb{y}}_i, \widetilde{\pmb{c}}_i): i = 1, 2, \ldots, N_s\}$ where each $(\widetilde{\pmb{y}}_i,\widetilde{\pmb{c}}_i)$ satisfies equation (\ref{eqn:linear_msrmt}).  Let $\widetilde{\pmb{c}}(\widetilde{\pmb{y}}_i;\pmb{\Theta})$ denote the DNN output given the input $\widetilde{\pmb{y}}_i.$ Then the loss function is defined as $\mathcal{L}(\pmb{\Theta}) \coloneqq \frac{1}{N_s}\sum_{i = 1}^{N_s} L\left(\widetilde{\pmb{c}}(\widetilde{\pmb{y}}_i;\pmb{\Theta}), \widetilde{\pmb{c}}_i\right)$ where $L\left(\widetilde{\pmb{c}}(\widetilde{\pmb{y}}_i;\pmb{\Theta}), \widetilde{\pmb{c}}_i\right)$ is the loss, or error, between the network output, $\widetilde{\pmb{c}}(\widetilde{\pmb{y}}_i;\pmb{\Theta})$, and the actual coefficients, $\widetilde{\pmb{c}}_i$. Common loss functions for image reconstruction neural networks include mean-squared error (MSE), \textcolor{black}{mean-absolute error (MAE)}, peak signal-to-noise ratio (PSNR), or SSIM~\cite{NN_loss}.

Algorithm unrolling creates a DNN by assigning the operations from each step $k$ of the iterative algorithm as the function $\pmb{f}_k$ defining layer $k$. That is, layer $k$ in the DNN should correspond to the output of $k$ iterations of the original iterative algorithm. Then parameters, $\pmb{\theta}_k$, on each step $k$ of the iterative algorithm parameterize $\pmb{f}_k$ in the DNN~\cite{learned_ISTA, algorithm_unrolling}. In training the unrolled DNN, each $\pmb{\theta}_k$ is learned, which optimizes the iterative algorithm to produce improved signal estimates.

\section{Iterative Algorithm (CG-LS)} \label{sec:interative_alg}

Let $h$ be the componentwise, invertible, nonlinear function in the CG prior and $f = h^{-1}$. Defining $A = \Psi\Phi$, we consider the RLS cost function and estimate given respectively by
  \begin{align}
        F(\pmb{u},\pmb{z}) = ||\pmb{y}-A(&\pmb{z}\odot\pmb{u})||_2^2 + \lambda ||\pmb{u}||_2^2 + \mu ||f(\pmb{z})||_2^2 \label{eqn:cost function} \\
          \begin{bmatrix}
  \pmb{u}^* &  \pmb{z}^*
  \end{bmatrix} &= \underset{[\pmb{u}\,\,\pmb{z}]}{\arg\min}\,\, F(\pmb{u},\pmb{z}).  \label{eqn:iterative_cost_func}
    \end{align}
Our CG-LS, given in Algorithm \ref{alg:CG-LS}, is an iterative algorithm that, approximately, solves (\ref{eqn:iterative_cost_func}). The cost function (\ref{eqn:cost function}) is a RLS where, as given by the CG prior, the coefficients are decomposed as $\pmb{c} = \pmb{z}\odot\pmb{u}$ and the regularization is taken to be $R(\pmb{c}) = R(\pmb{u},\pmb{z}) = \lambda ||\pmb{u}||_2^2 + \mu ||f(\pmb{z})||_2^2$ to enforce normality of $\pmb{u}$ and $\pmb{\rchi} = f(\pmb{z})$, a Gaussian tree process, as desired from the CG prior. We note that $R(\pmb{c})$ is not exactly proportional to $\log(p(\pmb{c}))$ as specified in the MAP estimate. Instead, $R(\pmb{c})$ is an approximation capturing important statistical properties of the CG prior while also simplifying the optimization.

Due to the explicit joint estimation in (\ref{eqn:iterative_cost_func}), we optimize by block coordinate descent~\cite{wright2015coordinate}, which on iteration $k$ produces
    \begin{align}
\pmb{z}_k 
    &=  \underset{\pmb{z}\in\mathcal{Z}^n}{\arg\min}\, ||\pmb{y} - A_{\pmb{u}_{k-1}}\pmb{z}||_2^2 + \mu ||f(\pmb{z})||_2^2 \label{eqn:iterative_z_update} \\
\pmb{u}_k 
    &= \underset{\pmb{u}\in\mathbb{R}^n}{\arg\min} \, ||\pmb{y} - A_{\pmb{z}_{k}} \pmb{u}||_2^2 + \lambda ||\pmb{u}||_2^2 \label{eqn:iterative_u_update} 
\end{align}
where $\mathcal{Z}\subseteq [0,\infty)$ is the domain of $f$. Convergence rates of block coordinate descent under different conditions for the cost function, such as convexity, have been proven~\cite{wright2015coordinate, beck2013convergence, grippo2000convergence}. As the optimization in (\ref{eqn:iterative_z_update}) cannot be solved analytically, for most choices of $f$, we implement the steepest descent method to iteratively and approximately solve (\ref{eqn:iterative_z_update}). Recall~\cite{boyd2004convex} that given a norm $||\cdot||$ on $\mathbb{R}^n$ and differentiable function, $g(\pmb{x}):\mathbb{R}^n\to\mathbb{R}$, the steepest descent vector, $\pmb{d}:\mathbb{R}^n\to\mathbb{R}^n$, is
    \begin{align}
    \pmb{d}(\pmb{x}) = ||\nabla g(\pmb{x})||_* \left(\underset{||\pmb{v}||= 1}{\arg\min}\,\,\, \nabla g(\pmb{x})^T\pmb{v}\right) \label{eqn:steepest descent direction}
    \end{align}
where $||\cdot||_*$ is the dual norm given by ${\displaystyle ||\pmb{w}||_* = \max_{||\pmb{v}||=1} \pmb{w}^T\pmb{v}.}$ For instance the Euclidean norm produces $\pmb{d}(\pmb{x}) = -\nabla g(\pmb{x})$.

 Applying steepest descent to (\ref{eqn:iterative_z_update}) we write $\pmb{z}_k^j$ as the estimate of $\pmb{z}$ on steepest descent step $j$ of iteration $k$. For generality, we assume that a different norm may define each steepest descent step as is the case in Newton's descent for a convex cost function with non-constant Hessian. Let $\pmb{d}_k^j = \pmb{d}_k^j(\pmb{z})$ denote the descent vector corresponding to norm $||\cdot||_{(k,j)}$, with dual norm $||\cdot||_{*(k,j)}$, for steepest descent step $j$ of iteration $k$. Thus, $\pmb{z}_k^j$ is given by
\[
    \pmb{z}_k^{j} = \pmb{z}_k^{j-1} + \eta_k^{(j)} \pmb{d}_k^j(\pmb{z}_k^{j-1})
\]
where $\eta_k^{(j)}$ is a step size determined by a backtracking line search~\cite{boyd2004convex}. Note, $\pmb{d}_k^j= \pmb{d}_k^j(\pmb{z}) = \pmb{d}_k^j(\pmb{z};\pmb{u}_{k-1},\pmb{y})$ as $\pmb{d}_k^j$ is parameterized by $\pmb{u}_{k-1}$ and $\pmb{y}$. Let $J$ be the maximum number of steepest descent steps; then, for notation, we take $\pmb{z}_{k}^J = \pmb{z}_k = \pmb{z}_{k+1}^0$ for all $k > 0$.

Equation (\ref{eqn:iterative_u_update}) \textcolor{black}{is} the well-known Tikhonov solution 
\begin{align}
\pmb{u}_k = (A_{\pmb{z}_k}^TA_{\pmb{z}_k} + \lambda I)^{-1}A_{\pmb{z}_k}^T \pmb{y} \hspace{.1cm}\textcolor{black}{\equiv A_{\pmb{z}_k}^T(A_{\pmb{z}_k}A_{\pmb{z}_k}^T + \lambda I)^{-1}\pmb{y}}  \label{eqn:Tikhonov solution}
\end{align}
\textcolor{black}{where the second equality results from the Woodbury matrix identity. Note that we do not calculate the inverse in (\ref{eqn:Tikhonov solution}) and instead solve a system of linear equations.}

  \begin{algorithm}[b]
\caption{Compound Gaussian Least Squares (CG-LS)}\label{alg:CG-LS}
\begin{algorithmic}[1]
\STATE $\pmb{z}_0 = \mathcal{P}_{a,b}(A^T\pmb{y})$ and $\pmb{u}_0 = (A_{\pmb{z}_{0}}^TA_{\pmb{z}_{0}} + \lambda I)^{-1}A_{\pmb{z}_{0}}^T\pmb{y}$ \label{line:initial}
\FOR{$k\in \{1,2,\ldots, K\}$}
\STATE  {\underline{$\pmb{z}$ \textsc{estimation}}:}
\STATE  $\pmb{z}_k^0 = \pmb{z}_{k-1}$
\FOR{$j\in \{1,2, \ldots, J\}$}
\IF{$||\nabla_{\bm{z}} F(\pmb{u}_{k-1},\pmb{z}_k^{j-1})||_{*(k,j)}^2 < \delta$}
    \STATE return $\pmb{z}_k = \pmb{z}_k^{j-1}$
\ENDIF
\STATE Compute step size $\eta_k^{(j)}$ (backtracking line search)
\STATE Compute descent vector $\pmb{d}_k^j = \pmb{d}_k^j(\pmb{z}_k^{j-1};\pmb{u}_{k-1},\pmb{y})$
\STATE  $\pmb{z}_{k}^{j} =
    \pmb{z}_k^{j-1} + \eta_k^{(j)}\pmb{d}_k^j(\pmb{z}_k^{j-1};\pmb{u}_{k-1},\pmb{y})$ \label{line:z update}
\ENDFOR
\STATE $\pmb{z}_{k} = \pmb{z}_k^{J}$ 
\STATE {\underline{$\pmb{u}$ \textsc{estimation}}:}
\STATE $ \pmb{u}_{k} = (A_{\pmb{z}_{k}}^TA_{\pmb{z}_{k}} + \lambda I)^{-1}A_{\pmb{z}_{k}}^T\pmb{y} $ \label{line:u update}
\ENDFOR
\ENSURE{$\pmb{c}^* = \pmb{z}_K\odot \pmb{u}_K$}
\end{algorithmic}
\end{algorithm}

Next, define the initial estimate of $\pmb{z}$ as $\pmb{z}_0 = \mathcal{P}_{a,b}(A^T\pmb{y})$ where the mReLU function $\mathcal{P}_{a,b}$ is applied elementwise to $A^T\pmb{y}$. We remark that $\mathcal{P}_{a,b}$ is a projection operator onto the interval $[\min\{a,b\},\max\{a,b\}]$. This eliminates negative values, as $\pmb{z}$ should have positive components, and limits the maximum values in the initial $\pmb{z}$ estimate for stability. The initial $\pmb{u}$ estimate, denoted as $\pmb{u}_0$, is given by (\ref{eqn:Tikhonov solution}). 

Finally, the gradient, $\nabla_{\bm{z}} F(\pmb{u},\pmb{z})$, and a user-chosen parameter $\delta > 0$ determine convergence of CG-LS. On each steepest descent step $j$ of iteration $k$, we check if $||\nabla_{\bm{z}} F(\pmb{u}_{k-1},\pmb{z}_k^{j-1})||_{*(k,j)} < \delta$. When this holds, we exit the steepest descent steps taking $\pmb{z}_k = \pmb{z}_k^{j-1}$. Once $||\nabla_{\bm{z}} F(\pmb{u}_{k-1},\pmb{z}_k^0)||_{*(k,1)} < \delta$ we say CG-LS has converged and return estimates $\pmb{u}_{k-1}$ and $\pmb{z}_{k-1}$. Otherwise, CG-LS terminates after a user-chosen maximum number of iterations $K$.

\subsection{Existence and Location of Minimizers} \label{sec:existence and location of minimizers}

Here, we discuss the existence of minimizers to (\ref{eqn:cost function}) along with details on their location that can be ensured through a sufficient scaling on the measurements, $\pmb{y}$, or proper choices of CG-LS parameters, $\lambda$ and $\mu$. We remark that (\ref{eqn:cost function}) is strongly convex in the $\pmb{u}$ block, with parameter at least $2\lambda$, and thus has no local maxima.

Throughout the remainder of this paper we will assume that $f:\mathcal{Z}\to\mathbb{R}$ is a $C^2$ function defined on $\mathcal{Z} = (z_{\min},\infty)$. While in theory $f$ should be an invertible function, we will not require this. Additionally, we assume that $f$ is coercive on $\mathcal{Z}$, that is $\underset{z\to z_{\min}}{\lim} f(z)\to\pm\infty$ and $\underset{z\to \infty}{\lim} f(z)\to\pm\infty$. Now, we give a necessary and sufficient condition that any stationary point of (\ref{eqn:cost function}) must satisfy.
\begin{lemma} \label{lemma:z root requirement}
Define $\pmb{v}:\mathcal{Z}^n\to\mathbb{R}^n$ and $\pmb{\mathcal{F}}:\mathcal{Z}^n\to\mathbb{R}^n$ by
\begin{align}
    \pmb{v}(\pmb{z}) &= A^T(A_{\pmb{z}}A_{\pmb{z}}^T+\lambda I)^{-1}\pmb{y} \label{eqn:v function z} \\
    \pmb{\mathcal{F}}(\pmb{z}) &= -2\lambda \pmb{z}\odot\pmb{v}(\pmb{z})\odot\pmb{v}(\pmb{z}) + 2\mu f'(\pmb{z})\odot f(\pmb{z}). \label{eqn:z root requirement}
\end{align} 
Then (\ref{eqn:cost function}) has a stationary point $[\pmb{u}^*,\pmb{z}^*]$ if and only if $\pmb{\mathcal{F}}(\pmb{z}^*) = \pmb{0}$ and $\pmb{u}^* = \pmb{z}^*\odot\pmb{v}(\pmb{z}^*).$
\end{lemma}
\begin{proof}
Solving $\nabla_{\pmb{u}} F(\pmb{u},\pmb{z}) = \pmb{0}$ and applying the Woodbury matrix identity gives $\pmb{u} = \pmb{z}\odot\pmb{v}(\pmb{z})$. Next, applying the Woodbury matrix identity to $\nabla_{\bm{z}} F(\pmb{z}\odot\pmb{v}(\pmb{z}),\pmb{z})$ produces (\ref{eqn:z root requirement}). Therefore, $\nabla_{\pmb{u}} F(\pmb{u},\pmb{z}) = \nabla_{\bm{z}} F(\pmb{u},\pmb{z}) = \pmb{0}$ if and only if $\pmb{z} = \pmb{z}^*$ is a root of (\ref{eqn:z root requirement}) and $\pmb{u} = \pmb{z}^*\odot\pmb{v}(\pmb{z}^*).$
\end{proof}

We remark that if $f(z)^2$ is strictly decreasing on an interval $\mathcal{I}$ then, from Lemma \ref{lemma:z root requirement}, $\pmb{z}^*\in\mathcal{Z}^n\setminus \mathcal{I}^n$ for any stationary point $[\pmb{u}^*,\pmb{z}^*]$ of (\ref{eqn:cost function}). A particular application is when $f$ is invertible and obtains a root, at $z_0$, on $\mathcal{Z}$ then $\pmb{z}^*\in [z_0,\infty)^n$.

Next, we assume that the measurements, $\pmb{y}$, have been scaled by a positive constant $s$. That is, $\pmb{y} = s\widetilde{\pmb{y}}$ for $\widetilde{\pmb{y}}$ given by (\ref{eqn:linear_msrmt}). Under certain conditions and choices of $s, \lambda,$ and $\mu$ we gain significant insight into the location of a minimizer of (\ref{eqn:cost function}).

 \begin{prop} \label{prop:existence of a minimizer}
Let $f^2(z)$ be strictly convex and obtain an interior local minimum, at $z_0$, on $[a, b]\subseteq \mathcal{Z}$. Then there exists positive $s$, $\lambda$, and $\mu$  such that (\ref{eqn:cost function}) has a non-degenerate local minimizer $[\pmb{u}^*,\pmb{z}^*]$ where $\pmb{z}^*\in [z_0,b]^n$ and $\pmb{u}^* = \pmb{z}^*\odot\pmb{v}(\pmb{z}^*)$.
\end{prop}

Proof of Proposition \ref{prop:existence of a minimizer} is given in Appendix \ref{apndx:prop existence of a minimizer}. We remark that Proposition \ref{prop:existence of a minimizer} extends optimization insights from the single variable function $f^2(z)$ to the multivariate cost function (\ref{eqn:cost function}) that involves $\pmb{z}$ and $\pmb{u}$. In particular, Proposition \ref{prop:existence of a minimizer} is useful for choosing the correct scaling constant to match the mReLU interval, which is informed by intervals of convexity and roots of $f^2(z)$, defining the initial estimate of $\pmb{z}$ in Algorithm \ref{alg:CG-LS}. Additionally, the convexity of $f^2(z)$ in conjunction with the scaling law revealed in Proposition \ref{prop:existence of a minimizer} guarantees convexity of (\ref{eqn:iterative_z_update}) and the existence of a unique stationary point within this convex region.

\subsection{Convergence of CG-LS} \label{sec:convergence}
For any norm, $||\cdot||$, on $\mathbb{R}^n$ we remark that a Euclidean lower bound exists. That is, there exists a constant $0 < \gamma \leq 1$ such that $||\cdot||\geq \gamma ||\cdot||_2.$ Thus, for every CG-LS steepest descent norm $||\cdot||_{(k,j)}$ with dual norm $||\cdot||_{*(k,j)}$ we let $\gamma_{(k,j)}$ and $\gamma_{*(k,j)}$ be the respective Euclidean bound constants.

To show Algorithm \ref{alg:CG-LS} converges, we first give a lower bound on the change in the cost function for a steepest descent step on $\pmb{z}$. To simplify notation, define the function $G_k(\pmb{z}) = F(\pmb{u}_{k-1},\pmb{z})$.
  \begin{prop} \label{prop:steepest descent step bound}
  For every $k,j\in \mathbb{N}$, there exists a positive constant, $c_{(k,j)}$, such that Algorithm CG-LS satisfies
  \begin{align}
  G_k\left(\pmb{z}_k^{j-1}\right) - G_k\left(\pmb{z}_k^j\right) \geq c_{(k,j)} \left|\left|\nabla G_k\left(\pmb{z}_k^{j-1}\right)\right|\right|_{*(k,j)}^2. \label{eqn:steepest descent lower bound}
  \end{align}
\end{prop}

A proof of Proposition \ref{prop:steepest descent step bound} is given in Appendix \ref{apndx:prop steepest descent step bound}. From Proposition \ref{prop:steepest descent step bound}, the sequence of cost function values, $\{G_k(\pmb{z}_k^j)\}_{j = 0}^\infty$, monotonically decreases, and when the gradient is large, we expect a large decrease in the cost function. Since $G_k(\pmb{z}) \geq 0$, we know the sequence $\{G_k(\pmb{z}_k^j)\}_{j = 0}^\infty$ converges. That is, each steepest descent process in CG-LS will converge. To show the convergence of CG-LS to a stationary point of (\ref{eqn:cost function}), we apply Proposition \ref{prop:steepest descent step bound} and similarly bound the change in the cost function over an iteration of CG-LS.

\begin{theorem} \label{thm:CG-LS convergence}
Assume the Euclidean bound sequences $\{\gamma_{(k,1)}\}_{k = 1}^\infty$ and $\{\gamma_{*(k,1)}\}_{k = 1}^\infty$ are bounded below by $\gamma > 0$. Then CG-LS converges to a stationary point of (\ref{eqn:cost function}). Furthermore, the sequence of minimum gradient dual norms satisfies
    \[
  \underset{1\leq k\leq {K}}{\min} ||\nabla F(\pmb{u}_k,\pmb{z}_k)||_{*(k,1)}^2 \leq \mathcal{O}\left(\frac{1}{K}\right).
    \]
\end{theorem}

A proof of Theorem \ref{thm:CG-LS convergence} is in Appendix \ref{apndx:thm CG-LS convergence} and shows that CG-LS generates a monotonic decreasing sequence of cost function values. Combining Theorem \ref{thm:CG-LS convergence} with the fact that (\ref{eqn:cost function}) has no local maxima, Algorithm \ref{alg:CG-LS} is guaranteed to converge to a local minimum or saddle point of (\ref{eqn:cost function}). As (\ref{eqn:cost function}) is, in general, not convex, we cannot guarantee that Algorithm \ref{alg:CG-LS} converges to a local nor global minimum. Although, if Proposition \ref{prop:existence of a minimizer} is also satisfied, we can guarantee convergence to a global minimizer of (\ref{eqn:cost function}) relative to $[z_0,b]^n\times \mathbb{R}^n$.

We further strengthen our convergence results when Proposition \ref{prop:existence of a minimizer} is satisfied. Let $\gamma$ be given as in Theorem \ref{thm:CG-LS convergence} and $c$ from (\ref{eqn:CG-LS constant c}).

\begin{theorem} \label{thm:CG-LS converges with convexity}
Let Proposition \ref{prop:existence of a minimizer} be satisfied with local minimizer $[\pmb{u}^*,\pmb{z}^*]$ corresponding to minimum value $F^*$. Then there exists a region $\mathcal{C}\subseteq \mathbb{R}^n\times\mathcal{Z}^n$ such that $[\pmb{u}^*,\pmb{z}^*]\in\mathcal{C}$ and (\ref{eqn:cost function}) is strongly convex on $\mathcal{C}$ with constant $\ell$. Furthermore, for any $[\pmb{u}_0,\pmb{z}_0]\in\mathcal{C}$ we have for all $k\geq 0$
\begin{align}
    F(\pmb{u}_k,\pmb{z}_k) - F^* \leq (1-2\ell \gamma^2 c)^k (F(\pmb{u}_0,\pmb{z}_0) - F^*). \label{eqn:strongly convex CG-LS convergence}
\end{align}
\end{theorem}

A proof of Theorem \ref{thm:CG-LS converges with convexity} is in Appendix \ref{apndx:thm CG-LS converges with convexity}. Theorem \ref{thm:CG-LS converges with convexity} shows a region around the local minimizer where (\ref{eqn:cost function}) is strongly convex exists and thus we can guarantee a linear rate of convergence of the cost function values in this region.

\subsection{Numerical Results} \label{sec:CG-LS numerical results}


\begin{figure}[t]
\centering
\begin{subfigure}{0.15\textwidth}
    \centering
    \includegraphics[width=\textwidth]{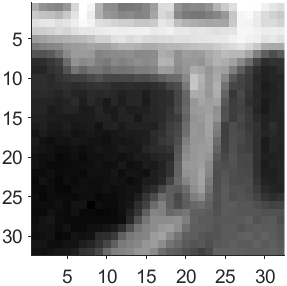}
    \caption{Original}
    \label{fig:exact_barb}
\end{subfigure}
\begin{subfigure}{0.15\textwidth}
    \centering
    \includegraphics[width=\textwidth]{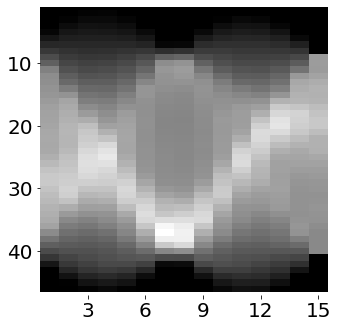}
    \caption{Radon Transform}
    \label{fig:radon_msmnt}
\end{subfigure}
\begin{subfigure}{0.15\textwidth}
    \centering
    \includegraphics[width=\textwidth]{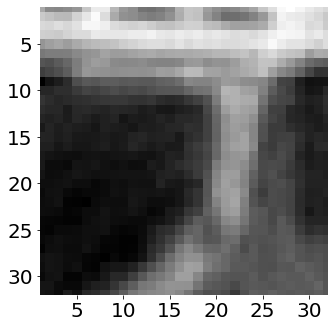}
    \caption{gCG-LS (0.952)}
    \label{fig:gCG-LS_barb}
\end{subfigure}
\begin{subfigure}{0.15\textwidth}
    \centering
    \includegraphics[width=\textwidth]{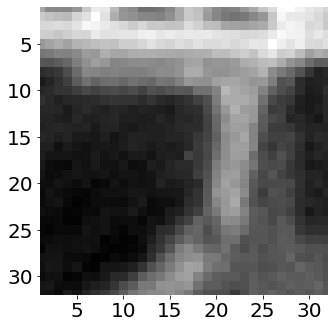}
    \caption{nCG-LS (0.953)}
    \label{fig:nCG-LS_barb}
\end{subfigure}
\begin{subfigure}{0.15\textwidth}
    \centering
    \includegraphics[width=\textwidth]{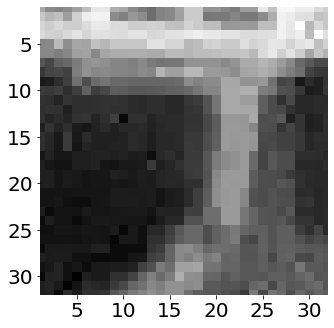}
    \caption{\scalebox{1}{\textcolor{black}{FISTA (0.896)}}}
    \label{fig:FISTA_barb}
\end{subfigure}
\begin{subfigure}{0.15\textwidth}
    \centering
    \includegraphics[width=\textwidth]{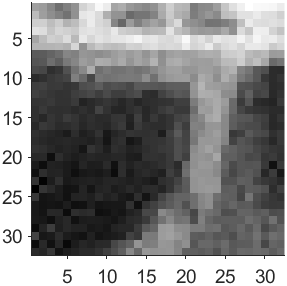}
    \caption{$\ell_1$-LS (0.849)}
    \label{fig:l1ls_barb}
\end{subfigure}
\begin{subfigure}{0.15\textwidth}
    \centering
    \includegraphics[width=\textwidth]{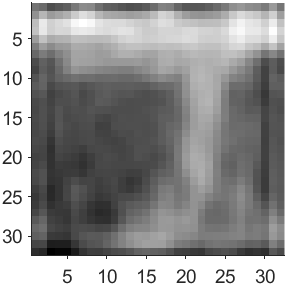}
    \caption{\textcolor{black}{FBP} (0.844)}
    \label{fig:fbp_barb}
\end{subfigure}
\begin{subfigure}{0.15\textwidth}
    \centering
    \includegraphics[width=\textwidth]{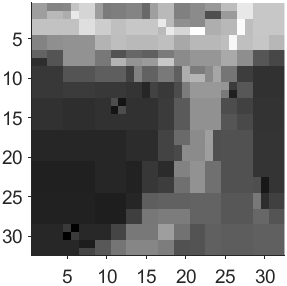}
    \caption{BCS (0.827)}
    \label{fig:bcs_barb}
\end{subfigure}
\begin{subfigure}{0.15\textwidth}
    \centering
    \includegraphics[width=\textwidth]{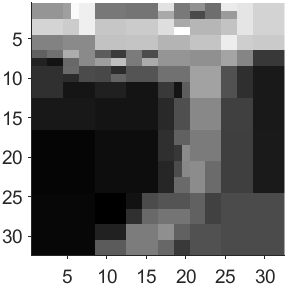}
    \caption{CoSaMP (0.838)}
    \label{fig:cosamp_barb}
\end{subfigure}
\caption{Image reconstructions (SSIM) using our gCG-LS, our nCG-LS, \textcolor{black}{FISTA}, $\ell_1$-LS, \textcolor{black}{FBP}, \textcolor{black}{BCS}, and CoSaMP. The input to each algorithm is a vectorized (\ref{fig:radon_msmnt}), which is \textcolor{black}{a Radon transform of (\ref{fig:exact_barb}) at 15 uniformly spaced angles with an SNR of 60dB.} We observe that CG-LS outperforms all methods with nCG-LS slightly outperforming gCG-LS.}
\label{fig:barbra_reconstruction}
\end{figure}


We test the CG-LS algorithm using gradient descent, which we denote by gCG-LS, and Newton descent, which we denote by nCG-LS, as the steepest descent method for updating $\pmb{z}$. Note, the gradient and Hessian of (\ref{eqn:cost function}) w.r.t $\pmb{z}$ are, respectively:
\begin{align}
    \nabla_{\bm{z}} F(\pmb{u},\pmb{z}) &= -2A_{\pmb{u}}^T(\pmb{y} - A_{\pmb{u}}\pmb{z}) + 2\mu f'(\pmb{z})\odot f(\pmb{z}) \label{eqn:grad z} \\
    H_{F;\pmb{z}}(\pmb{u},\pmb{z}) &= 2A^T_{\pmb{u}}A_{\pmb{u}} + 2\mu D\{\pmb{h}_f(\pmb{z})\} \label{eqn:Hess z}
\end{align}
where
\begin{align}
\pmb{h}_f(\pmb{z}) = f''(\pmb{z})\odot f(\pmb{z}) + f'(\pmb{z})\odot f'(\pmb{z}). \label{eqn:hf}
\end{align}


\begin{table*}[t]
\caption{Average SSIM/PSNR with $99\%$ confidence intervals for our gCG-LS and nCG-LS \textcolor{black}{and five comparative methods}. Each algorithm estimated two hundred, $32\times 32$ and $64\times 64$ images \textcolor{black}{along with 11, $128\times 128$ images all} from Radon transform measurements taken at 15, 10, or 6 uniformly spaced angles. The measurement noise is set at 60dB or 40dB SNR. We find that CG-LS performs best in all noise and sparse scenarios with nCG-LS, in \textbf{bold}, slightly outperforming gCG-LS, in \textit{italics}.}
\centering
\adjustbox{max width=\textwidth}{
\begin{tabular}{lc|c|c|c|c|c}
\multicolumn{7}{c}{SSIM ($\times 10^{2}$)/PSNR for $32\times 32$ \textcolor{black}{CIFAR10} images} \\
\hline
Angles & \multicolumn{2}{c|}{$15$} & \multicolumn{2}{c|}{$10$} & \multicolumn{2}{c}{$6$}  \\
SNR & \multicolumn{1}{c}{$60$ dB}  & \multicolumn{1}{c|}{$40$ dB} &  \multicolumn{1}{c}{$60$ dB}  & \multicolumn{1}{c|}{$40$ dB} &  \multicolumn{1}{c}{$60$ dB}  & \multicolumn{1}{c}{$40$ dB} \\
\hline
\scalebox{1}{nCG-LS} & $\bm{91.3}\pm \bm{.59}$/$\bm{28.0}\pm \bm{.39}$ & $\bm{87.0}\pm \bm{.97}$/$\bm{26.2}\pm \bm{.28}$ & $\bm{83.6}\pm \bm{.82}$/$\bm{24.8}\pm \bm{.39}$ & $\bm{80.6}\pm \bm{.94}$/$\bm{24.0}\pm \bm{.32}$ & $\bm{70.0}\pm \bm{1.1}$/$\bm{21.7}\pm \bm{.37}$ & $\bm{68.0}\pm \bm{1.2}$/$\bm{21.3}\pm \bm{.35}$ \\
\scalebox{1}{gCG-LS} & $\mathit{90.1}\pm \mathit{.57}$/$\mathit{27.1}\pm \mathit{.38}$ & $\mathit{85.9}\pm \mathit{.93}$/$\mathit{25.5}\pm \mathit{.28}$ & $\mathit{82.3}\pm \mathit{.78}$/$\mathit{24.3}\pm \mathit{.40}$ & $\mathit{80.1}\pm \mathit{.86}$/$\mathit{23.8}\pm \mathit{.34}$ & $\mathit{68.9}\pm \mathit{1.1}$/$\mathit{21.4}\pm \mathit{.42}$ & $\mathit{67.9}\pm \mathit{1.1}$/$\mathit{21.3}\pm \mathit{.40}$ \\
\textcolor{black}{FISTA} & \textcolor{black}{$86.6\pm 1.0$/$26.6\pm .29$} & \textcolor{black}{$78.6\pm 1.1$/$23.6\pm .27$} & \textcolor{black}{$81.4\pm .89$/$24.8\pm .39$} & \textcolor{black}{$69.4\pm .97$/$21.7\pm .31$} & \textcolor{black}{$67.9\pm 1.2$/$21.4\pm .39$} & \textcolor{black}{$53.3\pm 1.2$/$19.1\pm .38$} \\
$\ell_1$-LS &  $86.6\pm .58$/$26.0\pm .42$ & $72.7\pm 1.6$/$22.1\pm .35$ & $74.5\pm .93$/$22.8\pm .42$ & $67.2\pm 1.1$/$21.2\pm .29$ & $55.2\pm 1.3$/$19.4\pm .42$ & $52.5\pm 1.2$/$18.9\pm .36$ \\
FBP &  $85.7\pm .77$/$20.6\pm .36$ & $81.2\pm 1.2$/$20.2\pm .35$ & $72.4\pm 1.1$/$14.5\pm .42$ & $68.4\pm 1.3$/$14.3\pm .43$ & $53.4\pm 1.1$/$18.8\pm .46$ & $50.2\pm 1.2$/$18.7\pm .45$ \\
BCS &  $77.4\pm .96$/$22.7\pm .43$ & $74.1\pm .93$/$22.1\pm .34$ & $62.4\pm 1.2$/$19.9\pm .44$ & $60.4\pm 1.2$/$19.6\pm .42$ & $44.6\pm 1.5$/$17.3\pm .45$ & $44.1\pm 1.5$/$17.2\pm .44$ \\
\scalebox{1}{CoSaMP} &  $75.9\pm 1.0$/$22.4\pm .48$ & $71.2\pm .98$/$21.3\pm .34$ & $61.7\pm 1.2$/$19.5\pm .44$ & $58.6\pm 1.1$/$18.9\pm .36$ & $35.5\pm 1.5$/$14.8\pm .48$ & $32.4\pm 1.3$/$14.2\pm .43$
\end{tabular}
}
\adjustbox{max width=\textwidth}{
\begin{tabular}{lc|c|c|c|c|c}
\multicolumn{7}{c}{SSIM ($\times10^{2}$)/PSNR for $64\times 64$ \textcolor{black}{CalTech101} images} \\
\hline
Angles & \multicolumn{2}{c|}{$15$} & \multicolumn{2}{c|}{$10$} & \multicolumn{2}{c}{$6$}  \\
SNR & \multicolumn{1}{c}{$60$ dB}  & \multicolumn{1}{c|}{$40$ dB} &  \multicolumn{1}{c}{$60$ dB}  & \multicolumn{1}{c|}{$40$ dB} &  \multicolumn{1}{c}{$60$ dB}  & \multicolumn{1}{c}{$40$ dB} \\
\hline
\scalebox{1}{nCG-LS} & $\bm{67.5}\pm \bm{1.2}$/$\bm{22.7}\pm \bm{.45}$ & $\bm{63.2}\pm \bm{1.4}$/$\bm{22.0}\pm \bm{.41}$ & $\bm{58.3}\pm \bm{1.3}$/$\bm{21.0}\pm \bm{.43}$ & $\bm{56.0}\pm \bm{1.3}$/$\bm{20.7}\pm \bm{.41}$ & $\bm{46.8}\pm \bm{1.3}$/$\bm{19.0}\pm \bm{.42}$ & $\bm{45.7}\pm \bm{1.3}$/$\bm{18.9}\pm \bm{.42}$ \\
\scalebox{1}{gCG-LS} & $\mathit{65.5}\pm \mathit{1.1}$/$\mathit{22.0}\pm \mathit{.43}$ & $\mathit{61.3}\pm \mathit{1.3}$/$\mathit{21.4}\pm \mathit{.39}$ &  $\mathit{56.9}\pm \mathit{1.2}$/$\mathit{20.5}\pm \mathit{.44}$ & $\mathit{54.4}\pm \mathit{1.2}$/$\mathit{20.3}\pm \mathit{.42}$ &  $\mathit{45.4}\pm \mathit{1.4}$/$\mathit{18.5}\pm \mathit{.47}$ & $\mathit{44.0}\pm \mathit{1.2}$/$\mathit{18.4}\pm \mathit{.46}$ \\
\textcolor{black}{FISTA} &  \textcolor{black}{$63.5\pm 1.2$/$22.6\pm .44$} & \textcolor{black}{$51.2\pm 1.3$/$19.8\pm .40$} & \textcolor{black}{$56.5\pm 1.3$/$21.1\pm .44$} & \textcolor{black}{$42.9\pm 1.0$/$18.5\pm .39$} & \textcolor{black}{$45.5\pm 1.5$/$19.1\pm .44$} & \textcolor{black}{$31.1 \pm 1.0$/$16.5\pm .40$} \\
$\ell_1$-LS &  $61.6\pm 1.3$/$21.5\pm .49$ & $45.3\pm 1.7$/$18.3\pm .46$ & $48.7\pm 1.4$/$19.4\pm .46$ & $40.7\pm 1.1$/$18.0\pm .40$ & $33.7\pm 1.4$/$16.7\pm .44$ & $30.0\pm 1.0$/$16.2\pm .43$ \\
FBP &  $54.6\pm 1.3$/$15.3\pm .57$ & $48.1\pm 1.4$/$14.9\pm .56$ & $41.5\pm 1.1$/$9.33\pm .66$ & $36.2\pm 1.1$/$9.13\pm .65$ & $27.6\pm .81$/$14.4\pm .71$ & $23.6\pm .76$/$14.1\pm .70$ \\
BCS &  $54.3\pm 1.9$/$19.4\pm .49$ & $52.3\pm 1.8$/$19.1\pm .47$ & $40.7\pm 1.9$/$17.2\pm .47$ & $39.6\pm 1.9$/$17.0\pm .46$ & $29.1\pm 1.9$/$15.2\pm .49$ & $28.9\pm 1.9$/$15.2\pm .48$ \\
\scalebox{1}{CoSaMP} &  $54.2\pm 2.0$/$19.2\pm .49$ & $51.5\pm 1.7$/$18.6\pm .45$ & $41.1\pm 2.1$/$16.8\pm .48$ & $40.0\pm 1.9$/$16.6\pm .46$ & $24.8\pm 1.6$/$13.5\pm .47$ & $24.3\pm 1.5$/$13.4\pm .46$
\end{tabular}
}
{\color{black}\adjustbox{max width=\textwidth}{
\begin{tabular}{lc|c|c|c|c|c}
\multicolumn{7}{c}{SSIM ($\times10^{2}$)/PSNR for $128\times 128$ \textcolor{black}{Set11} images} \\
\hline
Angles & \multicolumn{2}{c|}{$15$} & \multicolumn{2}{c|}{$10$} & \multicolumn{2}{c}{$6$}  \\
SNR & \multicolumn{1}{c}{$60$ dB}  & \multicolumn{1}{c|}{$40$ dB} &  \multicolumn{1}{c}{$60$ dB}  & \multicolumn{1}{c|}{$40$ dB} &  \multicolumn{1}{c}{$60$ dB}  & \multicolumn{1}{c}{$40$ dB} \\
\hline
\scalebox{1}{nCG-LS} & $\bm{51.3}\pm \bm{7.7}$/$\bm{20.7}\pm \bm{2.0}$ & $\bm{46.9}\pm \bm{7.6}$/$\bm{19.9}\pm \bm{1.7}$ & $\bm{44.3}\pm \bm{7.4}$/$\bm{19.2}\pm \bm{1.5}$ & $\bm{40.4}\pm \bm{7.1}$/$\bm{18.6}\pm \bm{1.3}$ & $\bm{37.8}\pm \bm{7.9}$/$\bm{17.8}\pm \bm{1.4}$ & $\bm{35.1}\pm \bm{6.7}$/$\bm{17.6}\pm \bm{1.3}$ \\
\scalebox{1}{gCG-LS} & $\mathit{51.3}\pm \mathit{7.8}$/$\mathit{20.7}\pm \mathit{2.0}$ & $\mathit{46.9}\pm \mathit{7.6}$/$\mathit{19.9}\pm \mathit{1.7}$ &  $\mathit{44.3}\pm \mathit{7.4}$/$\mathit{19.2}\pm \mathit{1.5}$ & $\mathit{40.3}\pm \mathit{7.1}$/$\mathit{18.5}\pm \mathit{1.4}$ &  $\mathit{37.8}\pm \mathit{7.9}$/$\mathit{17.8}\pm \mathit{1.4}$ & $\mathit{35.1}\pm \mathit{6.7}$/$\mathit{17.6}\pm \mathit{1.3}$ \\
FISTA & $50.9\pm 7.4$/$20.6\pm 1.9$ & $33.0\pm 3.9$/$17.4\pm 1.3$ & $43.4\pm 7.1$/$19.1\pm 1.5$ & $27.8\pm 4.0$/$16.6\pm 1.2$ & $37.3\pm 8.3$/$17.8\pm 1.4$ & $26.4\pm 4.3$/$16.0\pm .97$ \\
$\ell_1$-LS & $42.6\pm 7.3$/$19.2\pm 1.9$ & $32.9\pm 4.0$/$17.4\pm 1.3$ & $33.5\pm 6.8$/$17.5\pm 1.6$ & $27.3\pm 4.1$/$16.5\pm 1.3$ & $25.6\pm 6.4$/$15.7\pm 1.4$ & $21.7\pm 4.5$/$15.1\pm 1.2$ \\
FBP &  $35.3\pm 4.0$/$11.2\pm .69$ & $27.4\pm 2.9$/$10.6\pm .62$ & $25.0\pm 3.2$/$5.46\pm .55$ & $18.7\pm 2.0$/$5.23\pm .55$ & $15.7\pm 2.5$/$0.85\pm .59$ & $11.2\pm 1.2$/$0.68\pm .59$ \\
BCS & $39.1\pm 11$/$17.7\pm 2.1$ & $37.6\pm 11$/$17.5\pm 1.8$ & $31.7\pm 10$/$16.3\pm 1.4$ & $31.4\pm 9.7$/$16.3\pm 1.4$ & $28.7\pm 10$/$15.1\pm 1.1$ & $28.5\pm 10$/$15.1\pm 1.1$  \\
\scalebox{1}{CoSaMP} &  $38.7\pm 11$/$17.3\pm 2.0$ & $38.0\pm 11$/$17.2\pm 1.9$ & $32.2\pm 10$/$15.6\pm 1.3$ & $32.0\pm 10$/$15.6\pm 1.4$ & $28.9\pm 10$/$14.6\pm 1.0$ & $28.8\pm 10$/$14.5\pm .95$
\end{tabular}
}
}
\label{table:CG-LS}
\end{table*}


We use $32\times 32$ images from the CIFAR10~\cite{CIFAR10} image dataset, $64\times 64$ images from the CalTech101~\cite{CalTech101} image dataset\textcolor{black}{, and the $11$ images from the Set11~\cite{MADUN} dataset resized to $128\times 128$.} Each image has been converted to a single-channel grayscale image, scaled down by the maximum pixel value, and vectorized. A Radon transform, at a specified number of uniformly spaced angles, is performed on each image to which white noise is added producing noisy measurements, $\pmb{y}$, at a specified signal-to-noise ratio (SNR). Finally, a biorthogonal wavelet transformation is applied to each image to produce the coefficients, $\pmb{c}$.

For all simulations, we use $f(z) = \ln(z)$, $\mu = 2$, $K = 1000$, $J = 1$, and $\delta = 10^{-6}.$ For measurements at an SNR of $60$dB and $40$dB, we take $\lambda = 0.3$ and $\lambda = 2$, respectively. \textcolor{black}{We remark that for the $128\times 128$ Set11 reconstructions we instead take $K = 100$ and, in the $40$dB case, $\lambda = 30$.} Using nCG-LS requires (\ref{eqn:Hess z}) to be positive definite in the $\pmb{z}$ variable, which for $f(z) = \ln(z)$, is guaranteed when $\pmb{z}\in (0,e)^n$. Informed by Proposition \ref{prop:existence of a minimizer}, we can guarantee a local minimizer lies in $[1,e)^n$ under sufficient scaling of the input data. Therefore, in each nCG-LS test using $32\times 32$ images, we scale the input measurement by a factor, chosen empirically, of $e^{-4}$. Similarly, in each nCG-LS test on larger images we scale the input measurement by $e^{-6}.$ Additionally, we use an eigendecomposition on (\ref{eqn:Hess z}) to find the closest positive semi-definite matrix that is then used in the Newton descent step. Alternatively, we remark that the mReLU function $\mathcal{P}_{1,e}$ may be applied at each $\pmb{z}$ update in line \ref{line:z update} of Algorithm \ref{alg:CG-LS} to ensure (\ref{eqn:Hess z}) is positive semi-definite. Finally, we choose $\mathcal{P}_{a,b} = \mathcal{P}_{1, e}$ for nCG-LS whereas for gCG-LS we use $\mathcal{P}_{a,b} = \mathcal{P}_{1, e^2}$.

Fig. \ref{fig:barbra_reconstruction} contains the reconstruction of a $32\times 32$ Barbara snippet \textcolor{black}{from seven} iterative algorithms: gCG-LS, nCG-LS, \textcolor{black}{Fast Iterative Shrinkage and Thresholding Algorithm (FISTA)~\cite{fast_ISTA}},  $\ell_1$-least squares ($\ell_1$-LS)~\cite{l1ls}, Fourier backprojection (FBP)~\cite{radontransform}, BCS~\cite{BCS}, and CoSaMP~\cite{CoSamp}. Each algorithm takes as input (\ref{fig:radon_msmnt}) vectorized, which is a 60dB SNR Radon transform at 15 uniformly spaced angles. In (\ref{fig:radon_msmnt}) the angle and distance are given on the $x$ and $y$ axes, respectively.
\begin{figure}[!b]
\centering
\begin{subfigure}{0.15\textwidth}
    \centering
    \includegraphics[width=\textwidth]{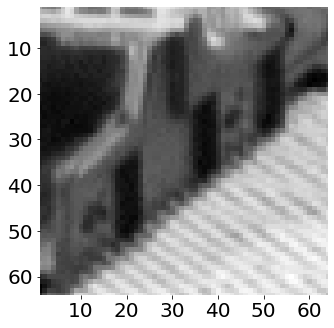}
    \caption{Original}
    \label{fig:exact_barb64}
\end{subfigure}
\begin{subfigure}{0.15\textwidth}
    \centering
    \includegraphics[width=\textwidth]{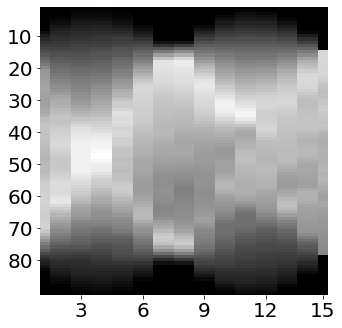}
    \caption{Radon Transform}
    \label{fig:radon_msmnt64}
\end{subfigure}
\begin{subfigure}{0.15\textwidth}
    \centering
    \includegraphics[width=\textwidth]{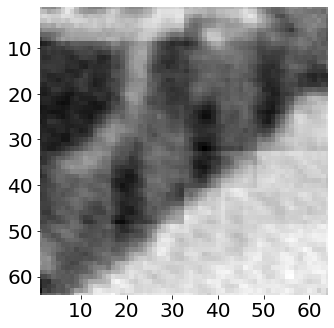}
    \caption{gCG-LS (0.809)}
    \label{fig:gCG-LS_barb64}
\end{subfigure}
\begin{subfigure}{0.15\textwidth}
    \centering
    \includegraphics[width=\textwidth]{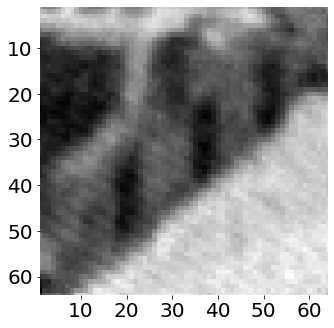}
    \caption{nCG-LS (0.832)}
    \label{fig:nCG-LS_barb64}
\end{subfigure}
\begin{subfigure}{0.15\textwidth}
    \centering
    \includegraphics[width=\textwidth]{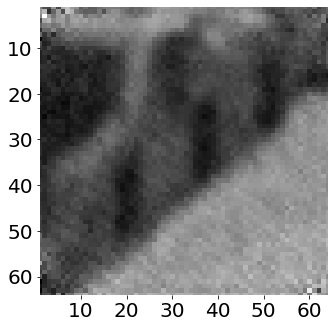}
    \caption{\scalebox{1}{\textcolor{black}{FISTA (0.765)}}}
    \label{fig:FISTA_barb64}
\end{subfigure}
\begin{subfigure}{0.15\textwidth}
    \centering
    \includegraphics[width=\textwidth]{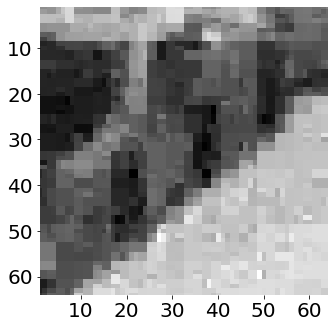}
    \caption{$\ell_1$-LS (0.692)}
    \label{fig:l1ls_barb64}
\end{subfigure}
\begin{subfigure}{0.15\textwidth}
    \centering
    \includegraphics[width=\textwidth]{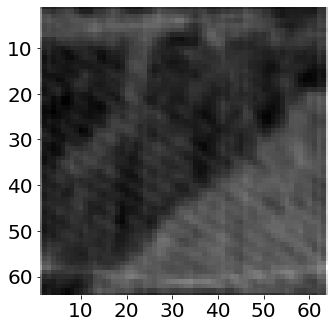}
    \caption{\textcolor{black}{FBP} (0.611)}
    \label{fig:fbp_barb64}
\end{subfigure}
\begin{subfigure}{0.15\textwidth}
    \centering
    \includegraphics[width=\textwidth]{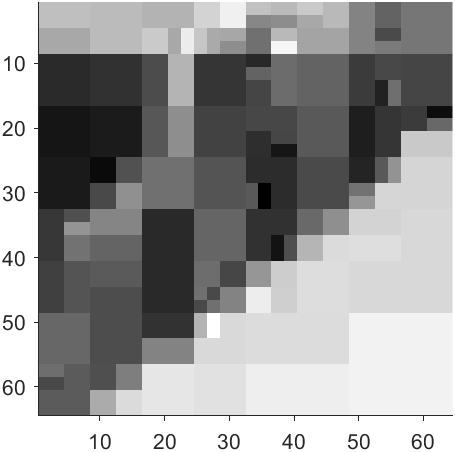}
    \caption{BCS (0.524)}
    \label{fig:bcs_barb64}
\end{subfigure}
\begin{subfigure}{0.15\textwidth}
    \centering
    \includegraphics[width=\textwidth]{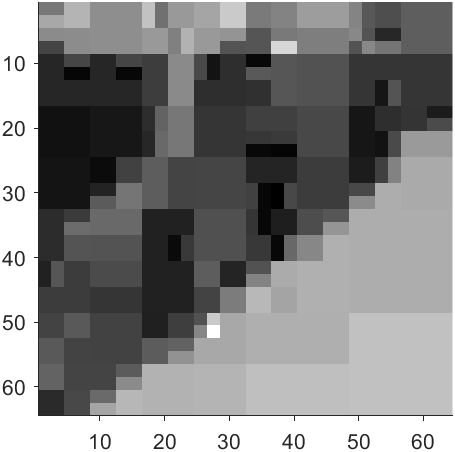}
    \caption{CoSaMP (0.526)}
    \label{fig:cosamp_barb64}
\end{subfigure}
\caption{Image reconstructions (SSIM) using our gCG-LS, our nCG-LS, \textcolor{black}{FISTA}, $\ell_1$-LS, \textcolor{black}{FBP}, \textcolor{black}{BCS}, and CoSaMP. The input to each algorithm is a vectorized (\ref{fig:radon_msmnt64}), \textcolor{black}{which is the Radon transform of (\ref{fig:exact_barb64}) at 15 uniformly spaced angles with an SNR of 60dB.} We observe that CG-LS outperforms all methods with nCG-LS slightly outperforming gCG-LS.}
\label{fig:barbra_reconstruction64}
\end{figure}

\begin{figure}[!b]
\centering
\begin{subfigure}{0.15\textwidth}
    \centering
    \includegraphics[width=\textwidth]{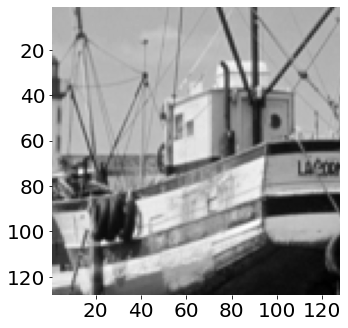}
    \caption{\textcolor{black}{Original}}
    \label{fig:exact_boat128}
\end{subfigure}
\begin{subfigure}{0.15\textwidth}
    \centering
    \includegraphics[width=\textwidth]{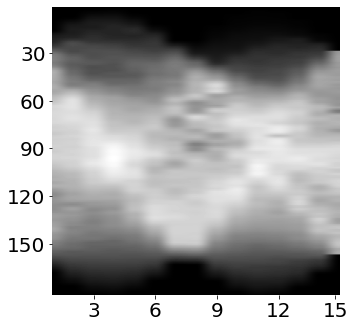}
    \caption{\textcolor{black}{Radon Transform}}
    \label{fig:radon_msmnt128}
\end{subfigure}
\begin{subfigure}{0.15\textwidth}
    \centering
    \includegraphics[width=\textwidth]{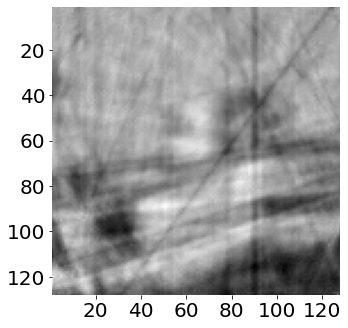}
    \caption{\textcolor{black}{gCG-LS (0.503)}}
    \label{fig:gCG-LS_boat128}
\end{subfigure}
\begin{subfigure}{0.15\textwidth}
    \centering
    \includegraphics[width=\textwidth]{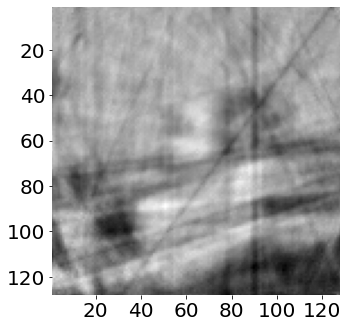}
    \caption{\textcolor{black}{nCG-LS (0.504)}}
    \label{fig:nCG-LS_boat128}
\end{subfigure}
\begin{subfigure}{0.15\textwidth}
    \centering
    \includegraphics[width=\textwidth]{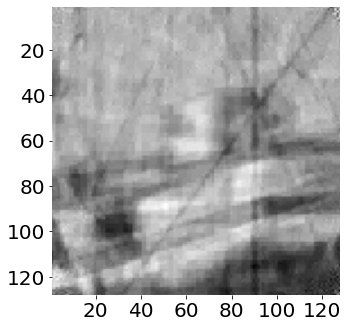}
    \caption{\scalebox{1}{\textcolor{black}{FISTA (0.487)}}}
    \label{fig:FISTA_boat128}
\end{subfigure}
\begin{subfigure}{0.15\textwidth}
    \centering
    \includegraphics[width=\textwidth]{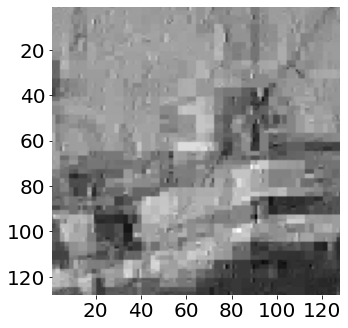}
    \caption{\textcolor{black}{$\ell_1$-LS (0.401)}}
    \label{fig:l1ls_boat128}
\end{subfigure}
\begin{subfigure}{0.15\textwidth}
    \centering
    \includegraphics[width=\textwidth]{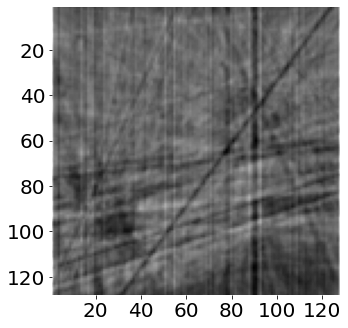}
    \caption{\textcolor{black}{FBP (0.348)}}
    \label{fig:fbp_boat128}
\end{subfigure}
\begin{subfigure}{0.15\textwidth}
    \centering
    \includegraphics[width=\textwidth]{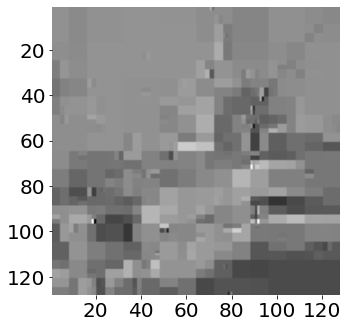}
    \caption{\textcolor{black}{BCS (0.342)}}
    \label{fig:bcs_boat128}
\end{subfigure}
\begin{subfigure}{0.15\textwidth}
    \centering
    \includegraphics[width=\textwidth]{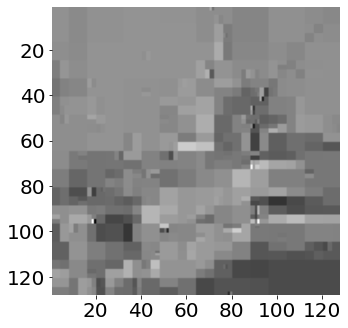}
    \caption{\textcolor{black}{CoSaMP (0.342)}}
    \label{fig:cosamp_boat128}
\end{subfigure}
\caption{\textcolor{black}{Image reconstructions (SSIM) using our gCG-LS, our nCG-LS, FISTA, $\ell_1$-LS, FBP, BCS, and CoSaMP. The input to each algorithm is a vectorized (\ref{fig:radon_msmnt128}), which is the Radon transform of (\ref{fig:exact_boat128}) at 15 uniformly spaced angles with an SNR of 60dB. We observe that CG-LS outperforms all methods with nCG-LS slightly outperforming gCG-LS.}}
\label{fig:boat_reconstruction128}
\end{figure}

Similarly, Fig. \ref{fig:barbra_reconstruction64} \textcolor{black}{and Fig. \ref{fig:boat_reconstruction128} respectively} contain the reconstruction of a $64\times 64$ Barbara snippet \textcolor{black}{and $128\times 128$ boat image} from a 60dB SNR Radon transform at 15 uniformly spaced angles. We see by visual inspection and SSIM that both gCG-LS and nCG-LS produce superior reconstructions to the other iterative algorithms with nCG-LS slightly outperforming gCG-LS.

The superiority of CG-LS is further highlighted in Table \ref{table:CG-LS}, which shows the average image reconstruction SSIM and PSNR along with $99\%$ confidence intervals. Note, larger SSIM and PSNR values correspond to image reconstructions that are visually closer to the original image. One reason our method outperforms the comparative methods may be attributed to the relatively low sparsity level of the wavelet coefficients, $\pmb{c}$, which CG-LS can manage while $\ell_1$-LS, BCS, and CoSaMP require a sufficiently high signal sparsity for superb performance. Although, CG-LS is slower, as shown in Table \ref{tab:computational time}. 

\textcolor{black}{Lastly, we remark that we tested a Kalman Filter approach~\cite{iglesias2022filter} and found that it underperformed in general inverse problems. For example, in CS experiments the Kalman approach underperformed while being relatively competitive to the prior art in Radon inversion problems. A thorough examination of Kalman based approaches to general inverse problems is a subject of future work.}

\section{CG-Net}

\begin{figure*}[!t]
    \centering
    \begin{subfigure}{\textwidth}
    \centering
    \includegraphics[scale = 0.65]{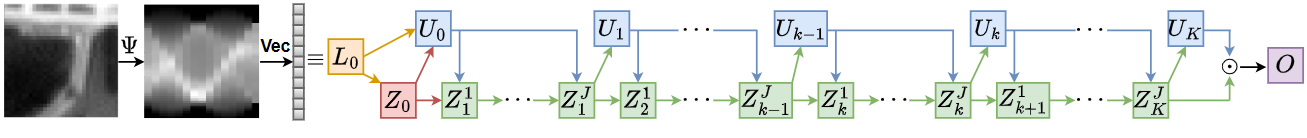}
    \caption{End-to-end network structure of CG-Net.}
    \label{fig:CG-Net}
\end{subfigure}
\begin{subfigure}{.24\textwidth}
    \centering
    \includegraphics[scale = 0.64]{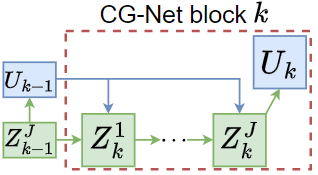}
    \caption{Block $k$ of CG-Net}
    \label{fig:CG-Net block}
\end{subfigure}
\begin{subfigure}{.75\textwidth}
    \centering
    \includegraphics[scale = 1]{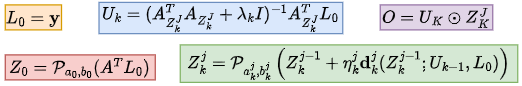}
    \caption{Mathematical descriptions of the CG-Net layers.}
    \label{fig:CG-Net Layers}
\end{subfigure}
\caption{End-to-end network structure for CG-Net, the unrolled deep neural network of Algorithm \ref{alg:CG-LS}, is shown in (\ref{fig:CG-Net}). A mathematical description of each layer is given in (\ref{fig:CG-Net Layers}). CG-Net consists of an input layer, $L_0$, initialization layer, $Z_0$, output layer, $O$, and $K$ CG-Net blocks (\ref{fig:CG-Net block}). Each CG-Net block, $k$, contains $J$ steepest descent layers, $Z_k^1, \ldots, Z_k^J$, and a single Tikhonov layer, $U_k$. Every layer takes the measurements, $L_0\equiv\pmb{y}$, as an input so these connections are omitted for clarity.}
\label{fig:CG-Net structure}
\end{figure*}

\subsection{Network Structure}
We create a DNN by applying algorithm unrolling to CG-LS in Algorithm \ref{alg:CG-LS}. CG-Net has a structure shown in Fig. \ref{fig:CG-Net structure} and consists of an input layer, $L_0$, an initialization layer, $Z_0$, an initial $\pmb{u}$ layer $U_0$, $K$ blocks given in Fig. \ref{fig:CG-Net block}, and an output layer, $O$. Two main functions,  $\pmb{f}_k:\mathbb{R}^{n}\times \mathbb{R}^m\to\mathbb{R}^n$ and $\pmb{g}_k^j:\mathbb{R}^{n}\times \mathbb{R}^n\times \mathbb{R}^m\to\mathbb{R}^n$, define the CG-Net layers. Each $\pmb{f}_k(\pmb{z},\pmb{y}) \equiv \pmb{f}_k(\pmb{z},\pmb{y};\lambda_k)$, which is parameterized by a scalar $\lambda_k$, corresponds to updating $\pmb{u}$ as
\[
    \pmb{f}_k(\pmb{z},\pmb{y}) \coloneqq \textcolor{black}{A_{\pmb{z}}^T(A_{\pmb{z}}A_{\pmb{z}}^T + \lambda_k I)^{-1}\pmb{y}.}
\]
Next, each $\pmb{g}_k^j(\pmb{z},\pmb{u},\pmb{y}) \equiv \pmb{g}_k^j(\pmb{z},\pmb{u},\pmb{y}; a_k^j, b_k^j, \eta_k^{(j)}, \pmb{d}_k^j)$, which is parameterized by a descent vector, $\pmb{d}_k^j$, step size, $\eta_k^{(j)}$ and mReLU parameters $a_k^j$ and $b_k^j$, corresponds to updating $\pmb{z}$ as
    \begin{align}
    \pmb{g}_k^j(\pmb{z},\pmb{u},\pmb{y}) \coloneqq \mathcal{P}_{a_k^j,b_k^j}\left(\pmb{z} + \eta_k^{(j)} \pmb{d}_k^j(\pmb{z};\pmb{u}, \pmb{y})\right). \label{eqn:z layer update}
    \end{align}
In CG-LS, the step size, $\eta_k^{(j)}$, is found by a backtracking line search, which we cannot implement in CG-Net.  Thus, the application of the mReLU activation function, $\mathcal{P}_{a,b}$, at each steepest descent step serves to guarantee the next step is not too large and stays within $\mathcal{Z}$.

Now, we mathematically detail the CG-Net layers:

\noindent\begin{tikzpicture}
\node[scale = .9, fill = peachfill, draw = darkorangeborder, thick] at (0, 0) {$\phantom{L}$};
\node[scale = 1] at (0, 0) {$L_0$};
\node[anchor = west] at (0.15,0) {$= \pmb{y}$ is the input measurements to the network};
\end{tikzpicture} 

\vspace*{-.05cm}

\noindent\begin{tikzpicture}
\node[scale = .9, fill = pinkfill, draw = darkredborder, thick] at (0, 0) {$\phantom{Z}$};
\node[scale = 1] at (0, 0) {$Z_0$};
\node[anchor = west] at (0.15,0) {$=\mathcal{P}_{a_0,b_0}(\hat{A}^T\pmb{y})$\textcolor{black}{, for $\hat{A} = A/||A||_2$,} is the initial};
\node[anchor = west] at (0.15,-.45) {estimate of $\pmb{z}$ from line \ref{line:initial} of Algorithm \ref{alg:CG-LS}.};
\end{tikzpicture} 

\vspace*{-.15cm}

\noindent\begin{tikzpicture}
\node[scale = .9, fill = bluefill, draw = blueborder, thick] at (0, 0) {$\phantom{U}$};
\node[scale = 1] at (0, 0) {$U_0$};
\node[anchor = west] at (0.15,0) {$ = \pmb{f}_0(Z_0,\pmb{y})$ is the initial estimate of $\pmb{u}$ corresponding};
\node[anchor = west] at (0.15,-.45) {to line \ref{line:initial} of Algorithm \ref{alg:CG-LS}.};
\end{tikzpicture}

\vspace*{-.15cm}

\noindent The $k$th CG-Net block, shown in Fig. \ref{fig:CG-Net block}, consists of layers:

\hspace*{-.16cm}\begin{tikzpicture}
\node[scale = 1.1, fill = greenfill, draw = greenborder, thick] at (0, 0) {$\phantom{Z}$};
\node[scale = 1] at (0, 0) {$Z_k^j$};
\node[anchor = west] at (0.2,0) {\scalebox{.95}{$= \pmb{g}_k^j(Z_k^{j-1},U_{k-1},\pmb{y})$} is $\pmb{z}$ on steepest descent step $j$ of};
\node[anchor = west] at (0.2,-.45) {iteration $k$, corresponding to line \ref{line:z update} in Algorithm \ref{alg:CG-LS}.};
\end{tikzpicture}

\vspace*{-.15cm}

\hspace*{-.16cm}\begin{tikzpicture}
\node[scale = .9, fill = bluefill, draw = blueborder, thick] at (0, 0) {$\phantom{U}$};
\node[scale = 1] at (0, 0) {$U_k$};
\node[anchor = west] at (0.2,0) {$= \pmb{f}_k(Z_k^J, \pmb{y})$ is $\pmb{u}$ on iteration $k$, corresponding to line};
\node[anchor = west] at (0.2,-.45) {\ref{line:u update} in Algorithm \ref{alg:CG-LS}.};
\end{tikzpicture}

\vspace*{-.15cm}

\noindent\begin{tikzpicture}
\node[scale = .9, fill = purplefill, draw = purpleborder, thick] at (0, 0) {$\phantom{O}$};
\node[scale = 1] at (0, 0) {$O$};
\node[anchor = west] at (0.15,0) {$= U_K\odot Z_K^J$ is the estimated wavelet coefficients};
\node[anchor = west] at (0.15,-.45) {produced by CG-Net.};
\end{tikzpicture}

Note, to simplify notation, we let $Z_k^0 = Z_{k-1}^J$. In total, CG-Net has $K(J+1)+4$ layers: $K+1$ $\pmb{u}$ update layers, $KJ$ steepest descent $\pmb{z}$ layers, one input, output, and initialization layer.

\textcolor{black}{By incorporating the CG prior, CG-Net differentiates from previous unrolled DNNs in two key ways. First, the Tikhonov updates of $\pmb{u}$ provide non-linear transformation layers of $\pmb{z}$. These layers impose a significant structure via data consistency in equating measurements and measured estimated signals. Second, instead of estimating the original signal of interest directly, CG-Net simultaneously estimates two separate signals and formulates its output as a Hadamard product, which can be viewed as a unique output activation function. These network attributes, inspired by the theory in Sections \ref{sec:existence and location of minimizers} and \ref{sec:convergence}, are shown to be advantageous empirically in Section \ref{sec:NN numerical results}.}

\subsection{Network Parameters and Loss Functions}

We further detail the parameters that will be learned by CG-Net. For every $k = 0, 1, \ldots, K$ the layer $U_k$ is parameterized by regularization scalar, $\lambda_k > 0$. In Algorithm \ref{alg:CG-LS}, every $\lambda_k$ is taken to be the same constant, $\lambda$ from (\ref{eqn:cost function}), but we increase the trainability of CG-Net by allowing different $\lambda_k$ at each layer updating $\pmb{u}$. The initialization layer, $Z_0$, is parameterized by two positive real numbers $a_0 > z_{\min}$ and $b_0 > z_{\min}$, which are applied through the mReLU function, $\mathcal{P}_{a_0,b_0}$.

Next, for each $Z_k^j$ layer, defined by (\ref{eqn:z layer update}), we implement the steepest descent vector $\pmb{d}_k^j(\pmb{z};\pmb{u}) = -B_k^j \nabla_{\bm{z}} F(\pmb{z};\pmb{u})$ for a positive definite matrix $B_k^j$ that will be learned in CG-Net. Note, this descent vector is the steepest descent based upon the quadratic norm $||\cdot||_{(B_k^j)^{-1}}^2$. Thus, CG-Net can be understood as learning the quadratic norm defining every steepest descent step that optimally traverses the landscape of the cost function in (\ref{eqn:iterative_z_update}). For matrix $L$, let $Q$ and $\Lambda$ be the eigendecomposition of $(L+L^T)/2$. That is, $(L+L^T)/2 = Q\Lambda Q^T$. Define, for small $\epsilon > 0$, the diagonal matrix $\Lambda_{\epsilon}$ as $[\Lambda_{\epsilon}]_{ii} = \max\{\Lambda_{ii},\epsilon\}$ and let
\begin{align}
P_{\epsilon}(L) = Q\Lambda_{\epsilon}Q^T. \label{eqn:eigendecomp}
\end{align}
That is, $P_{\epsilon}(L)$ can be viewed as the closest symmetric, real-valued matrix with minimum eigenvalue of $\epsilon$ to $(L+L^T)/2$ as measured by the Frobenius norm. 

We enforce $B_k^j$ to be positive definite by learning a lower triangular matrix $L_k^j$ and setting $B_k^j = P_{\epsilon}(L_k^j).$ Therefore, CG-Net layer $Z_k^j$ is parameterized by a lower triangular matrix $L_k^j$ defining the steepest descent vector
\[
\pmb{d}_k^j(\pmb{z};\pmb{u}) = -P_{\epsilon}(L_k^j) \nabla_{\bm{z}} F(\pmb{u},\pmb{z}).
\]
\textcolor{black}{Note, while the entire matrix $L_k^j$ could be learned, we set CG-Net to learn only the diagonal and sub-diagonal in $L_k^{j}$ constraining $B_k^j$ to be a tridiagonal matrix.}

Additionally, as $\nabla_{\bm{z}} F(\pmb{u},\pmb{z})$, given in (\ref{eqn:grad z}), depends on the regularization scalar $\mu$, layer $Z_k^j$ is parameterized by regularization scalar $\mu_k^j$. Furthermore, layer $Z_k^j$ is parameterized by the step size $\eta_k^{(j)}$, which we take to be a diagonal matrix. That is, instead of learning a single constant to scale the steepest descent vector $\pmb{d}_k^j$ we learn a different constant to scale each component of $\pmb{d}_k^j$ separately. Finally, layer $Z_k^j$ learns positive real numbers $a_k^j > z_{\min}$ and $b_k^j > z_{\min}$, which are applied through the mReLU activation function, $\mathcal{P}_{a_k^j,b_k^j}$, in (\ref{eqn:z layer update}).

Fix a small real-valued $\epsilon> 0$. We remark that to ensure $\lambda_k > 0$ and $a_0, b_0, a_k^j, b_k^j > z_{\min}$ in implementation we use $\max\{\lambda_k, \epsilon\}$ in place of $\lambda_k$, $\max\{a_0, z_{\min}+\epsilon\}$ in place of $a_0$, and similarly for $b_0, a_k^j$, and $b_k^j.$

\begin{table*}[t]
\caption{Average SSIM/PSNR with $99\%$ confidence intervals for \textcolor{black}{ten} machine learning-based image reconstruction methods. Each method reconstructed two hundred, $32\times 32$ CIFAR10~\cite{CIFAR10} and $64\times 64$ CalTech101~\cite{CalTech101} images after training on a \textbf{\textcolor{black}{set of only 20 samples}}. Sensing \textcolor{black}{matrix}, $\Psi$, \textcolor{black}{is} a Radon transform at 15, 10, or 6 uniformly spaced angles and the dictionary, $\Phi$, is a biorthogonal wavelet transform. Each measurement has an SNR of 60dB or 40dB SNR. In all cases, our method CG-Net, highlighted in \textbf{bold}, outperforms the other approaches. Samples from these tests are visualized in Fig. \ref{fig:dog reconstructions}, Fig. \ref{fig:truck reconstructions}, and Fig. \ref{fig:barb reconstructions}.}
\centering
\adjustbox{max width=\textwidth}{
\begin{tabular}{lc|c|c|c|c|c}
\multicolumn{7}{c}{SSIM($\times 10^{2}$)/PSNR for $32\times 32$ CIFAR10 images} \\
\hline
Angles & \multicolumn{2}{c|}{$15$} & \multicolumn{2}{c|}{$10$} & \multicolumn{2}{c}{$6$}  \\
SNR & \multicolumn{1}{c}{$60$ dB}  & \multicolumn{1}{c|}{$40$ dB} &  \multicolumn{1}{c}{$60$ dB}  & \multicolumn{1}{c|}{$40$ dB} &  \multicolumn{1}{c}{$60$ dB}  & \multicolumn{1}{c}{$40$ dB} \\
\hline
\scalebox{.85}{CG-Net} & $\bm{89.9}\pm \bm{.56}$/$\bm{27.7}\pm \bm{.38}$ & $\bm{84.4}\pm \bm{.86}$/$\bm{25.5}\pm \bm{.28}$ & $\bm{81.5}\pm \bm{.83}$/$\bm{24.8}\pm \bm{.40}$ & $\bm{77.4}\pm \bm{.83}$/$\bm{23.7}\pm \bm{.34}$ & $\bm{67.6}\pm \bm{1.2}$/$\bm{21.8}\pm \bm{.40}$ & $\bm{66.1}\pm \bm{1.2}$/$\bm{21.6}\pm \bm{.37}$ \\
\textcolor{black}{\scalebox{.85}{LEARN$^{++}$}} & \textcolor{black}{$79.1\pm .98$/$22.2\pm .32$} & \textcolor{black}{$78.7\pm 1.0$/$22.2\pm .32$} & \textcolor{black}{$72.7\pm 1.1$/$21.1\pm .33$} & \textcolor{black}{$72.4\pm 1.0$/$21.1\pm .31$} & \textcolor{black}{$59.2\pm 1.4$/$19.7\pm .33$} & \textcolor{black}{$59.1\pm 1.3$/$19.6\pm .32$}  \\
\textcolor{black}{LPD} &  \textcolor{black}{$86.8\pm .64$/$25.1\pm .34$} & \textcolor{black}{$84.3\pm .77$/$24.4\pm .30$} & \textcolor{black}{$76.5\pm .93$/$22.5\pm .36$} & \textcolor{black}{$75.9\pm .90$/$22.4\pm .33$} & \textcolor{black}{$61.0\pm 1.2$/$19.7\pm .32$} & \textcolor{black}{$59.6\pm 1.2$/$19.4\pm .31$}  \\
\textcolor{black}{\scalebox{.8}{FBPConvNet}} & \textcolor{black}{$62.6\pm 2.2$/$17.2\pm .41$} & \textcolor{black}{$56.5\pm 1.7$/$15.4\pm .32$} & \textcolor{black}{$54.1\pm 1.6$/$16.3\pm .31$} & \textcolor{black}{$53.6\pm 2.1$/$16.3\pm .37$} & \textcolor{black}{$41.9\pm 1.4$/$14.8\pm .31$} & \textcolor{black}{$40.7\pm 1.6$/$14.5\pm .31$} \\
\scalebox{.85}{MADUN} &  $82.6\pm .74$/$24.3\pm .33$ & $80.2\pm .89$/$23.8\pm .29$ & $71.3\pm .86$/$21.9\pm .35$ & $69.9\pm .90$/$21.7\pm .33$ & $55.3\pm 1.2$/$19.5\pm .36$ & $54.6\pm 1.2$/$19.4\pm .35$ \\
\textcolor{black}{\scalebox{.85}{FISTA-Net}} &  \textcolor{black}{$83.8\pm .81$/$23.2\pm .30$} & \textcolor{black}{$81.9\pm 1.0$/$22.8\pm .31$} & \textcolor{black}{$75.8\pm .94$/$21.5\pm .29$} & \textcolor{black}{$74.9\pm 1.0$/$21.4\pm .29$} & \textcolor{black}{$61.9\pm 1.3$/$19.6\pm .29$} & \textcolor{black}{$61.6\pm 1.3$/$19.5\pm .28$} \\
\textcolor{black}{\scalebox{.85}{iPiano-Net}} & \textcolor{black}{$84.2\pm .63$/$23.1\pm .44$} & \textcolor{black}{$82.1\pm .72$/$22.8\pm .40$} & \textcolor{black}{$75.3\pm .91$/$21.4\pm .43$} & \textcolor{black}{$74.2\pm .91$/$21.2\pm .42$} & \textcolor{black}{$64.3\pm 1.2$/$19.5\pm .46$} & \textcolor{black}{$61.8\pm 1.3$/$19.4\pm .43$} \\
\scalebox{.83}{ISTA-Net$^+$} &  $86.4\pm .62$/$24.5\pm .39$ & $83.6\pm .83$/$24.1\pm .36$ & $74.1\pm 1.0$/$21.4\pm .39$ & $73.6\pm .99$/$21.4\pm .37$ & $55.4\pm 1.4$/$18.9\pm .33$ & $55.3\pm 1.4$/$18.7\pm .33$ \\
\textcolor{black}{\scalebox{.8}{iRadonMAP}} &  \textcolor{black}{$21.4\pm 1.5$/$14.6\pm .33$} & \textcolor{black}{$20.7\pm 1.4$/$13.9\pm .37$} & \textcolor{black}{$20.8\pm 1.4$/$14.4\pm .35$} & \textcolor{black}{$20.5\pm 1.4$/$14.3\pm .34$} & \textcolor{black}{$18.4\pm 1.3$/$14.1\pm .37$} & \textcolor{black}{$18.2\pm 1.3$/$14.0\pm .37$}  \\
\scalebox{.85}{ReconNet} &  $19.6\pm 1.4$/$15.8\pm .32$ & $19.0\pm 1.5$/$15.2\pm .31$ & $18.7\pm 1.6$/$14.9\pm .32$ & $18.3\pm 1.6$/$15.0\pm .32$ & $17.7\pm 1.5$/$14.6\pm .32$ & $17.1\pm 1.6$/$14.5\pm .31$  \\
\multicolumn{7}{c}{} \\
\multicolumn{7}{c}{SSIM($\times 10^{2}$)/PSNR for $64\times 64$ CalTech101 images} \\
\hline
Angles & \multicolumn{2}{c|}{$15$} & \multicolumn{2}{c|}{$10$} & \multicolumn{2}{c}{$6$}  \\
SNR & \multicolumn{1}{c}{$60$ dB}  & \multicolumn{1}{c|}{$40$ dB} &  \multicolumn{1}{c}{$60$ dB}  & \multicolumn{1}{c|}{$40$ dB} &  \multicolumn{1}{c}{$60$ dB}  & \multicolumn{1}{c}{$40$ dB} \\
\hline
\scalebox{.85}{CG-Net} & $\bm{65.4}\pm \bm{1.2}$/$\bm{22.8}\pm \bm{.46}$ & $\bm{60.6}\pm \bm{1.3}$/$\bm{22.0}\pm \bm{.41}$ &  $\bm{56.6}\pm \bm{1.3}$/$\bm{21.1}\pm \bm{.44}$ & $\bm{53.1}\pm \bm{1.2}$/$\bm{20.7}\pm \bm{.41}$ &  $\bm{45.7}\pm \bm{1.3}$/$\bm{19.1}\pm \bm{.43}$ & $\bm{43.1}\pm \bm{1.2}$/$\bm{18.9}\pm \bm{.41}$ \\
\textcolor{black}{\scalebox{.85}{LEARN$^{++}$}} & \textcolor{black}{$53.4\pm 1.4$/$19.3\pm .36$} & \textcolor{black}{$52.3\pm 1.4$/$19.2\pm .35$} & \textcolor{black}{$49.8\pm 1.5$/$19.0\pm .37$} & \textcolor{black}{$48.7\pm 1.5$/$18.9\pm .37$} & \textcolor{black}{$38.6\pm 1.3$/$17.1\pm .35$} & \textcolor{black}{$37.1\pm 1.3$/$17.0\pm .35$} \\
\textcolor{black}{LPD} & \textcolor{black}{$64.4\pm 1.2$/$21.0\pm .37$} & \textcolor{black}{$60.7\pm 1.1$/$20.7\pm .36$} & \textcolor{black}{$53.9\pm 1.2$/$19.5\pm .36$} & \textcolor{black}{$51.1\pm 1.2$/$19.1\pm .35$} & \textcolor{black}{$44.2\pm 1.3$/$17.5\pm .36$} & \textcolor{black}{$43.5\pm 1.3$/$17.6\pm .35$} \\
\textcolor{black}{\scalebox{.8}{FBPConvNet}} & \textcolor{black}{$45.0\pm 1.6$/$15.9\pm .50$} & \textcolor{black}{$42.2\pm 1.5$/$15.6\pm .43$} & \textcolor{black}{$36.1\pm 1.5$/$14.2\pm .38$} & \textcolor{black}{$32.3\pm 1.4$/$13.7\pm .39$} & \textcolor{black}{$30.4\pm 1.2$/$12.6\pm .45$} & \textcolor{black}{$24.4\pm 1.1$/$11.9\pm .40$} \\
\scalebox{.85}{MADUN} &  $57.8\pm 1.2$/$21.0\pm .46$ & $56.5\pm 1.2$/$20.8\pm .43$ & $49.1\pm 1.2$/$19.8\pm .42$ & $47.9\pm 1.1$/$19.6\pm .40$ & $38.8\pm 1.4$/$18.1\pm .42$ & $38.2\pm 1.4$/$18.0\pm .42$ \\
\textcolor{black}{\scalebox{.85}{FISTA-Net}} & \textcolor{black}{$63.7\pm 1.1$/$21.7\pm .38$} & \textcolor{black}{$59.6\pm 1.2$/$21.0\pm .35$} & \textcolor{black}{$55.2\pm 1.3$/$20.2\pm .36$} & \textcolor{black}{$53.4\pm 1.2$/$19.4\pm .34$} & \textcolor{black}{$45.6\pm 1.3$/$17.9\pm .38$} & \textcolor{black}{$44.4\pm 1.3$/$17.9\pm .38$} \\
\textcolor{black}{\scalebox{.85}{iPiano-Net}} & \textcolor{black}{$63.9\pm 1.3$/$20.4\pm .49$} & \textcolor{black}{$61.0\pm 1.2$/$19.7\pm .44$} & \textcolor{black}{$56.3\pm 1.6$/$19.1\pm .56$} & \textcolor{black}{$54.0\pm 1.6$/$18.7\pm .54$} & \textcolor{black}{$45.7\pm 1.6$/$17.7\pm .49$} & \textcolor{black}{$45.1\pm 1.6$/$17.5\pm .48$} \\
\scalebox{.83}{ISTA-Net$^+$} &  $60.0\pm 1.6$/$20.3\pm .42$ & $58.9\pm 1.5$/$20.2\pm .41$ & $54.5\pm 1.7$/$19.6\pm .45$ & $53.9\pm 1.7$/$19.4\pm .44$ & $45.6\pm 1.5$/$18.3\pm .42$ & $44.5\pm 1.4$/$18.1\pm .42$ \\
\textcolor{black}{\scalebox{.8}{iRadonMAP}} &  \textcolor{black}{$12.9\pm 1.1$/$12.5\pm .34$} & \textcolor{black}{$12.8\pm 1.1$/$12.4\pm .34$} & \textcolor{black}{$12.4\pm 1.1$/$12.3\pm .35$} & \textcolor{black}{$12.3\pm 1.1$/$12.3\pm .35$} & \textcolor{black}{$11.5\pm .93$/$11.4\pm .36$} & \textcolor{black}{$11.3\pm .92$/$11.4\pm .35$}  \\
\scalebox{.85}{ReconNet} & $12.9\pm .97$/$14.0\pm .38$ & $11.3\pm .92$/$13.5\pm .36$ & $11.5\pm .94$/$13.7\pm .37$ & $11.0\pm .91$/$13.6\pm .38$ & $10.6\pm .96$/$13.4\pm .38$ & $9.40\pm .90$/$13.1\pm .38$ 
\end{tabular}
}
\label{table:CG-Net}
\end{table*}

\begin{figure*}[!t]
\centering
\begin{subfigure}{0.32\textwidth}
    \centering
    \includegraphics[width=\textwidth]{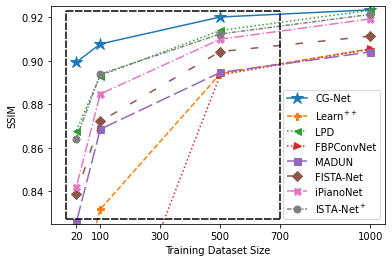}
\end{subfigure}
\begin{subfigure}{0.32\textwidth}
    \centering
    \includegraphics[width=\textwidth]{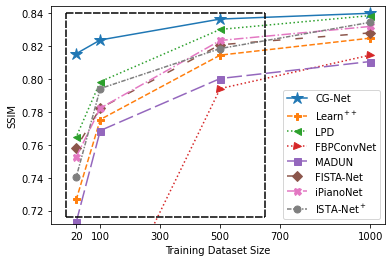}
\end{subfigure}
\begin{subfigure}{0.32\textwidth}
    \centering
    \includegraphics[width=\textwidth]{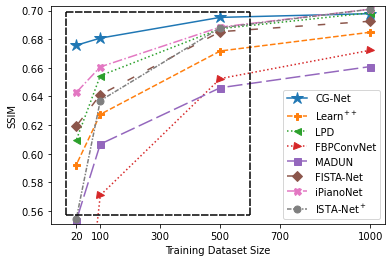}
\end{subfigure}
\begin{subfigure}{0.32\textwidth}
    \centering
    \includegraphics[width=\textwidth]{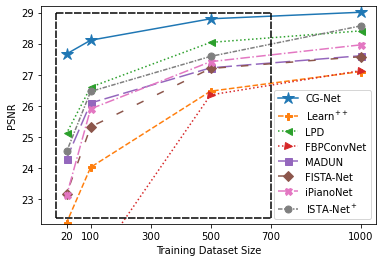}
    \caption{\textcolor{black}{Fifteen uniformly spaced angles.}}
    \label{fig:compare_15_60}
\end{subfigure}
\begin{subfigure}{0.32\textwidth}
    \centering
    \includegraphics[width=\textwidth]{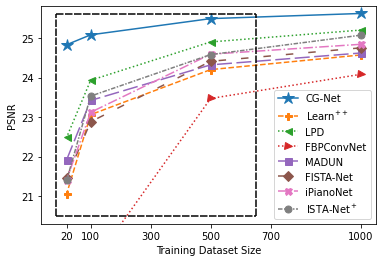}
    \caption{\textcolor{black}{Ten uniformly spaced angles.}}
    \label{fig:compare_10_60}
\end{subfigure}
\begin{subfigure}{0.32\textwidth}
    \centering
    \includegraphics[width=\textwidth]{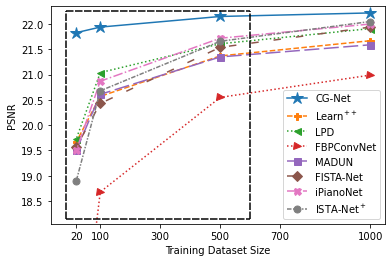}
    \caption{\textcolor{black}{Six uniformly spaced angles.}}
    \label{fig:compare_6_60}
\end{subfigure}
\caption{Average test image reconstruction SSIM \textcolor{black}{and PSNR} when varying the amount of CIFAR10~\cite{CIFAR10} data in training \textcolor{black}{eight} machine learning-based image reconstruction methods. Here, the sensing matrices, $\Psi$, are a Radon transform at 15, 10, or 6 uniformly spaced angles and the sparsity dictionary, $\Phi$, is a biorthogonal wavelet transform. \textcolor{black}{Measurements in training and testing have an SNR of 60dB.} Our method, CG-Net, significantly outperforms all comparative methods, as highlighted in the boxed region, in the low training regime. The 20 training samples scenario is further detailed in Table \ref{table:CG-Net}.}
\label{fig:compare}
\end{figure*}

\textcolor{black}{When only the diagonal and sub-diagonal in $L_k^j$ are learned then CG-Net has $K(J(3n+2)+1)+3$ parameters}
\[
\pmb{\Theta} = \left\{\lambda_0, a_0, b_0, \lambda_k, \mu_k^j, L_k^j, \eta_k^{(j)}, a_k^j, b_k^j\right\}_{k = 1, 2, \ldots, K}^{j = 1, 2, \ldots, J}
\]
where $n$ is the image size. The CG-Net parameters are trained by minimizing a loss function involving the SSIM image quality metric~\cite{NN_loss}, namely for $\mathcal{B}\subset\mathcal{D}$ a batch of data points
\begin{align*}
    \mathcal{L}_{\mathcal{B}}(\pmb{\Theta}) = \frac{1}{|\mathcal{B}|}\sum_{(\widetilde{\pmb{y}}_i, \widetilde{\pmb{c}}_i)\in \mathcal{B}} \left(1-\text{SSIM}(\Phi\widetilde{\pmb{c}}(\widetilde{\pmb{y}}_i; \pmb{\Theta}), \Phi\widetilde{\pmb{c}}_i)\right).
\end{align*}
\textcolor{black}{We note that the MAE loss function is an adequate alternative providing nearly identical results for CG-Net.} The SSIM loss function is optimized through adaptive moment estimation (Adam)~\cite{ADAM}, which is a stochastic gradient-based optimizer. The gradient $\nabla_{\bm{\Theta}} \mathcal{L}_{\mathcal{B}}$ for Adam is calculated via backpropagation through the network, which we implement with automatic differentiation~\cite{auto_diff} using TensorFlow.

\subsection{Numerical Results} \label{sec:NN numerical results}

Given a sensing matrix, $\Psi$, dictionary, $\Phi$, and noise level in SNR, we create a set of training and testing measurement-coefficient pairs, $(\pmb{y},\pmb{c})$, as in Section \ref{sec:CG-LS numerical results}, that we use to train and evaluate a CG-Net. Measurements are formed from $32\times 32$ CIFAR10~\cite{CIFAR10} images and $64\times 64$ CalTech101~\cite{CalTech101} images. For network size, CG-Net running on $32\times32$ or $64\times64$ image measurements use $(K,J) = (20,1)$ or $(K,J) = (5,1),$ respectively. The network sizes were chosen empirically such that the time to complete one image reconstruction was reasonably quick while still producing excellent reconstructions on a validation set of test images. We let $f(z) = \ln(z)$ and initialize $a_0 = 1, b_0 = \exp(2)$, every $\eta_k^{(j)} = \frac{1}{2}I$, all $\mu_k^j = 2$, each $a_k^j = \frac{8}{10}$, all $b_k^j = \exp(3)$, and every $L_k^j = I$. Finally, we initialize $\lambda_k = 0.3$ for 60dB SNR noise level and $\lambda_k = 2$ all other noise levels. Each CG-Net was trained for 30 epochs using a learning rate of $10^{-3}$ with early stopping.

\textcolor{black}{We compare CG-Net against nine state-of-the-art methods: memory augmented deep unfolding network (MADUN)~\cite{MADUN}, ISTA-Net$^+$~\cite{zhang2018ista}, FISTA-Net~\cite{xiang2021FISTANet}, iPiano-Net~\cite{su2020iPianoNet}, ReconNet~\cite{reconnet}, LEARN$^{++}$~\cite{zhang2022learn++}, Learned Primal-Dual (LPD)~\cite{adler2018LPD}, FBPConvNet~\cite{jin2017FBPConvNet}, and iRadonMAP~\cite{he2020iRadonMAP}. Additionally, memory augmented proximal unrolled network (MAPUN)~\cite{song2023MAPUN} was considered, but due to the similarity in performance to MADUN only MADUN results are shown. Note, LEARN$^{++}$, LPD, FBPConvNet, and iRadonMAP are CT-specific reconstruction methods relying on the structure of the CT sinogram measurements. Instead MADUN, ISTA-Net$^+$, FISTA-Net, iPiano-Net, and ReconNet are linear inverse problem reconstruction methods with particular application in image compressive sensing. Furthermore, we remark that MADUN, ISTA-Net$^+$, FISTA-Net, iPiano-Net, LEARN$^{++}$, and LPD are DNNs formed by algorithm unrolling while ReconNet, iRadonMAP, and FBPConvNet are instead standard DNNs.}

\begin{figure*}[!t]
\centering
\begin{subfigure}{0.32\textwidth}
    \centering
    \includegraphics[width=\textwidth]{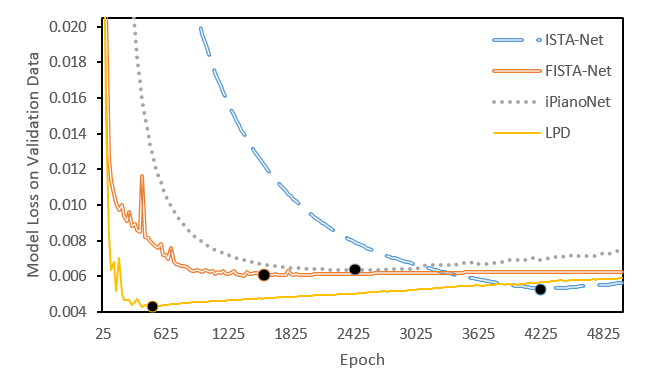}
\end{subfigure}
\begin{subfigure}{0.32\textwidth}
    \centering
    \includegraphics[width=\textwidth]{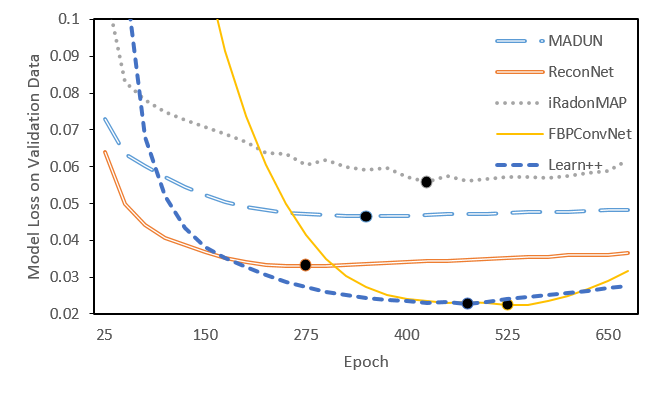}
\end{subfigure}
\begin{subfigure}{0.32\textwidth}
    \centering
    \includegraphics[width=\textwidth]{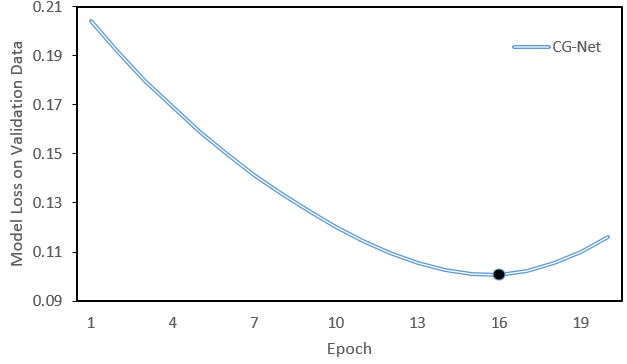}
\end{subfigure}
\caption{\textcolor{black}{Model loss curves on a validation dataset when training each deep learning-based method on Radon inversion from 15 uniformly spaced angles with 20 training samples. The point on each model's loss curve represents the epoch in which that model achieved its best loss on a validation dataset and overfits on subsequent epochs. Only these best loss epoch results are presented for each method. Note, for the plots above, CG-Net uses an SSIM loss function, MADUN uses a mean absolute error loss function, and all other methods use a mean square error loss function (possibly with some additional regularization). As the network loss functions are not equivalent, these plots do not contribute a comparison between the methods and only illustrate that each method is maximally trained for every supplied training dataset.}}
\label{fig:loss curves}
\end{figure*}

\textcolor{black}{For every set of training data, each method was trained using early stopping. That is, as shown in Fig. \ref{fig:loss curves}, training was conducted until the model initially overfits as compared to a validation dataset. In doing so, we ensure every model is sufficiently trained while also not being over trained thereby presenting the best performance for each model for the provided set of training data.}

Shown in Fig. \ref{fig:compare} is the average SSIM quality, over 200 test image reconstructions, for CG-Net \textcolor{black}{and the comparison methods} when each is trained on a varying amount of training data. We consider small sets of training data specifically of size 1000, 500, 100, and 20 measurement, coefficient pairs. \textcolor{black}{Note, ReconNet, iRadonMAP, and some instances of FBPConvNet perform significantly lower and are thus omitted from Fig. \ref{fig:compare}.}
In Fig. \ref{fig:compare_15_60}, \ref{fig:compare_10_60}, and \ref{fig:compare_6_60} we see when reconstructing images from Radon transforms -- at 15, 10, or 6 uniformly spaced angles, respectively -- with an SNR of 60dB that CG-Net outperforms all comparative methods and does so appreciably in low training.

\textcolor{black}{The smallest training dataset, of size 20,} is further detailed in Table \ref{table:CG-Net}, which displays the average SSIM and PSNR plus $99\%$ confidence intervals over 200 test image reconstructions for CG-Net \textcolor{black}{and each comparison method.} We see \textcolor{black}{in Table \ref{table:CG-Net}} that CG-Net significantly outperforms all comparative methods \textcolor{black}{in this low training scenario. While the values in Table \ref{table:CG-Net} may seem low, they are a result of training the models only on a very small training dataset. To this point, while CG-Net does not appreciably outperform CG-LS in the lowest training scenario it will do so when provided enough training data. In particular, we see in Fig. \ref{fig:compare} and in Table \ref{table:CG-Net Extension} that with over 100 training samples CG-Net will outperform CG-LS. Nevertheless, as highlighted in Section \ref{sec:time reqs}, CG-Net provides a significant reconstruction time improvement over CG-LS while still providing comparable reconstruction quality even in low training.
} 

\textcolor{black}{We believe the high performance of CG-Net in low training scenarios is due to a couple of factors. First, the initialization of CG-Net corresponds precisely to CG-LS and, unlike other algorithm unrolling methods, we do not replace any part of the optimization with a CNN or other subnetwork structure. As these optimization-replacing subnetworks are learned completely from scratch, they can overfit quickly in low training. Second, CG-Net naturally incorporates the powerful statistical CG prior through the Tikhonov estimates and Hadamard product output, which provides beneficial data-consistency solution structure for low training. However, with greater noise and higher training, CG-Net can perform lower than the best performing comparative methods, possibly due to the enforced structure that makes CG-Net so successful in low training. Another consideration, as shown in Table \ref{table:parameters}, is that the model complexity of CG-Net may need to be increased in higher training by increasing the number of unrolled iterations.}

\begin{figure*}[!t]
\centering
\begin{subfigure}{0.16\textwidth}
    \centering
    \includegraphics[width=\textwidth]{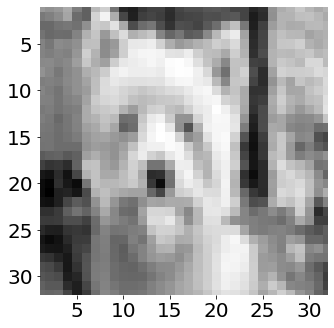}
    \caption{Original}
    \label{fig:dog original}
\end{subfigure}
\begin{subfigure}{0.16\textwidth}
    \centering
    \includegraphics[width=\textwidth]{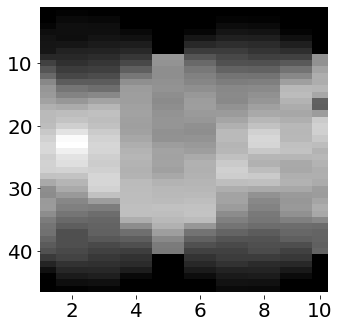}
    \caption{Radon Transform}
    \label{fig:dog msmnt}
\end{subfigure}
\begin{subfigure}{0.16\textwidth}
    \centering
    \includegraphics[width=\textwidth]{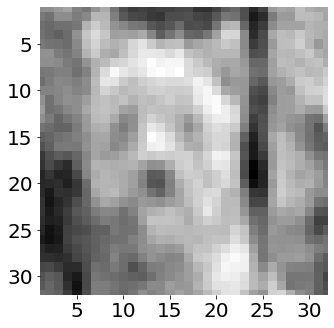}
    \caption{CG-Net (0.870)}
    \label{fig:gCG-Net dog}
\end{subfigure}
\begin{subfigure}{0.16\textwidth}
    \centering
    \includegraphics[width=\textwidth]{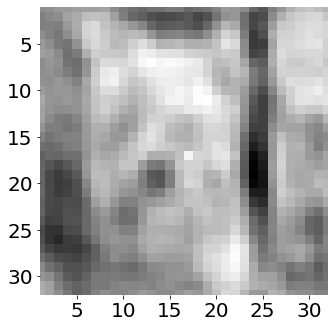}
    \caption{\textcolor{black}{\scalebox{.9}{LEARN$^{++}$ (0.763)}}}
    \label{fig:LEARN dog}
\end{subfigure}
\begin{subfigure}{0.16\textwidth}
    \centering
    \includegraphics[width=\textwidth]{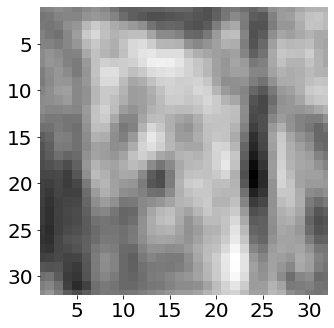}
    \caption{\textcolor{black}{LPD (0.796)}}
    \label{fig:LPD dog}
\end{subfigure}
\begin{subfigure}{0.16\textwidth}
    \centering
    \includegraphics[width=\textwidth]{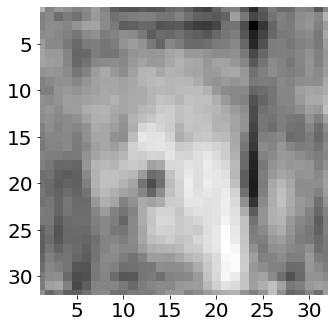}
    \caption{\textcolor{black}{\scalebox{.9}{FBPConvNet (0.583)}}}
    \label{fig:FBPConvNet dog}
\end{subfigure}
\begin{subfigure}{0.16\textwidth}
    \centering
    \includegraphics[width=\textwidth]{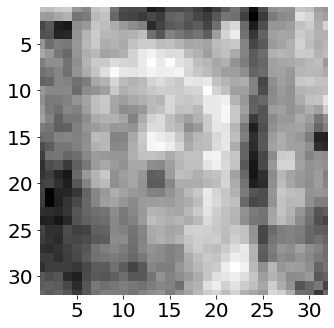}
    \caption{MADUN (0.730)}
    \label{fig:MADUN dog}
\end{subfigure}
\begin{subfigure}{0.16\textwidth}
    \centering
    \includegraphics[width=\textwidth]{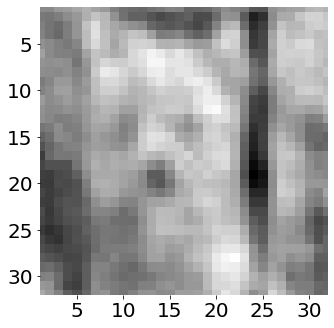}
    \caption{\textcolor{black}{\scalebox{.9}{FISTA-Net (0.803)}}}
    \label{fig:FISTANet dog}
\end{subfigure}
\begin{subfigure}{0.16\textwidth}
    \centering
    \includegraphics[width=\textwidth]{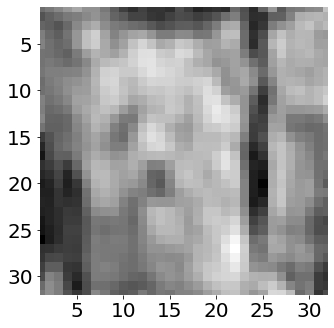}
    \caption{\textcolor{black}{\scalebox{.9}{iPiano-Net (0.791)}}}
    \label{fig:iPianoNet dog}
\end{subfigure}
\begin{subfigure}{0.16\textwidth}
    \centering
    \includegraphics[width=\textwidth]{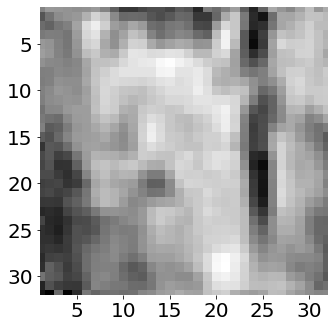}
    \caption{ISTA-Net$^+$ (0.735)}
    \label{fig:ISTA-Net dog}
\end{subfigure}
\begin{subfigure}{0.16\textwidth}
    \centering
    \includegraphics[width=\textwidth]{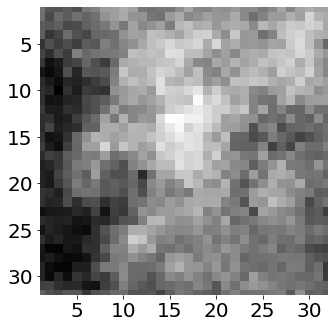}
    \caption{\textcolor{black}{\scalebox{.9}{iRadonMAP (0.173)}}}
    \label{fig:iRadonMap dog}
\end{subfigure}
\begin{subfigure}{0.16\textwidth}
    \centering
    \includegraphics[width=\textwidth]{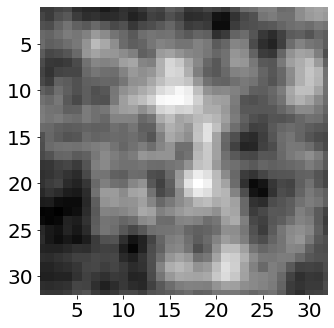}
    \caption{ReconNet (0.415)}
    \label{fig:ReconNet dog}
\end{subfigure}
\caption{Test image reconstructions (SSIM) of a $32\times 32$ dog image. Here, $\Psi$ is a Radon transform at 10 uniformly spaced angles, $\Phi$ is a biorthogonal wavelet transformation, and each measurement has an SNR of 60dB. Our method CG-Net (\ref{fig:gCG-Net dog}) performs best.}
\label{fig:dog reconstructions}
\end{figure*}

\begin{figure*}[!t]
\centering
\begin{subfigure}{0.16\textwidth}
    \centering
    \includegraphics[width=\textwidth]{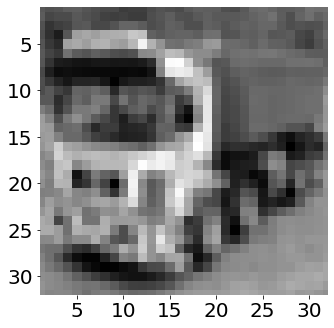}
    \caption{Original}
    \label{fig:truck}
\end{subfigure}
\begin{subfigure}{0.16\textwidth}
    \centering
    \includegraphics[width=\textwidth]{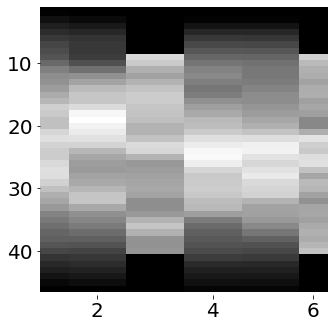}
    \caption{Radon Transform}
    \label{fig:truck msmnt}
\end{subfigure}
\begin{subfigure}{0.16\textwidth}
    \centering
    \includegraphics[width=\textwidth]{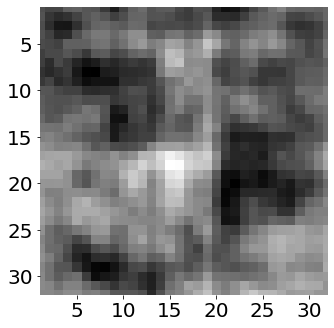}
    \caption{CG-Net (0.602)}
    \label{fig:gCG-Net truck}
\end{subfigure}
\begin{subfigure}{0.16\textwidth}
    \centering
    \includegraphics[width=\textwidth]{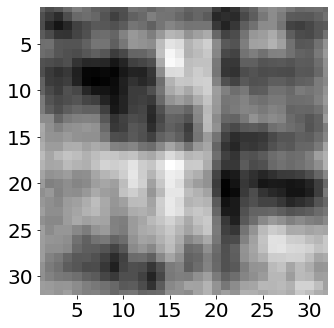}
    \caption{\textcolor{black}{\scalebox{.9}{LEARN$^{++}$ (0.496)}}}
    \label{fig:LEARN truck}
\end{subfigure}
\begin{subfigure}{0.16\textwidth}
    \centering
    \includegraphics[width=\textwidth]{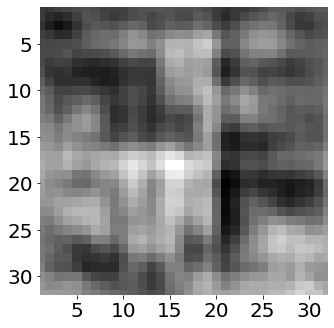}
    \caption{\textcolor{black}{LPD (0.556)}}
    \label{fig:LPD truck}
\end{subfigure}
\begin{subfigure}{0.16\textwidth}
    \centering
    \includegraphics[width=\textwidth]{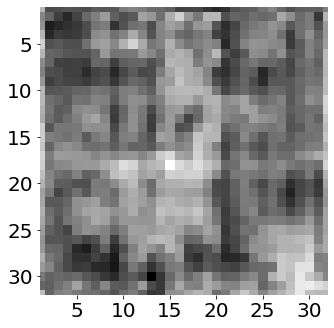}
    \caption{\textcolor{black}{\scalebox{.9}{FBPConvNet (0.430)}}}
    \label{fig:FBPConvNet truck}
\end{subfigure}
\begin{subfigure}{0.16\textwidth}
    \centering
    \includegraphics[width=\textwidth]{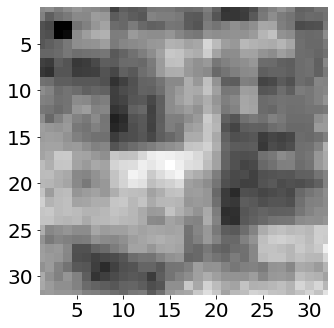}
    \caption{MADUN (0.510)}
    \label{fig:MADUN truck}
\end{subfigure}
\begin{subfigure}{0.16\textwidth}
    \centering
    \includegraphics[width=\textwidth]{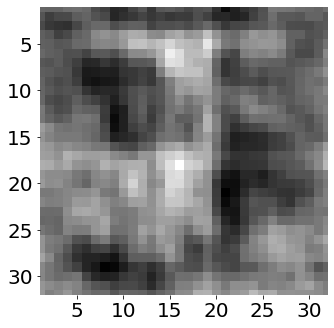}
    \caption{\textcolor{black}{\scalebox{.9}{FISTA-Net (0.569)}}}
    \label{fig:FISTANet truck}
\end{subfigure}
\begin{subfigure}{0.16\textwidth}
    \centering
    \includegraphics[width=\textwidth]{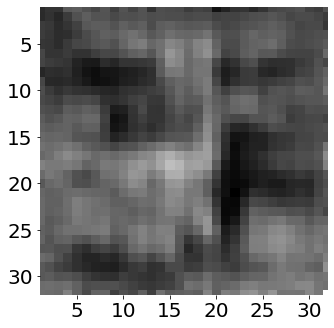}
    \caption{\textcolor{black}{\scalebox{.9}{iPiano-Net (0.565)}}}
    \label{fig:iPianoNet truck}
\end{subfigure}
\begin{subfigure}{0.16\textwidth}
    \centering
    \includegraphics[width=\textwidth]{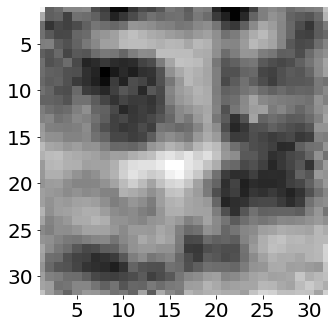}
    \caption{ISTA-Net$^+$ (0.492)}
    \label{fig:ISTA-Net truck}
\end{subfigure}
\begin{subfigure}{0.16\textwidth}
    \centering
    \includegraphics[width=\textwidth]{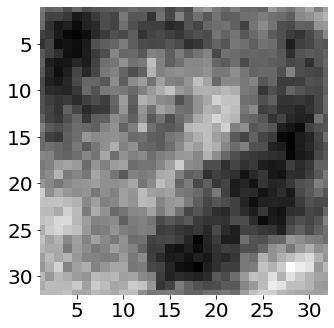}
    \caption{\textcolor{black}{\scalebox{.9}{iRadonMAP (0.090)}}}
    \label{fig:iRadonMap truck}
\end{subfigure}
\begin{subfigure}{0.16\textwidth}
    \centering
    \includegraphics[width=\textwidth]{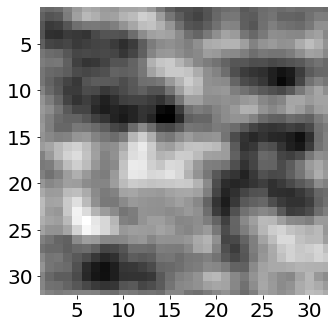}
    \caption{ReconNet (0.321)}
    \label{fig:ReconNet truck}
\end{subfigure}
\caption{Test image reconstructions (SSIM) for a $32\times 32$ truck image. Here, $\Psi$ is a 6 uniformly spaced angle Radon transform, $\Phi$ is a biorthogonal wavelet transformation, and measurement have an SNR of 60dB. Our method CG-Net (\ref{fig:gCG-Net truck}) performs best.}
\label{fig:truck reconstructions}
\end{figure*}

\begin{figure*}[!t]
\centering
\begin{subfigure}{0.16\textwidth}
    \centering
    \includegraphics[width=\textwidth]{Images/barb_32x32_exact.png}
    \caption{Original}
    \label{fig:actual_barb32}
\end{subfigure}
\begin{subfigure}{0.16\textwidth}
    \centering
    \includegraphics[width=\textwidth]{Images/msmnt_plot.png}
    \caption{Radon Transform}
    \label{fig:msmnt}
\end{subfigure}
\begin{subfigure}{0.16\textwidth}
    \centering
    \includegraphics[width=\textwidth]{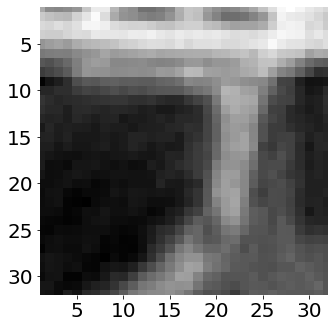}
    \caption{CG-Net (0.954)}
    \label{fig:gCG-Net barb32}
\end{subfigure}
\begin{subfigure}{0.16\textwidth}
    \centering
    \includegraphics[width=\textwidth]{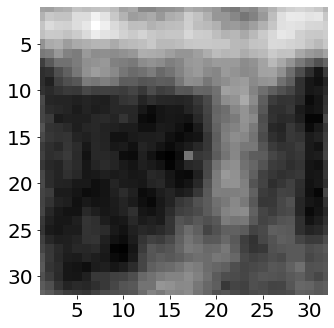}
    \caption{\textcolor{black}{\scalebox{.9}{LEARN$^{++}$ (0.764)}}}
    \label{fig:LEARN barb32}
\end{subfigure}
\begin{subfigure}{0.16\textwidth}
    \centering
    \includegraphics[width=\textwidth]{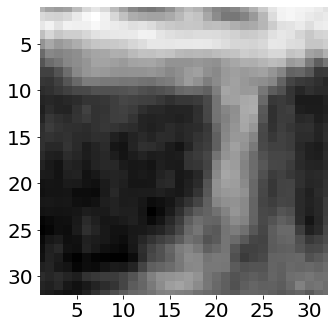}
    \caption{\textcolor{black}{LPD (0.862)}}
    \label{fig:LPD barb32}
\end{subfigure}
\begin{subfigure}{0.16\textwidth}
    \centering
    \includegraphics[width=\textwidth]{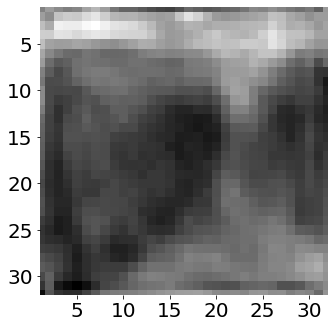}
    \caption{\textcolor{black}{\scalebox{.9}{FBPConvNet (0.618)}}}
    \label{fig:FBPConvNet barb32}
\end{subfigure}
\begin{subfigure}{0.16\textwidth}
    \centering
    \includegraphics[width=\textwidth]{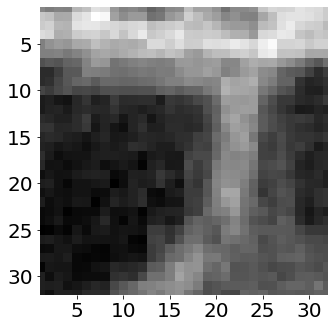}
    \caption{MADUN (0.829)}
    \label{fig:MADUN barb32}
\end{subfigure}
\begin{subfigure}{0.16\textwidth}
    \centering
    \includegraphics[width=\textwidth]{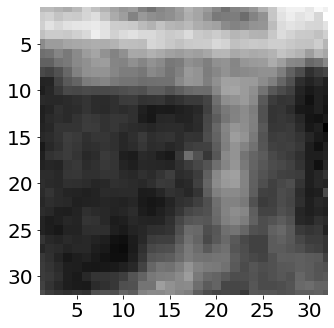}
    \caption{\textcolor{black}{\scalebox{.9}{FISTA-Net (0.832)}}}
    \label{fig:FISTANet barb32}
\end{subfigure}
\begin{subfigure}{0.16\textwidth}
    \centering
    \includegraphics[width=\textwidth]{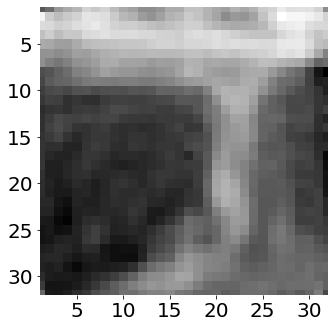}
    \caption{\textcolor{black}{\scalebox{.9}{iPiano-Net (0.864)}}}
    \label{fig:iPianoNet barb32}
\end{subfigure}
\begin{subfigure}{0.16\textwidth}
    \centering
    \includegraphics[width=\textwidth]{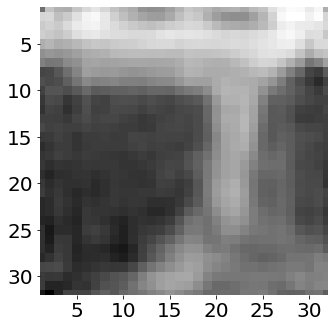}
    \caption{ISTA-Net$^+$ (0.908)}
    \label{fig:ISTA-Net barb32}
\end{subfigure}
\begin{subfigure}{0.16\textwidth}
    \centering
    \includegraphics[width=\textwidth]{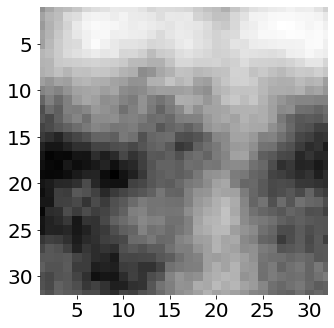}
    \caption{\textcolor{black}{\scalebox{.9}{iRadonMAP (0.501)}}}
    \label{fig:iRadonMap barb32}
\end{subfigure}
\begin{subfigure}{0.16\textwidth}
    \centering
    \includegraphics[width=\textwidth]{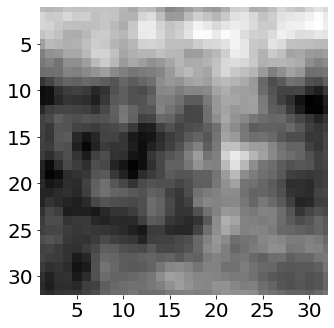}
    \caption{ReconNet (0.526)}
    \label{fig:ReconNet barb32}
\end{subfigure}

\begin{subfigure}{0.16\textwidth}
    \centering
    \includegraphics[width=\textwidth]{Images/64x64_barb_orig.png}
    \caption{Original}
    \label{fig:actual_barb64}
\end{subfigure}
\begin{subfigure}{0.16\textwidth}
    \centering
    \includegraphics[width=\textwidth]{Images/64x15_barb_msmnt.png}
    \caption{Radon Transform}
    \label{fig:msmnt64}
\end{subfigure}
\begin{subfigure}{0.16\textwidth}
    \centering
    \includegraphics[width=\textwidth]{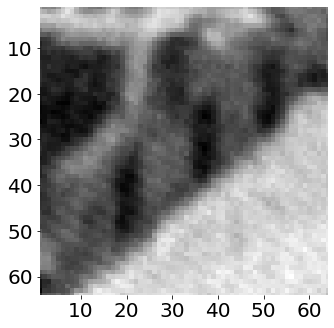}
    \caption{CG-Net (0.829)}
    \label{fig:gCG-Net barb64}
\end{subfigure}
\begin{subfigure}{0.16\textwidth}
    \centering
    \includegraphics[width=\textwidth]{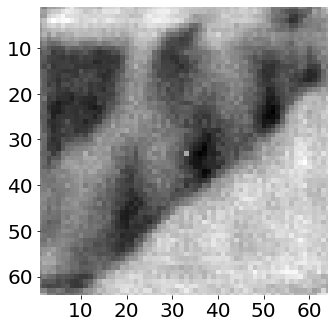}
    \caption{\textcolor{black}{\scalebox{.9}{LEARN$^{++}$ (0.627)}}}
    \label{fig:LEARN barb64}
\end{subfigure}
\begin{subfigure}{0.16\textwidth}
    \centering
    \includegraphics[width=\textwidth]{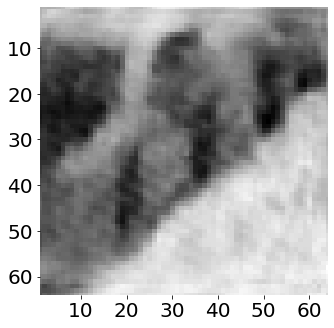}
    \caption{\textcolor{black}{LPD (0.724)}}
    \label{fig:LPD barb64}
\end{subfigure}
\begin{subfigure}{0.16\textwidth}
    \centering
    \includegraphics[width=\textwidth]{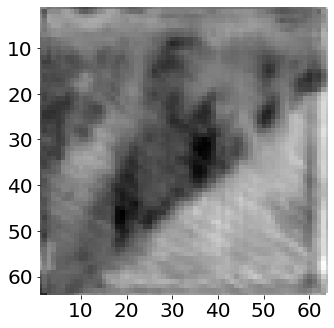}
    \caption{\textcolor{black}{\scalebox{.9}{FBPConvNet (0.584)}}}
    \label{fig:FBPConvNet barb64}
\end{subfigure}
\begin{subfigure}{0.16\textwidth}
    \centering
    \includegraphics[width=\textwidth]{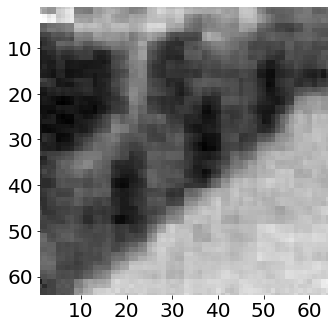}
    \caption{MADUN (0.718)}
    \label{fig:MADUN barb64}
\end{subfigure}
\begin{subfigure}{0.16\textwidth}
    \centering
    \includegraphics[width=\textwidth]{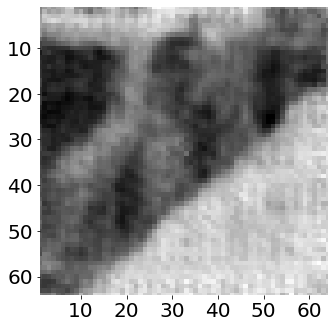}
    \caption{\textcolor{black}{\scalebox{.9}{FISTA-Net (0.721)}}}
    \label{fig:FISTANet barb64}
\end{subfigure}
\begin{subfigure}{0.16\textwidth}
    \centering
    \includegraphics[width=\textwidth]{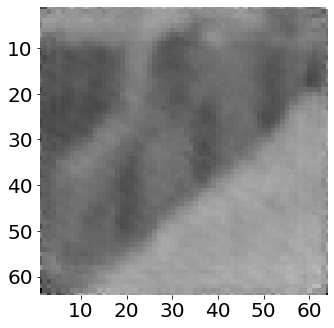}
    \caption{\textcolor{black}{\scalebox{.9}{iPiano-Net (0.787)}}}
    \label{fig:iPianoNet barb64}
\end{subfigure}
\begin{subfigure}{0.16\textwidth}
    \centering
    \includegraphics[width=\textwidth]{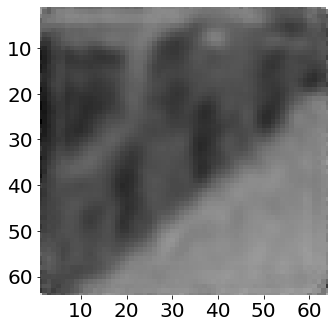}
    \caption{\scalebox{.9}{ISTA-Net$^+$ (0.751)}}
    \label{fig:ISTA-Net barb64}
\end{subfigure}
\begin{subfigure}{0.16\textwidth}
    \centering
    \includegraphics[width=\textwidth]{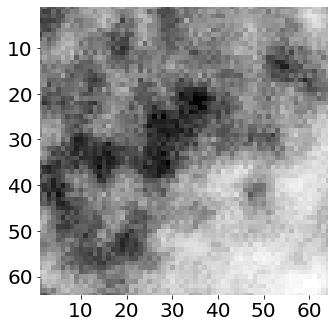}
    \caption{\textcolor{black}{\scalebox{.9}{iRadonMAP (0.185)}}}
    \label{fig:iRadonMap barb64}
\end{subfigure}
\begin{subfigure}{0.16\textwidth}
    \centering
    \includegraphics[width=\textwidth]{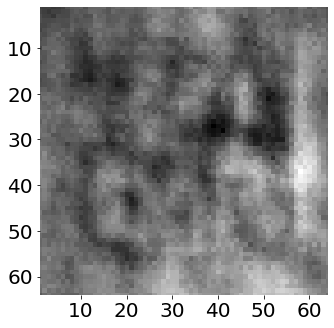}
    \caption{ReconNet (0.186)}
    \label{fig:ReconNet barb64}
\end{subfigure}
\caption{Image reconstructions (SSIM) using CG-Net \textcolor{black}{and nine other deep learning methods} for $32\times 32$ and $64\times 64$ Barbara images. The sensing matrix, $\Psi$, is a Radon transform at 15 uniformly spaced angles, the dictionary, $\Phi$, is a biorthogonal wavelet transformation, and each measurement has an SNR of 60dB. Our method CG-Net, shown in (\ref{fig:gCG-Net barb32}) and (\ref{fig:gCG-Net barb64}), performs best.}
\label{fig:barb reconstructions}
\end{figure*}

\textcolor{black}{For a visual comparison of the methods, Fig. \ref{fig:dog reconstructions} shows the reconstructions, plus SSIM values, of a $32\times 32$ dog image from a 60dB SNR Radon transform at 10 uniformly spaced angles. Next,  Fig. \ref{fig:truck reconstructions} shows the reconstructions of a $32\times 32$ truck image from a 60dB SNR Radon transform at 6 uniformly spaced angles. Finally, Fig. \ref{fig:barb reconstructions} shows the reconstructions of a $32\times 32$ and $64\times 64$ Barbara snippet each from a 60dB SNR Radon transform at 15 uniformly spaced angles. All reconstructions are performed after training on a dataset of only 20 samples. In all figures, we see CG-Net producing higher quality images visually and by SSIM than the \textcolor{black}{nine} comparison methods.}

To further highlight the applicability of CG-Net we consider alternative sensing matrices, $\Psi$, and dictionaries, $\Phi$ as summarized in Table \ref{table:CG-Net Extension}. As typical in CS applications, we consider $\Psi\in\mathbb{R}^{m\times n}$ as a Gaussian matrix where $\frac{m}{n}$ is the sampling ratio. Also, we consider a DCT as an alternative representation dictionary. The training dataset for each experiment consists of 2000 measurement, coefficient pairs. Note, for each CS problem, we initialize every $\lambda_k = 10^2$.

Table \ref{table:CG-Net Extension} provides the average SSIM and PSNR, across 200 test image reconstructions, with $99\%$ confidence intervals for various $\Psi$ and $\Phi$. \textcolor{black}{As LEARN$^{++}$, LPD, FBPConvNet, and iRadonMAP are CT-specific reconstructions the CS problem is not directly applicable and thus these comparisons are omitted from Table \ref{table:CG-Net Extension}.} Again, CG-Net outperforms or performs comparably to the competitive methods in these alternative sensing matrices and dictionary schemes.
\textcolor{black}{We remark that an advantage of CG-Net is its applicability to any general linear inverse problem while many of the comparison methods are only for the specific problem of image estimation. It remains a point of future work to study CG-Net and the comparative methods for non-image estimation tasks.}

\begin{table*}[t]
\caption{Average SSIM/PSNR with $99\%$ confidence intervals for \textcolor{black}{six} deep learning image reconstruction methods when training and testing on alternative sensing matrices, $\Psi$, and dictionaries, $\Phi$. Here, $\Psi\in\mathbb{R}^{m\times n}$ is a Radon transform, at 15 or 10 uniformly spaced angles, or a Gaussian matrix, with 0.5 or 0.3 sampling ratio $\left(\textnormal{defined as } \frac{m}{n}\right)$, as in compressive sensing. Additionally, $\Phi$ is a biorthogonal wavelet or discrete cosine transformation. Each measurement is corrupted with noise at an SNR of 60dB. Every method reconstructed two hundred $32\times 32$ CIFAR10 images after training on a set of 2000 samples. In all cases, our method CG-Net, highlighted in \textbf{bold}, performs better or comparably to all other approaches.}
\adjustbox{max width=\textwidth}{
\begin{tabular}{lc|c|c|c|c|c}
\multicolumn{7}{c}{SSIM($\times 10^2)$/PSNR for $32\times 32$ CIFAR10 images} \\
\hline
$\Phi$ & \multicolumn{2}{c|}{Discrete Cosine Transformation} & \multicolumn{2}{c|}{Biorthogonal Wavelet Dictionary} & \multicolumn{2}{c}{Discrete Cosine transform}  \\
$\Psi$ & \multicolumn{1}{c}{15 Angles}  & \multicolumn{1}{c|}{10 Angles} &  \multicolumn{1}{c}{0.5 sampling ratio}  & \multicolumn{1}{c|}{0.3 sampling ratio} &  \multicolumn{1}{c}{0.5 sampling ratio}  & \multicolumn{1}{c}{0.3 sampling ratio} \\
\hline
\scalebox{.85}{CG-Net} & $\bm{92.6}\pm \bm{.44}$/$\bm{29.2}\pm \bm{.44}$ & $\bm{84.3}\pm \bm{.80}$/$\bm{25.7}\pm \bm{.44}$ & $\bm{84.4}\pm \bm{.80}$/$\bm{25.3}\pm \bm{.49}$ & $\bm{70.7}\pm \bm{1.1}$/$\bm{22.2}\pm \bm{.48}$ & $\bm{89.4}\pm \bm{.72}$/$\bm{26.9}\pm \bm{.37}$ & $\bm{78.8}\pm \bm{1.1}$/$\bm{23.2}\pm \bm{.36}$ \\
\scalebox{.85}{MADUN} &  $90.9\pm .53$/$28.1\pm .42$ & $82.5\pm .84$/$25.1\pm .42$ & $72.0\pm .98$/$22.2\pm .38$ & $62.8\pm 1.2$/$20.9\pm .41$ & $79.2\pm .95$/$23.5\pm .37$ & $70.6\pm 1.2$/$21.9\pm .41$ \\
\textcolor{black}{\scalebox{.85}{FISTA-Net}} & \textcolor{black}{$91.6\pm .53$/$27.8\pm .41$} & \textcolor{black}{$83.5\pm .92$/$25.0\pm .41$} & \textcolor{black}{$75.8\pm 1.1$/$23.0\pm .39$} & \textcolor{black}{$66.5\pm 1.4$/$21.4\pm .42$} & \textcolor{black}{$75.8\pm 1.1$/$23.0\pm .39$} & \textcolor{black}{$66.5\pm 1.4$/$21.4\pm .42$} \\
\textcolor{black}{iPiano-Net} & \textcolor{black}{$92.9\pm .57$/$28.9\pm .49$} & \textcolor{black}{$85.0\pm .88$/$25.5\pm .42$} & \textcolor{black}{$88.7\pm .64$/$26.5\pm .39$} & \textcolor{black}{$79.5\pm 1.0$/$24.1\pm .44$} & \textcolor{black}{$88.7\pm .64$/$26.5\pm .39$} & \textcolor{black}{$79.5\pm 1.0$/$24.1\pm .44$} \\
\scalebox{.83}{ISTA-Net$^+$} &  $92.7\pm .48$/$28.7\pm .44$ & $85.2\pm .85$/$25.6\pm .42$ & $85.5\pm .72$/$25.6\pm .39$ & $80.1\pm .99$/$24.2\pm .43$ & $85.5\pm .72$/$25.6\pm .39$ & $80.1\pm .99$/$24.2\pm .43$ \\
\scalebox{.85}{ReconNet} &  $68.0\pm 1.3$/$22.0\pm .35$ & $64.2\pm 1.5$/$21.5\pm .38$ & $77.4\pm 1.0$/$23.4\pm .40$ & $65.3\pm 1.3$/$21.2\pm .39$ & $77.4\pm 1.0$/$23.4\pm .40$ & $65.3\pm 1.3$/$21.2\pm .39$ \\
\end{tabular}
}
\label{table:CG-Net Extension}
\end{table*}  

\textcolor{black}{Lastly, as shown in Fig. \ref{fig:compare}, Table \ref{table:CG-Net}, and Table \ref{table:CG-Net Extension} the unrolled DNN methods have an advantage over standard DNN methods in low training scenarios. This is perhaps unsurprising given that the unrolled methods enforce some solution structure, through data consistency layers (typically a gradient step on a data fidelity measure), providing better initialization over standard DNN methods with no such required structure. That is, with no training the reconstructions from an unrolled DNN method, generally, are closer to the actual signal of interest as compared to the reconstructions from a standard DNN method. This likely leads to unrolled methods requiring less learning for ample performance and thus succeeding in low training scenarios. To this point, the excellent performance of our CG-Net method in low training, as compared to state-of-the-art DNN and standard DNN methods, can be expected as CG-Net enforces an appreciable solution structure by incorporating the CG prior through the Tikhonov update layers and Hadamard product output. However, given enough training data, standard DNN methods could outperform unrolled methods, which may now have hindered learning capacities due to the required solution structure. As full coverage of a comparison between unrolled and standard DNN methods is beyond the scope of this paper; we leave it to a future study.} 
\textcolor{black}{As full coverage of a comparison between unrolled and standard DNN methods is beyond the scope of this paper, we leave it to a future study.
}

\subsection{Ablation Study}

We consider the effect of removing the learned steepest descent matrix $B_k^j$ from CG-Net by fixing it during training. Two possibilities are employed, a gradient CG-Net (gCG-Net) where $B_k^j = I$ and a Newton CG-Net (nCG-Net) where $B_k^j = \left(H_{F;\pmb{z}}(\pmb{u}_{k-1},\pmb{z}_k^{j-1})\right)^{-1}$ where $H_{F;\pmb{z}}$ is given in (\ref{eqn:Hess z}). All other aspects of training are identical to CG-Net in Section \ref{sec:NN numerical results} except that for nCG-Net we scale the input measurements by $e^{-4}$ and initialize $a_0 = a_k^{(j)} = 1$ and $b_0 = b_k^{(j)} = e$.

Shown in Table \ref{tab:ablation} is the average SSIM of 200 test image reconstructions -- from Radon transforms at 15, 10, and 6 uniformly spaced angles with an SNR of 60dB or 40dB -- when varying the amount of training data. With fewer than 100 training data samples both gCG-Net and nCG-Net structures perform comparably or outperform the fully general version of CG-Net. Likely, this is due to gCG-Net and nCG-Net having fewer parameters to be fit for these cases and thus avoiding overfitting. With more than 100 training data samples CG-Net outperforms both gCG-Net and nCG-Net.

\begin{table}[!ht]
    \caption{Ablation study for CG-Net. Average SSIM ($\times 10^2)$ and PSNR for $32\times 32$ image reconstructions from a Radon transform, at several different amounts of uniformly spaced angles, with a set SNR. We find that gCG-Net and nCG-Net outperform the full version of CG-Net in the lowest training scenario since fewer parameters must be fit for these cases.}
    \label{tab:ablation}
    \centering
    \begin{tabular}{c|c|c|c|c|c|c|c}
    && \multicolumn{6}{c}{Training Dataset Size} \\
     Method    & (Angles,SNR) & \multicolumn{2}{c|}{20} & \multicolumn{2}{c|}{100} & \multicolumn{2}{c}{500} \\
     && SSIM & PSNR & SSIM & PSNR & SSIM & PSNR \\
     \hline
      gCG-Net   & & 90.0 & 27.7 & 90.8 & 28.2 & 91.2 & 28.5 \\
      nCG-Net & (15,60) & 90.7& 28.1 & 90.7 & 28.2 & 90.8 & 28.2\\
      CG-Net & &89.9&27.7 & 90.8 & 28.1 & 92.0 & 28.8 \\
           \hline
      gCG-Net   & &81.6 & 24.9 & 82.4 & 25.1 & 82.8 & 25.3 \\
      nCG-Net & (10,60) & 81.7 & 24.9 & 81.8 &24.9 & 82.0 & 25.0\\
      CG-Net & & 81.5&24.8 & 82.4&25.1 & 83.6 & 25.5\\
           \hline
      gCG-Net   && 67.6 & 21.8 & 68.1 & 21.8 & 68.7 & 22.0 \\
      nCG-Net & (6,60) & 67.1 & 21.7 & 67.2 & 21.7 & 67.9 & 21.8\\
      CG-Net & & 67.6 &21.8 & 68.1 &21.9& 69.5 & 22.1 \\
                 \hline
      gCG-Net   && 66.1 & 21.6 & 66.4 & 21.6 & 66.7 & 21.6\\
      nCG-Net & (6,40) & 65.5 & 21.5 & 65.6 & 21.5 & 66.2 & 21.6 \\
      CG-Net & & 66.1 &21.6 & 66.4 &21.6& 67.7 & 21.7
    \end{tabular}
\end{table}

\section{\textcolor{black}{Time Requirements \& Parameters}} \label{sec:time reqs}

Table \ref{tab:computational time} lists the average computational time per image, in \textcolor{black}{milli}seconds, across 200 test image reconstructions running on a 64-bit Intel(R) Xeon(R) CPU E5-2690. We see that CG-LS is slower than the comparative iterative methods although it has yet to be optimized for computational efficiency and speed. Fortunately, the required computational time is reduced by more than a factor of 100 for CG-Net while producing the same or slightly improved quality image reconstructions \textcolor{black}{as compared to CG-LS. This reduced computational time of CG-Net is one of the primary advantages over using CG-LS.} 

\begin{table}[!ht]
\caption{Average reconstruction time of $32\times 32$ images from Radon transform measurements at 15 uniform angles.}
    \centering
    \adjustbox{max width=\columnwidth}{
    \begin{tabular}{c|c|c|c|c|c|c|c}
    Method     &  nCG-LS & gCG-LS & \textcolor{black}{FISTA} &  $\ell_1$-LS & FBP & BCS & CoSaMP \\
    Time \textcolor{black}{(ms)}     & $1.5\times 10^5$ & $8.7\times 10^4$ & \textcolor{black}{$410$} & $180$ & $1.5$ & 622 & 407 \\
    \hline
        Method  & CG-Net & ReconNet  & \textcolor{black}{LPD} & \multicolumn{2}{c|}{\textcolor{black}{LEARN$^{++}$}}  & \multicolumn{2}{c}{\textcolor{black}{iRadonMAP}}  \\
    Time \textcolor{black}{(ms)}  & 765 & $0.65$ & \textcolor{black}{$4.7$} & \multicolumn{2}{c|}{\textcolor{black}{10.6}} & \multicolumn{2}{c}{\textcolor{black}{4.5}} \\
    \hline
       \textcolor{black}{Method}  & \textcolor{black}{iPiano-Net} & \textcolor{black}{FISTA-Net}  & MADUN & \multicolumn{2}{c|}{ISTA-Net$^+$}  & \multicolumn{2}{c}{\textcolor{black}{FBPConvNet}}  \\
    \textcolor{black}{Time (ms)}  & \textcolor{black}{13.0} & \textcolor{black}{6.4} & $26.0$ & \multicolumn{2}{c|}{$7.9$} & \multicolumn{2}{c}{\textcolor{black}{2.0}}
    \end{tabular}
    }
    \label{tab:computational time}
\end{table}

\textcolor{black}{Nevertheless,} CG-Net lags in computational time against the comparative DNN methods that take a fraction of the time to reconstruct an image. This is likely an outcome of the required linear solver to calculate the inverse in (\ref{eqn:Tikhonov solution}) to update $\pmb{u}$ and the required eigendecomposition in (\ref{eqn:eigendecomp}). \textcolor{black}{Instead, the comparative methods} solely consist of convolutions or matrix, vector products that are appreciably faster to implement than linear solvers.

\textcolor{black}{Finally, Table \ref{table:parameters} provides the number of parameters for CG-Net and all nine comparative deep learning-based methods for linear inverse problems. We remark that CG-Net has the fewest parameters to be trained out of the compared methods, which may be a contributing factor to the success of CG-Net when small training datasets are used.} 

\textcolor{black}{Increasing the number of unrolled iterations in CG-Net to produce a DNN with a model complexity matching the average of the compared methods remains a point of future work. In particular, with larger training datasets, CG-Net can be outperformed by the best-performing comparative method where we may be required to increase the number of unrolled iterations in CG-Net, thereby increasing the number of learned parameters, to more closely match the comparative methods.} 

\begin{table}[!ht]
\caption{\textcolor{black}{Parameter count for our CG-Net and each compared deep learning method when reconstructing a $32\times 32$ image from Radon transform measurements at 15 uniformly spaced angles.}}
    \centering
    {\color{black}
    \adjustbox{max width=\columnwidth}{
    \begin{tabular}{c|c||cccc||c|c}
        Method  & Parameters ($\times 10^5$) & & & & & Method & Parameters ($\times 10^5$) \\
        \hline\hline
        CG-Net & 0.62 & & & & & ISTA-Net$^+$ & 3.37 \\
        LPD  & 2.53 & & & & & MADUN & 29.7\\
        LEARN$^{++}$ & 12.0 & & & & & FISTA-Net & 0.75\\
        iRadonMAP & 8.33 & & & & & iPiano-Net & 19.3\\
        FBPConvNet & 7.09 & & & & & ReconNet & 7.30
    \end{tabular}
    }
    }
    \label{table:parameters}
\end{table}

\section{Conclusion and Future Work}

Informed by the powerful statistical representation of signal coefficients through the compound Gaussian prior, we developed a novel iterative signal reconstruction algorithm, named CG-LS, that enforces the CG prior. CG-LS is based upon a regularized least squares estimate of the signal coefficients where the regularization, equivalent to the negative log prior from a MAP estimate, is chosen to capture the fundamental statistics of the CG prior. We conducted a rigorous theoretical characterization of CG-LS, which gave important insights into the implementation of the algorithm. Numerical validation of CG-LS was conducted, which showed a significant improvement over other state-of-the-art image reconstruction algorithms.

Furthermore, we have applied algorithm unrolling to CG-LS, creating a deep neural network named CG-Net. To the best of our knowledge, CG-Net is the first unrolled DNN for natural and tomographic image reconstruction to be fundamentally informed by a CG prior. Multiple datasets were used to train and test CG-Net where, after each training, CG-Net was shown to outperform other unrolled DNNs as well as standard DNNs. In particular, CG-Net significantly outperforms the comparative methods in low training for both tomographic imaging and CS applications. Finally, a comparison of the computational time to reconstruct a single image from each iterative and DNN method was discussed. CG-Net significantly improved upon the necessary time over CG-LS, but still falls short of the speed other DNN methods achieve for image reconstruction. 

Improving the speed of CG-Net serves as one direction of future work for which we suggest a technique. When training CG-Net the eigendecomposition in (\ref{eqn:eigendecomp}) must be implemented within each $Z_k^j$ layer call as the entries of $L_k^j$ are actively being updated and thus, we need to actively ensure $(L_k^j+(L_k^j)^T)/2$ stays positive definite. Using CG-Net post-training, the eigendecomposition only needs to be implemented once upon instantiating the model with pre-trained network parameters. 

\textcolor{black}{Analyzing the performance of CG-Net for larger image reconstructions also serves as a key point of future work. Here we presented fundamental development, theory, and results for a CG-inspired iterative reconstruction algorithm and unrolled DNN. As CG-LS continues to outperform competitive iterative methods when reconstructing larger images, we anticipate CG-Net will similarly outperform or perform comparably to competitive methods in low training scenarios when applied to larger image reconstructions. For larger images, the training time will be costly, however, so optimization of the CG-Net implementation may be required. Alternatively, a technique, common in CS applications, of splitting an image into disjoint blocks of small size and then measuring and reconstructing separately~\cite{MADUN, song2023MAPUN} could be employed for future experimental evaluation.}

\textcolor{black}{As we focus on laying the fundamental groundwork for solving linear inverse problems while incorporating a CG prior, further future work could extend CG-LS and CG-Net to non-linear inverse problems where matrix $A$ is replaced by a non-linear function $\mathcal{F}_A$. Alternatively, a linearization of a non-linear forward operator, which is an adequate approximation for many non-linear inverse problems under suitable conditions (e.g. radar imaging~\cite{munson1983tomographic}), could be used. Therefore, the theoretical and empirical groundwork laid in this paper can serve as the basis for future applications of our CG-based inverse problem methodology to CT, radar, or other experimental data.}

Another open question remains as to the generalizability of the CG-Net model through replacing aspects of CG-based optimization with relevant neural network structures. For instance, the CG prior depends on the choice of nonlinearity $h$, or the corresponding choice of its inverse $f$. A future implementation of CG-Net could learn $f$ by approximating it with a sub-network embedded inside of CG-Net. Such an extension of CG-Net could expand its applicability by both no longer requiring a user-specified function $f$, as well as providing CG-Net with a greater learning capacity.

\textcolor{black}{Finally, further empirical comparison between standard DNN approaches and algorithm-unrolled-based DNN approaches for solving inverse problems stands as another crucial goal of future work for deep-learning-based inverse problems as a whole. Briefly discussed in Section \ref{sec:NN numerical results}, unrolling approaches, such as CG-Net, appear to have a clear advantage when small training data sets are available. However, given enough training data, standard DNN approaches could have an advantage over unrolled approaches in a couple of ways. One is that unrolling approaches have required solution structure and restriction through data consistency layers. Additionally, unrolled approaches often share trained weights across network layers leading to a recurrent network structure, which can suffer from vanishing and exploding gradient problems.}

\section{Appendix} \label{apndx:proofs}

\subsection{Proposition \ref{prop:existence of a minimizer} Details} \label{apndx:prop existence of a minimizer}

First, a lemma on the eigenvalues of a $2\times 2$ block matrix.

\begin{lemma} \label{lemma:eigenvalues}
For $\bm{a}, \bm{b}, \bm{c}\in\mathbb{R}^n$ let $A = D\{\bm{a}\}, B = D\{\bm{b}\},$ and $C = D\{\bm{c}\}$. The eigenvalues of $M = \begin{bmatrix}
A & B \\
B & C
\end{bmatrix}$
are
\begin{align*}
\lambda^{\pm}_i = (a_i+c_i\pm \sqrt{(a_i-c_i)^2+4b_i^2}\,)/2, \hspace{.5cm}  i = 1, \ldots, n.
\end{align*}
\end{lemma}
\begin{proof}
Note $B$ and $C-r I$ commute. Thus, using~\cite{silvester2000determinants}
\[
\det(M-rI) = \det((A-rI)(C-rI)-B^2),
\]
which is zero for $r = \lambda_i^{\pm}$.
\end{proof}
We now prove Proposition \ref{prop:existence of a minimizer}.
\begin{proof}[Proof of Proposition \ref{prop:existence of a minimizer}]
Let $\pmb{\rho}:\mathbb{R}^n\times\mathcal{Z}^n\to\mathbb{R}^n$ be given by
\begin{align}
    \pmb{\rho}(\pmb{u},\pmb{z}) &= A^T(AD\{\pmb{z}\}\pmb{u}-\pmb{y}). \label{eqn:rho}
\end{align}
Let $i\in\{1,2,\ldots,n\}$. Define $\mathcal{Z}_i(z) \coloneqq \{\pmb{z}\in[z_0,b]^n : z_i = z\}$ for $z\in [z_0,b]$. Since $f^2(z)$ is differentiable and achieves a local minimum at $z_0 > 0$, then $f'(z_0)f(z_0) = 0$. Then
\begin{align*}
    \mathcal{F}_i(\pmb{z}) = -2\lambda z_0 v_i(\pmb{z})^2 \leq 0 \hspace{.2cm} \textnormal{ for all } \hspace{.15cm} \pmb{z}\in \mathcal{Z}_i(z_0)
\end{align*}
as $\lambda > 0$ and $\pmb{v}(\pmb{z}) = [v_1(\pmb{z}),\ldots,v_n(\pmb{z})]^T$ and $\pmb{\mathcal{F}}(\pmb{z}) = [\mathcal{F}_1(\pmb{z}),\ldots, \mathcal{F}_n(\pmb{z})]^T$ are given in (\ref{eqn:v function z}) and (\ref{eqn:z root requirement}), respectively. Writing $\pmb{y} = s\widetilde{\pmb{y}}$ we define $\widetilde{\pmb{v}}(\pmb{z}) = A^T(A_{\pmb{z}}A_{\pmb{z}}^T+\lambda I)^{-1}\widetilde{\pmb{y}}$ and
\[
\widetilde{v}_{\max}(b) =  \underset{1\leq i\leq n}{\max}\,\,\, \underset{\pmb{z}\in \mathcal{Z}_i(b)}{\max}\,\, |\widetilde{v}_i(\pmb{z})|.
\]
Let $\pmb{z}_b \in \mathcal{Z}_i(b)$. Note, $s^2\left(\widetilde{v}_{\max}(b)\right)^2 \geq s^2\widetilde{v}_i(\pmb{z}_b)^2 = v_i(\pmb{z}_b)^2$. Hence, if
\begin{align}
\frac{1}{\left(\widetilde{v}_{\max}(b)\right)^{2}}\frac{f'(b)f(b)}{b} \geq \frac{\lambda}{\mu} s^2 \label{eqn:scale for stationary point}
\end{align}
then $\mu f'(b)f(b) \geq \lambda b s^2 \left(\widetilde{v}_{\max}(b)\right)^2 \geq \lambda b v_i(\pmb{z}_b)^2$. Thus,
\[
\mathcal{F}_i(\pmb{z}) = -\lambda b v_i(\pmb{z})^2 + \mu f'(b)f(b) \geq 0 \hspace{.2cm} \textnormal{ for all } \hspace{.15cm} \pmb{z}\in\mathcal{Z}_i(b).
\]
By the Poincare-Miranda theorem~\cite{kulpa1997poincare}, when $s, \lambda$, and $\mu$ satisfy (\ref{eqn:scale for stationary point}) there exists a $\pmb{z}^*\in [z_0,b]^n$ such that $\pmb{\mathcal{F}}(\pmb{z}^*) = \pmb{0}$. Thus, by Lemma \ref{lemma:z root requirement}, $[\pmb{u}^*,\pmb{z}^*]$ with $\pmb{u}^* = \pmb{z}^*\odot\pmb{v}(\pmb{z}^*)$ is a stationary point of (\ref{eqn:cost function}).

Next, we assume that (\ref{eqn:scale for stationary point}) is satisfied and show the Hessian, $H_F = H_F(\pmb{u},\pmb{z})$, of (\ref{eqn:cost function}) at $[\pmb{u}^*, \pmb{z}^*]$ is positive definite. Note
\begin{align}
\setlength\arraycolsep{4pt}
    \scalebox{1}{${\displaystyle\frac{H_F}{2} = \begin{bmatrix}
    A_{\pmb{z}}^T \\
    A_{\pmb{u}}^T
    \end{bmatrix} \begin{bmatrix}
    A_{\pmb{z}} & A_{\pmb{u}} 
    \end{bmatrix} + \begin{bmatrix}
    \lambda I & D\{\pmb{\rho}(\pmb{u},\pmb{z})\} \\
    D\{\pmb{\rho}(\pmb{u},\pmb{z})\} & \mu D\{\pmb{h}_f(\pmb{z})\}
    \end{bmatrix}}$} \label{eqn:cost function Hessian}
\end{align}
for $\pmb{h}_f$ and $\pmb{\rho}$ given in (\ref{eqn:hf}) and (\ref{eqn:rho}), respectively. Observe
\begin{align*}
    \scalebox{.98}{${\displaystyle \pmb{\rho}(\pmb{z}\odot\pmb{v}(\pmb{z}), \pmb{z}) = A^T(A_{\pmb{z}} A_{\pmb{z}}^T (A_{\pmb{z}}A_{\pmb{z}}^T + \lambda I)^{-1}-I)\pmb{y} = -\lambda \pmb{v}(\pmb{z}).}$}
\end{align*}
Thus, at $\pmb{x}^* = [\pmb{u}^*,\pmb{z}^*]$, using that $\pmb{u}^* = \pmb{z}^*\odot\pmb{v}(\pmb{z}^*)$ we have
\begin{align*}
\setlength\arraycolsep{3.25pt}
  \scalebox{1}{${\displaystyle  \begin{bmatrix}
    \lambda I & D\{\pmb{\rho}(\pmb{x}^*)\} \\
    D\{\pmb{\rho}(\pmb{x}^*)\} & \mu D\{\pmb{h}_f(\pmb{z}^*)\}
    \end{bmatrix} = \begin{bmatrix}
    \lambda I & -\lambda D\{\pmb{v}(\pmb{z}^*)\} \\
    -\lambda D\{\pmb{v}(\pmb{z}^*)\} & \mu D\{\pmb{h}_f(\pmb{z}^*)\}
    \end{bmatrix}}$}
\end{align*}
which, by Lemma \ref{lemma:eigenvalues}, has eigenvalues 
\[
\lambda_i^{\pm} = \lambda(1+c_i\pm \sqrt{(1-c_i)^2+4v_i(\pmb{z}^*)^2}\,)/2
\]
for $c_i \coloneqq \frac{\mu}{\lambda}[\pmb{h}_f(\pmb{z}^*)]_i = \frac{\mu}{\lambda}\left(f''(z_i^*)f(z_i^*)+f'(z_i^*)^2\right)$. Since $f^2(z)$ is strictly convex on $[a,b]$, then $f''(z_i^*)f(z_i^*)+f'(z_i^*)^2 > 0$ implying $c_i > 0$ and thus $\lambda_i^+ > 0$ for all $i = 1, 2, \ldots, n$. 

Now $\lambda_i^- > 0$ if and only if $c_i > v_i(\pmb{z}^*)^2.$ Define
\begin{align*}
    h_{f\min} &= \min_{z\in [z_0,b]}\,\,\, f''(z)f(z)+f'(z)^2 \\
    \widetilde{v}_{\max} &= \max_{1\leq i\leq n}\,\,\, \max_{\pmb{z}\in [z_0,b]^n}\,\, |\widetilde{v}_i(\pmb{z})|. 
\end{align*}
Note $s^2\widetilde{v}^2_{\max} \geq s^2 \widetilde{v}_i(\pmb{z}^*)^2 = v_i(\pmb{z}^*)^2$ and $[\pmb{h}_f(\pmb{z}^*)]_i \geq h_{f\min}$ for all $i = 1, 2, \ldots, n$. Hence, if
\begin{align}
    \frac{h_{f\min}}{\widetilde{v}^2_{\max}} > \frac{\lambda}{\mu}s^2 \label{eqn:scale for PD}
\end{align}
then $c_i = \frac{\mu}{\lambda}[\pmb{h}_f(\pmb{z}^*)]_i \geq \frac{\mu}{\lambda} h_{f\min} > s^2\widetilde{v}^2_{\max} \geq v_i(\pmb{z}^*)^2$ and thus $\lambda_i^- > 0.$ Therefore, when $s$, $\lambda$, and $\mu$ satisfy (\ref{eqn:scale for PD}) then (\ref{eqn:cost function Hessian}) at $[\pmb{u}^*, \pmb{z}^*]$ is the sum of a positive semi-definite and positive definite matrix, implying the Hessian is positive definite. Lastly, note $f'(b)f(b) > 0$ and $h_{f\min} > 0$ as $f^2(z)$ is strictly convex on $[a,b]$ and achieves a minimizer $z_0\in (a,b)$. Thus (\ref{eqn:scale for stationary point}) and (\ref{eqn:scale for PD}) can be satisfied for positive $\lambda, \mu$, and $s$.
\end{proof}

Comparing (\ref{eqn:scale for stationary point}) and (\ref{eqn:scale for PD}) we note $\widetilde{v}^2_{\max}\geq \left(\widetilde{v}_{\max}(b)\right)^2$. Hence, if an approximation of $\widetilde{v}_{\max}$, which we denote by $\widehat{v}_{\max}$, is obtained then we can satisfy both (\ref{eqn:scale for stationary point}) and (\ref{eqn:scale for PD}) choosing  $\min\{f'(b)f(b)/b, h_{f\min}\}/\widehat{v}^2_{\max} > \lambda s^2/\mu.$ Note, for a given $f$, both $f'(b)f(b)/b$ and $h_{f\min}$ can easily be obtained.
\subsection{Proposition \ref{prop:steepest descent step bound} Details} \label{apndx:prop steepest descent step bound}
For initial estimates $\pmb{u}_0$ and $\pmb{z}_0$, define the sublevel set 
    \begin{align*}
   S(\pmb{u}_0,\pmb{z}_0) = \{(\pmb{u},\pmb{z})\in \mathbb{R}^n\times \mathcal{Z}^n: F(\pmb{u},\pmb{z})\leq F(\pmb{u}_0,\pmb{z}_0)\}.
\end{align*}
Since $F$ is continuous then $S(\pmb{u}_0,\pmb{z}_0)$ is closed and since $F$ is coercive in both $\pmb{u}$ and $\pmb{z}$ then $S(\pmb{u}_0,\pmb{z}_0)$ is bounded. Thus, $S(\pmb{u}_0,\pmb{z}_0)$ is compact, which implies that $F$ has Lipschitz continuous gradient on $S(\pmb{u}_0,\pmb{z}_0)$. Let  $L \coloneqq L_{\pmb{z}}(\pmb{u}_0,\pmb{z}_0)$ be the Lipschitz constant, on $S(\pmb{u}_0,\pmb{z}_0)$, for $\nabla_{\bm{z}} F$.

\begin{proof}[Proof of Proposition \ref{prop:steepest descent step bound}]
Let $\pmb{\zeta} =\pmb{z}_k^{j}$, $\pmb{\xi} = \pmb{z}_k^{j-1}$, and $\eta = \eta_k^{(j)}$. As $\nabla G_k$ is Lipschitz continuous with constant $L$, then
\begin{align}
    \scalebox{.98}{${\displaystyle G_{k}\left(\pmb{\zeta}\right) \leq G_{k}\left(\pmb{\xi}\right) + \eta \nabla G_k\left(\pmb{\xi}\right)^T \pmb{d}_k^j\left(\pmb{\xi}\right) + \frac{L}{2} \left|\left|\eta\pmb{d}_k^j\left(\pmb{\xi}\right)\right|\right|_2^2.}$} \label{eqn:quadratic Lipschitz bound}
\end{align}
 for sufficiently small $\eta$. Using (\ref{eqn:steepest descent direction}) note, $\nabla G_k\left(\pmb{x}\right)^T\pmb{d}_k^j(\pmb{x}) = -||\nabla G_k(\pmb{x})||^2_{*(k,j)}$ and for the Euclidean bound constant $\gamma_{(k,j)}$ of $||\cdot||_{(k,j)}$ we have $\gamma^2_{(k,j)} ||\pmb{d}_k^j(\pmb{x})||_2^2 \leq ||\nabla G_k(\pmb{x})||^2_{*(k,j)}$. Combining with (\ref{eqn:quadratic Lipschitz bound}) gives
\begin{align}
 \scalebox{.95}{${\displaystyle  G_{k}\left(\pmb{\zeta}\right) \leq  G_{k}\left(\pmb{\xi}\right)- \eta\left(1-\frac{\eta L}{2\gamma^2_{(k,j)}}\right) \left|\left|\nabla G_k\left(\pmb{\xi}\right)\right|\right|^2_{*(k,j)}.}$} \label{eqn:z step bound A}
\end{align}

From~\cite{boyd2004convex} a backtracking line employs two user-chosen parameters $\alpha \in (0,1/2]$ and $\beta\in (0,1)$ where the step size $\eta$ is chosen to be a multiple of $\beta$ satisfying
\begin{align}
  G_{k}\left(\pmb{\zeta}\right) &\leq 
   G_{k}\left(\pmb{\xi}\right)- \alpha \eta \left|\left|\nabla G_k\left(\pmb{\xi}\right)\right|\right|^2_{*(k,j)}. \label{eqn:z step bound B}
\end{align}
Note, (\ref{eqn:z step bound A}) implies (\ref{eqn:z step bound B}) when $\eta \leq \gamma_{(k,j)}^2/L$ and thus, from~\cite{boyd2004convex}, the backtracking line search step size satisfies $\eta \geq \min\{1,\beta \gamma^2_{(k,j)}/L\} > 0$. Combining with (\ref{eqn:z step bound A}) and (\ref{eqn:z step bound B}) produces (\ref{eqn:steepest descent lower bound}) when $c_{(k,j)} =\alpha \min\{1, \beta\gamma^2_{(k,j)}/L\}.$
\end{proof}

\subsection{Theorem \ref{thm:CG-LS convergence} Details} \label{apndx:thm CG-LS convergence}

Recall $G_k(\pmb{z}) = F(\pmb{u}_{k-1},\pmb{z})$. We now prove Theorem \ref{thm:CG-LS convergence}.
\begin{proof}[Proof of Theorem \ref{thm:CG-LS convergence}]
Define $\pmb{v}_k = [\pmb{u}_k, \pmb{z}_k]$, $\pmb{v}_{k-\frac{1}{2}} = [\pmb{u}_{k-1}, \pmb{z}_k]$  and $\pmb{v}_{k-\frac{1}{2}}^j = [\pmb{u}_{k-1}, \pmb{z}_k^j]$. We let $F(\pmb{v}_k) = F(\pmb{u}_k,\pmb{z}_k)$ and similarly for $F(\pmb{v}_{k-\frac{1}{2}}^j)$ and $F(\pmb{v}_{k-\frac{1}{2}}).$ Note, for all $k\geq 1$, $\pmb{v}_{k-\frac{1}{2}}^{0} = \pmb{v}_{k-1}$ and $\pmb{v}_{k-\frac{1}{2}}^J = \pmb{v}_{k-\frac{1}{2}}$. Summing over $j$ in (\ref{eqn:steepest descent lower bound}) 
\begin{align}
    F(\pmb{v}_{k-1}) - F(\pmb{v}_{k-\frac{1}{2}}) &\geq \sum_{j =1}^{J} c_{(k,j)} \left|\left|\nabla_{\bm{z}} F\left(\pmb{v}_{k-\frac{1}{2}}^{j-1}\right)\right|\right|_{*(k,j)}^2 \nonumber \\
      &\geq c_{(k,1)} ||\nabla_{\bm{z}} F(\pmb{v}_{k-1})||_{*(k,1)}^2. \label{eqn:CG-LSIA cost function decrease}
\end{align}
First, note  $\nabla_{\pmb{u}} F(\pmb{v}_{k-1}) = \pmb{0}$ since $\pmb{u}_{k-1}$ is a global minimizer of (\ref{eqn:cost function}) with $\pmb{z}$ fixed at $\pmb{z}_{k-1}$. Hence, we have $||\nabla_{\bm{z}} F(\pmb{v}_{k-1})||_{*(k,1)}^2 = ||\nabla F(\pmb{v}_{k-1})||_{*(k,1)}^2$. Second, note $F(\pmb{v}_k) \leq F(\pmb{v}_{k-\frac{1}{2}})$ since $\pmb{u}_k$ is a global minimizer of (\ref{eqn:cost function}) with $\pmb{z}$ fixed at $\pmb{z}_{k}$. Hence,  $-F(\pmb{v}_k) \geq -F(\pmb{v}_{k-\frac{1}{2}}).$ Third, define
\begin{align}
    c\coloneqq  \alpha \min\left\{1, \beta \gamma^2/L\right\} \label{eqn:CG-LS constant c}
\end{align}
and note, since $\gamma_{(k,1)}\geq \gamma$, then, from Proposition \ref{prop:steepest descent step bound} $c_{(k,1)}\geq c >0.$ 
Combining these three results with (\ref{eqn:CG-LSIA cost function decrease}) gives
\begin{align}
    F(\pmb{v}_{k-1}) - F(\pmb{v}_{k}) &\geq c ||\nabla F(\pmb{v}_{k-1})||_{*(k,1)}^2. \label{eqn:CG-LS cost function iteration dec}
\end{align}
Thus, CG-LS generates a monotonic decreasing sequence of cost function values, which are bounded below by $0$. Hence, $\{F(\pmb{v}_{k})\}_{k = 0}^\infty$ converges to a value $F^*$. Therefore, taking an infinite sum of (\ref{eqn:CG-LS cost function iteration dec}) w.r.t $k$ and using $\gamma_{*(k,1)}\geq \gamma$ gives
\begin{align*}
    \scalebox{.95}{${\displaystyle F(\pmb{v}_{0}) - F^* \geq c\sum_{k = 1}^\infty ||\nabla F(\pmb{v}_{k-1})||_{*(k,1)}^2 \geq \gamma^2c\sum_{k = 0}^\infty ||\nabla F(\pmb{v}_{k})||_2^2.}$}
\end{align*}
Implying that $\sum_{k = 0}^\infty ||\nabla F(\pmb{v}_{k})||_2^2$ converges. Thus, ${\displaystyle \lim_{k\to\infty} ||\nabla F(\pmb{v}_{k})||_2^2 \to 0}$ and so ${\displaystyle \lim_{k\to\infty} \nabla F(\pmb{v}_{k}) \to \pmb{0}}$. Finally, taking an average of (\ref{eqn:CG-LS cost function iteration dec}) gives
\[
    \frac{F(\pmb{v}_0)-F(\pmb{v}_K)}{c} \frac{1}{K} \geq \min_{1\leq k\leq K} ||\nabla F(\pmb{v}_{k})||_{*(k,1)}^2. \qedhere
\]
\end{proof}
\subsection{Theorem \ref{thm:CG-LS converges with convexity} Details} \label{apndx:thm CG-LS converges with convexity}
First, define
\[
 \mathcal{B}_{\delta}(\widehat{\pmb{u}},\widehat{\pmb{z}}) = \left\{[\pmb{u},\pmb{z}]\in\mathbb{R}^n\times\mathcal{Z}^n : \left|\left|[\pmb{u},\pmb{z}] - [\widehat{\pmb{u}},\widehat{\pmb{z}}]\right|\right|_2<\delta \right\}.
\]
for $\delta > 0$, $\widehat{\pmb{u}}\in\mathbb{R}^n,$ and $\widehat{\pmb{z}}\in\mathcal{Z}^n$. We now prove Theorem \ref{thm:CG-LS converges with convexity}.
\begin{proof}[Proof of Theorem \ref{thm:CG-LS converges with convexity}]
The non-degenerate local minimizer $[\pmb{u}^*, \pmb{z}^*]$ in Proposition \ref{prop:existence of a minimizer} satisfies, for all $\pmb{w}\in\mathbb{R}^{2n}$, $\pmb{w}^TH_F(\pmb{u}^*, \pmb{z}^*)\pmb{w} = \ell_0(\pmb{w}) \geq \lambda_{\min} > 0$ where $\lambda_{\min}$ is the minimum eigenvalue of the Hessian $H_F(\pmb{u}^*,\pmb{z}^*).$ Since (\ref{eqn:cost function}) is twice continuously differentiable then, for fixed $\pmb{w}\in \mathbb{R}^{2n}$, $Q(\pmb{u},\pmb{z}) \coloneqq \pmb{w}^TH_F(\pmb{u}, \pmb{z})\pmb{w}$ is a continuous function. Fix $\ell\in (0,\lambda_{\min})$ and let $\epsilon = \lambda_{\min}-\ell > 0$. By continuity of $Q$ there exists a $\delta > 0$ such that for all $[\pmb{u},\pmb{z}]\in \mathcal{B}_{\delta}(\pmb{u}^*,\pmb{z}^*)$ we have $|Q(\pmb{u},\pmb{z})-Q(\pmb{u}^*,\pmb{z}^*)| < \epsilon = \lambda_{\min}-\ell,$ which implies $Q(\pmb{u},\pmb{z}) \geq \ell$. Therefore, (\ref{eqn:cost function}) is strongly convex with constant $\ell$ on $\mathcal{C} = \mathcal{B}_{\delta}(\pmb{u}^*,\pmb{z}^*).$

Next, strong convexity of (\ref{eqn:cost function}) with constant $\ell$ implies that
\begin{align}
    F(\pmb{v}_i)-F^*\leq (2\ell)^{-1} ||\nabla F(\pmb{v}_i)||_2^2 \label{eqn:strong convexity implication}
\end{align}
for all $i\geq 0$. Rearranging (\ref{eqn:CG-LS cost function iteration dec}), subtracting $F^*$ from both sides, and applying $||\cdot||_{*(k,1)}\geq \gamma ||\cdot||_2$ gives
\[
 F(\pmb{v}_k) - F^* \leq F(\pmb{v}_{k-1}) - F^* - \gamma^2 c ||\nabla F(\pmb{v}_{k-1})||_2^2,
\]
which combined with (\ref{eqn:strong convexity implication}), when $i = k-1$, produces
 \begin{align}
 F(\pmb{v}_k) - F^* &\leq (1-2\ell\gamma^2 c)\left(F(\pmb{v}_{k-1})-F^*\right). \label{eqn:strongly convex CG-LS one step}
 \end{align}
 Applying (\ref{eqn:strongly convex CG-LS one step}) recursively gives (\ref{eqn:strongly convex CG-LS convergence}).
\end{proof}
\subsection{Compound Gaussian Representations} \label{apndx:compound gaussian representations}
Here, we discuss the generality of the CG prior.
\begin{prop} \label{prop:CG subsumes}
The generalized Gaussian, student's $t$, $\alpha$-stable, and symmetrized Gamma distributions are special cases of the compound Gaussian distribution. 
\end{prop}
The result of Proposition \ref{prop:CG subsumes} is found in~\cite{Scale_Mixtures, Wavelet_Trees}.  Now, we give an explicit nonlinearity, $h$, such that the CG prior reduces to a Laplace prior.
\begin{prop} \label{prop:Laplace as CG}
Let $\Upsilon$ be the cumulative distribution function (CDF) of a standard Gaussian random variable. Define
\[
h(x) = \left(-2\lambda^2\ln(1-\Upsilon(x))\right)^{1/2}.
\]
Then $\pmb{c} = h(\pmb{x})\odot \pmb{u}$, for $\pmb{x}\sim\mathcal{N}(\pmb{0}, I)$ and $\pmb{u}\sim\mathcal{N}(\pmb{0}, I)$, has a Laplace distribution. That is, $\pmb{c}\sim  \lambda\exp(-\lambda ||\pmb{c}||_1)/2$. 
\end{prop}
\begin{proof}
A Laplace random variable $X\sim \lambda\exp(-\lambda x)/2$ is decomposed as $X = \sqrt{Z}U$, for $U\sim\mathcal{N}(0,1)$ and $Z\sim \textnormal{Exp}(1/2\lambda^2)$. Let $F_Z(z) = 1-e^{-z/(2\lambda^2)}$ be the CDF of $Z$. Due to independence in the components of $\pmb{c}$, we only need to show that $h(x_i)^2 \sim Z$ for $x_i\sim \mathcal{N}(0,1)$, which is easily observed since
\[
 \mathbb{P}(h(x_i)^2\leq x) = \mathbb{P}\left(x_i\leq \Upsilon^{-1}\left(F_Z(x)\right)\right) = F_Z(x). \qedhere
\]
\end{proof}

\section{\textcolor{black}{Acknowledgements}}
\textcolor{black}{We would like to thank all anonymous reviewers for their thoughtful comments, which helped us improve this manuscript.}

\bibliographystyle{IEEEtran}
\bibliography{main}

\begin{thebibliography}{10}
\providecommand{\url}[1]{#1}
\csname url@samestyle\endcsname
\providecommand{\newblock}{\relax}
\providecommand{\bibinfo}[2]{#2}
\providecommand{\BIBentrySTDinterwordspacing}{\spaceskip=0pt\relax}
\providecommand{\BIBentryALTinterwordstretchfactor}{4}
\providecommand{\BIBentryALTinterwordspacing}{\spaceskip=\fontdimen2\font plus
\BIBentryALTinterwordstretchfactor\fontdimen3\font minus \fontdimen4\font\relax}
\providecommand{\BIBforeignlanguage}[2]{{%
\expandafter\ifx\csname l@#1\endcsname\relax
\typeout{** WARNING: IEEEtran.bst: No hyphenation pattern has been}%
\typeout{** loaded for the language `#1'. Using the pattern for}%
\typeout{** the default language instead.}%
\else
\language=\csname l@#1\endcsname
\fi
#2}}
\providecommand{\BIBdecl}{\relax}
\BIBdecl

\bibitem{Compressive_Sensing}
S.~Foucart and H.~Rauhut, \emph{A Mathematical Introduction to Compressive Sensing}.\hskip 1em plus 0.5em minus 0.4em\relax Springer New York, 2013.

\bibitem{fast_ISTA}
A.~Beck and M.~Teboulle, ``A {F}ast {I}terative {S}hrinkage-{T}hresholding {A}lgorithm for {L}inear {I}nverse {P}roblems,'' \emph{SIAM {J}ournal on {I}maging {S}ciences}, vol.~2, no.~1, pp. 183--202, Jan 2009.

\bibitem{BCS}
S.~Ji, Y.~Xue, and L.~Carin, ``Bayesian {C}ompressive {S}ensing,'' \emph{IEEE Transactions on Signal Processing}, vol.~56, no.~6, pp. 2346--2356, Jun 2008.

\bibitem{CoSamp}
D.~Needell and J.~A. Tropp, ``Co{S}a{MP}: Iterative signal recovery from incomplete and inaccurate samples,'' \emph{Applied and Computational Harmonic Analysis}, vol.~26, no.~3, pp. 301--321, May 2009.

\bibitem{ISTA}
I.~Daubechies, M.~Defrise, and C.~De~Mol, ``An {I}terative {T}hresholding {A}lgorithm for {L}inear {I}nverse {P}roblems with a {S}parsity {C}onstraint,'' \emph{Commun. Pure Appl. Math}, vol.~57, no.~11, pp. 1413--1457, Nov 2004.

\bibitem{l1ls}
S.-J. Kim, K.~Koh, M.~Lustig, S.~Boyd, and D.~Gorinevsky, ``An {I}nterior-{P}oint {M}ethod for {L}arge-{S}cale $\ell_1$-{R}egularized {L}east {S}quares,'' \emph{IEEE Journal of Selected Topics in Signal Processing}, vol.~1, no.~4, pp. 606--617, Dec 2007.

\bibitem{needell2010signal}
D.~Needell and R.~Vershynin, ``Signal {R}ecovery {F}rom {I}ncomplete and {I}naccurate {M}easurements {V}ia {R}egularized {O}rthogonal {M}atching {P}ursuit,'' \emph{IEEE Journal of Selected Topics in Signal Processing}, vol.~4, no.~2, pp. 310--316, 2010.

\bibitem{HB-MAP}
R.~G. Raj, ``A hierarchical {B}ayesian-{MAP} approach to inverse problems in imaging,'' \emph{Inverse Problems}, vol.~32, no.~7, p. 075003, Jul 2016.

\bibitem{fast_HB-MAP}
J.~McKay, R.~G. Raj, and V.~Monga, ``Fast stochastic hierarchical {B}ayesian map for tomographic imaging,'' in \emph{51st Asilomar Conference on Signals, Systems, and Computers}.\hskip 1em plus 0.5em minus 0.4em\relax IEEE, Oct 2017, pp. 223--227.

\bibitem{Scale_Mixtures}
M.~J. Wainwright and E.~P. Simoncelli, ``Scale {M}ixtures of {G}aussians and the {S}tatistics of {N}atural {I}mages.'' in \emph{Advances in Neural Information Processing Systems}, vol.~12.\hskip 1em plus 0.5em minus 0.4em\relax MIT Press, 1999, pp. 855--861.

\bibitem{Wavelet_Trees}
M.~J. Wainwright, E.~P. Simoncelli, and A.~S. Willsky, ``Random {C}ascades on {W}avelet {T}rees and {T}heir {U}se in {A}nalyzing and {M}odeling {N}atural {I}mages,'' \emph{Applied and Computational Harmonic Analysis}, vol.~11, no.~1, pp. 89--123, 2001.

\bibitem{chance2011information}
Z.~Chance, R.~G. Raj, and D.~J. Love, ``Information-theoretic structure of multistatic radar imaging,'' in \emph{IEEE RadarCon (RADAR)}, 2011, pp. 853--858.

\bibitem{reconnet}
K.~Kulkarni, S.~Lohit, P.~Turaga, R.~Kerviche, and A.~Ashok, ``Recon{N}et: {N}on-{I}terative {R}econstruction of {I}mages {F}rom {C}ompressively {S}ensed {M}easurements,'' in \emph{Proceedings of the IEEE Conference on Computer Vision and Pattern Recognition}, 2016, pp. 449--458.

\bibitem{he2020iRadonMAP}
J.~He, Y.~Wang, and J.~Ma, ``{Radon Inversion via Deep Learning},'' \emph{{IEEE Transactions on Medical Imaging}}, vol.~39, no.~6, pp. 2076--2087, 2020.

\bibitem{jin2017FBPConvNet}
K.~H. Jin, M.~T. McCann, E.~Froustey, and M.~Unser, ``{Deep Convolutional Neural Network for Inverse Problems in Imaging},'' \emph{{IEEE Transactions on Image Processing}}, vol.~26, no.~9, pp. 4509--4522, 2017.

\bibitem{wang2020deep}
G.~Wang, J.~C. Ye, and B.~De~Man, ``Deep learning for tomographic image reconstruction,'' \emph{Nature Machine Intelligence}, vol.~2, no.~12, pp. 737--748, Dec 2020.

\bibitem{qin2018convolutional}
C.~Qin, J.~Schlemper, J.~Caballero, A.~N. Price, J.~V. Hajnal, and D.~Rueckert, ``Convolutional {R}ecurrent {N}eural {N}etworks for {D}ynamic {MR} {I}mage {R}econstruction,'' \emph{IEEE Transactions on Medical Imaging}, vol.~38, no.~1, pp. 280--290, Jan 2019.

\bibitem{ganbora}
A.~Bora, A.~Jalal, E.~Price, and A.~G. Dimakis, ``Compressed {S}ensing using {G}enerative {M}odels,'' in \emph{Proceedings of the International Conference on Machine Learning}, vol.~70, 2017, pp. 537--546.

\bibitem{liang2020deep}
D.~Liang, J.~Cheng, Z.~Ke, and L.~Ying, ``Deep {M}agnetic {R}esonance {I}mage {R}econstruction: {I}nverse {P}roblems {M}eet {N}eural {N}etworks,'' \emph{IEEE Signal Processing Magazine}, vol.~37, no.~1, pp. 141--151, 2020.

\bibitem{lucas2018using}
A.~Lucas, M.~Iliadis, R.~Molina, and A.~K. Katsaggelos, ``Using {D}eep {N}eural {N}etworks for {I}nverse {P}roblems in {I}maging: {B}eyond {A}nalytical {M}ethods,'' \emph{IEEE Signal Processing Magazine}, vol.~35, no.~1, pp. 20--36, 2018.

\bibitem{learned_ISTA}
K.~Gregor and Y.~LeCun, ``Learning fast approximations of sparse coding,'' in \emph{Proceedings of the 27th International Conference on Machine Learning}, 2010, pp. 399--406.

\bibitem{algorithm_unrolling}
V.~Monga, Y.~Li, and Y.~C. Eldar, ``Algorithm {U}nrolling: Interpretable, efficient deep learning for signal and image processing,'' \emph{IEEE Signal Processing Magazine}, vol.~38, no.~2, pp. 18--44, Mar 2021.

\bibitem{MADUN}
J.~Song, B.~Chen, and J.~Zhang, ``Memory-{A}ugmented {D}eep {U}nfolding {N}etwork for {C}ompressive {S}ensing,'' in \emph{Proceedings of the 29th ACM International Conference on Multimedia}, Oct 2021, pp. 4249--4258.

\bibitem{zhang2018ista}
J.~Zhang and B.~Ghanem, ``{ISTA}-{N}et: {I}nterpretable {O}ptimization-{I}nspired {D}eep {N}etwork for {I}mage {C}ompressive {S}ensing,'' in \emph{Proceedings of the IEEE Conference on Computer Vision and Pattern Recognition}, 2018, pp. 1828--1837.

\bibitem{xiang2021FISTANet}
J.~Xiang, Y.~Dong, and Y.~Yang, ``{FISTA-Net: Learning a Fast Iterative Shrinkage Thresholding Network for Inverse Problems in Imaging},'' \emph{{IEEE Transactions on Medical Imaging}}, vol.~40, no.~5, pp. 1329--1339, May 2021.

\bibitem{learned_proximal_operators}
T.~Meinhardt, M.~Moller, C.~Hazirbas, and D.~Cremers, ``Learning {P}roximal {O}perators: {U}sing {D}enoising {N}etworks for {R}egularizing {I}nverse {I}maging {P}roblems,'' in \emph{Proceedings of the IEEE International Conference on Computer Vision}, 2017, pp. 1781--1790.

\bibitem{deep_priors}
S.~Diamond, V.~Sitzmann, F.~Heide, and G.~Wetzstein, ``Unrolled {O}ptimization with {D}eep {P}riors,'' \emph{arXiv preprint arXiv:1705.08041}, 2019.

\bibitem{zhang2022learn++}
Y.~Zhang, H.~Chen, W.~Xia, Y.~Chen, B.~Liu, Y.~Liu, H.~Sun, and J.~Zhou, ``{LEARN++: Recurrent Dual-Domain Reconstruction Network for Compressed Sensing CT},'' \emph{{IEEE Transactions on Radiation and Plasma Medical Sciences}}, 2022.

\bibitem{su2020iPianoNet}
Y.~Su and Q.~Lian, ``{iPiano-Net: Nonconvex optimization inspired multi-scale reconstruction network for compressed sensing},'' \emph{{Signal Processing: Image Communication}}, vol.~89, p. 115989, 2020.

\bibitem{adler2018LPD}
J.~Adler and O.~{\"O}ktem, ``{Learned Primal-Dual Reconstruction},'' \emph{{IEEE Transactions on Medical Imaging}}, vol.~37, no.~6, pp. 1322--1332, 2018.

\bibitem{APSIPAlyonsrajcheney}
C.~Lyons, R.~G. Raj, and M.~Cheney, ``{CG}-{N}et: {A} {C}ompound {G}aussian {P}rior {B}ased {U}nrolled {I}maging {N}etwork,'' in \emph{2022 IEEE Asia-Pacific Signal and Information Processing Association Annual Summit and Conference}, 2022, pp. 623--629.

\bibitem{Asilomarlyonsrajcheney}
C.~Lyons, R.~G. Raj, and M.~Cheney, ``{A Deep Compound Gaussian Regularized Unfoled Imaging Network},'' in \emph{2022 56th Asilomar Conference on Signals, Systems, and Computers}, 2022, pp. 940--947.

\bibitem{bertero1988linear}
M.~Bertero, C.~De~Mol, and E.~R. Pike, ``Linear inverse problems with discrete data: {II}. {S}tability and regularisation,'' \emph{Inverse Problems}, vol.~4, no.~3, pp. 573--594, Aug 1988.

\bibitem{ying2004tikhonov}
L.~Ying, D.~Xu, and Z.-P. Liang, ``On {T}ikhonov regularization for image reconstruction in parallel {MRI},'' in \emph{Proceedings of the 26th Annual International Conference of the IEEE Engineering in Medicine and Biology Society}, vol.~1.\hskip 1em plus 0.5em minus 0.4em\relax IEEE, 2004, pp. 1056--1059.

\bibitem{bioucas2007new}
J.~M. Bioucas-Dias and M.~A.~T. Figueiredo, ``A {N}ew {T}w{IST}: {T}wo-{S}tep {I}terative {S}hrinkage/{T}hresholding {A}lgorithms for {I}mage {R}estoration,'' \emph{IEEE Transactions on Image Processing}, vol.~16, no.~12, pp. 2992--3004, 2007.

\bibitem{dahl2010algorithms}
J.~Dahl, P.~C. Hansen, S.~H. Jensen, and T.~L. Jensen, ``Algorithms and software for total variation image reconstruction via first-order methods,'' \emph{Numerical Algorithms}, vol.~53, no.~1, pp. 67--92, Jan 2010.

\bibitem{zhang2022stochastic}
Y.~Zhang and C.~Chen, ``Stochastic asymptotical regularization for linear inverse problems,'' \emph{{Inverse Problems}}, vol.~39, no.~1, p. 015007, 2022.

\bibitem{compound_gaussian}
J.~Wang, A.~Dogand\v{z}i\'{c}, and A.~Nehorai, ``Maximum {L}ikelihood {E}stimation of {C}ompound-{G}aussian {C}lutter and {T}arget {P}arameters,'' \emph{IEEE Transactions on Signal Processing}, vol.~54, no.~10, pp. 3884--3898, 2006.

\bibitem{waveform_opt}
Z.~Idriss, R.~G. Raj, and R.~M. Narayanan, ``Waveform {O}ptimization for {M}ultistatic {R}adar {I}maging {U}sing {M}utual {I}nformation,'' \emph{IEEE Transactions on Aerospace and Electronics Systems}, vol.~57, no.~4, pp. 2410--2425, Aug 2021.

\bibitem{portilla2003image}
J.~Portilla, V.~Strela, M.~J. Wainwright, and E.~P. Simoncelli, ``Image {D}enoising {U}sing {S}cale {M}ixtures of {G}aussians in the {W}avelet {D}omain,'' \emph{IEEE Transactions on Image Processing}, vol.~12, no.~11, pp. 1338--1351, Nov 2003.

\bibitem{huang2021deep}
T.~Huang, W.~Dong, X.~Yuan, J.~Wu, and G.~Shi, ``Deep {G}aussian {S}cale {M}ixture {P}rior for {S}pectral {C}ompressive {I}maging,'' in \emph{Proceedings of the IEEE/CVF Conference on Computer Vision and Pattern Recognition}, 2021, pp. 16\,216--16\,225.

\bibitem{goodfellow2016deep}
I.~Goodfellow, Y.~Bengio, and A.~Courville, \emph{Deep {L}earning}.\hskip 1em plus 0.5em minus 0.4em\relax MIT press, 2016.

\bibitem{NN_loss}
H.~Zhao, O.~Gallo, I.~Frosio, and J.~Kautz, ``Loss {F}unctions for {I}mage {R}estoration {W}ith {N}eural {N}etworks,'' \emph{IEEE Transactions on Computational Imaging}, vol.~3, no.~1, pp. 47--57, Mar 2017.

\bibitem{wright2015coordinate}
S.~J. Wright, ``Coordinate descent algorithms,'' \emph{Mathematical Programming}, vol. 151, no.~1, pp. 3--34, Jun 2015.

\bibitem{beck2013convergence}
A.~Beck and L.~Tetruashvili, ``On the {C}onvergence of {B}lock {C}oordinate {D}escent {T}ype {M}ethods,'' \emph{SIAM Journal on Optimization}, vol.~23, no.~4, pp. 2037--2060, Jan 2013.

\bibitem{grippo2000convergence}
L.~Grippo and M.~Sciandrone, ``On the convergence of the block nonlinear {G}auss--{S}eidel method under convex constraints,'' \emph{Operations Research Letters}, vol.~26, no.~3, pp. 127--136, 2000.

\bibitem{boyd2004convex}
S.~Boyd and L.~Vandenberghe, \emph{Convex {O}ptimization}.\hskip 1em plus 0.5em minus 0.4em\relax Cambridge University Press, Mar 2004.

\bibitem{CIFAR10}
A.~Krizhevsky, ``Learning {M}ultiple {L}ayers of {F}eatures from {T}iny {I}mages,'' University of Toronto, Tech. Rep., 2009.

\bibitem{CalTech101}
L.~Fei-Fei, R.~Fergus, and P.~Perona, ``Learning {G}enerative {V}isual {M}odels from {F}ew {T}raining {E}xamples: {A}n {I}ncremental {B}ayesian {A}pproach {T}ested on 101 {O}bject {C}ategories,'' in \emph{Conference on Computer Vision and Pattern Recognition Workshop}, 2004, pp. 178--178.

\bibitem{radontransform}
S.~R. Deans, \emph{The {R}adon {T}ransform and {S}ome of {I}ts {A}pplications}.\hskip 1em plus 0.5em minus 0.4em\relax Dover, 2007.

\bibitem{iglesias2022filter}
M.~A. Iglesias, K.~Lin, S.~Lu, and A.~M. Stuart, ``{Filter Based Methods for Statistical Linear Inverse Problems},'' \emph{arXiv preprint arXiv:1512.01955}, 2022.

\bibitem{ADAM}
D.~P. Kingma and J.~L. Ba, ``Adam: A {M}ethod for {S}tochastic {O}ptimization,'' \emph{International Conference on Learning Representations}, 2015.

\bibitem{auto_diff}
A.~G. Baydin, B.~A. Pearlmutter, A.~A. Radul, and J.~M. Siskind, ``Automatic {D}ifferentiation in {M}achine {L}earning: a {S}urvey,'' \emph{Journal of Machine Learning Research}, vol.~18, 2018.

\bibitem{song2023MAPUN}
J.~Song, B.~Chen, and J.~Zhang, ``{Deep Memory-Augmented Proximal Unrolling Network for Compressive Sensing},'' \emph{{International Journal of Computer Vision}}, vol. 131, no.~6, pp. 1477--1496, Jun 2023.

\bibitem{munson1983tomographic}
D.~C. Munson, J.~D. O'Brien, and W.~K. Jenkins, ``{A Tomographic Formulation of Spotlight-Mode Synthetic Aperture Radar},'' \emph{{Proceedings of the IEEE}}, vol.~71, no.~8, pp. 917--925, Aug 1983.

\bibitem{silvester2000determinants}
J.~R. Silvester, ``Determinants of {B}lock {M}atrices,'' \emph{The Mathematical Gazette}, vol.~84, no. 501, pp. 460--467, 2000.

\bibitem{kulpa1997poincare}
W.~Kulpa, ``The {P}oincar{\'e}-{M}iranda {T}heorem,'' \emph{The American Mathematical Monthly}, vol. 104, no.~6, pp. 545--550, 1997.

\end{thebibliography}

\end{document}